\documentclass[%
 reprint,
 amsmath,amssymb,
 aps,
]{revtex4-2}

\usepackage{psfrag}
\usepackage{color}
\usepackage{subfigure}
\usepackage[utf8]{inputenc}
\usepackage{times}
\usepackage{gensymb}
\usepackage{graphicx}
\usepackage{dcolumn}
\usepackage{bm}
\usepackage{url}
\usepackage{morefloats}
\usepackage[hidelinks]{hyperref}
\usepackage{changes}
\usepackage{comment}
\usepackage{xcolor}
\begin{document}

\preprint{APS/123-QED}

\title{\Large{Deciphering the nature of temperature-induced phases of MAPbBr\textsubscript{3}\\ by {\it ab initio} molecular dynamics}}
\thanks{lavanya@barc.gov.in, vardha@iiserb.ac.in}%

\author
{Sayan Maity$^{1}$, Suraj Verma$^{1}$, Lavanya M. Ramaniah$^{2}$, Varadharajan Srinivasan$^{1}$}

\affiliation{$^{1}$Department of Chemistry, Indian Institute of Science Education and Research Bhopal, Bhopal 462 066, India}
\affiliation{$^{2}$High Pressure and Synchrotron Radiation Physics Division, Bhabha Atomic Research Centre, Trombay, Mumbai 400085, India}






\begin{abstract}
We present an \textit{ab initio} molecular dynamics study of the temperature-induced phases of methylammonium  lead bromide (MAPbBr$_3$). We confirm that the low-temperature phase is not ferroelectric, and rule out the existence of any overall polarization arising from the motion of the individual sub-lattices. Our simulations at room temperature resulted in a cubic \textit{Pm-3m} phase with no discernible local orthorhombic distortions. At low temperatures, such distortions are shown to originate from octahedral scissoring modes,  but they vanish at room temperature. The predicted timescales of MA motion agree very well with experimental estimates, establishing dynamic disordering of the molecular dipoles over several orientational minima at room temperature. We also identify the key modes of the inorganic and organic sub-lattices that are coupled at all temperatures mainly through the N-H$\cdots$Br hydrogen-bonds. Estimated lifetimes of the H-bonds correlate well with MA dynamics indicating a strong connection between these two aspects of organic inorganic hybrid perovskites. We also confirm that, in addition to disordering of MA orientations, the transition to the cubic phase is also associated with displacive characteristics arising from both MA as well as Br ions in the lattice.
\end{abstract}


\pacs{Valid PACS appear here}
\maketitle

\section{Introduction}
Organic-inorganic hybrid perovskites (OIHPs) indicated by chemical formula ABX\textsubscript{3} form an interesting class of systems where an inorganic lattice (B cation and X anion) hosts an organic cation (A) filling space inside the lattice. Recently, OIHPs, including MAPbBr\textsubscript{3} (MA=CH$_3$NH$_3^+$) have shown great promise in solar cell technology due to high photo conversion efficiency (PCE) up to to ~20\% \cite{kojima2006novel,kojima2009organometal,im20116,kim2012lead,lee2012efficient,etgar2012mesoscopic,chen2014planar,jeng2013ch3nh3pbi3,liu2013efficient, noh2013chemical,park2013organometal,liu2014perovskite,kogo2018amorphous}. 
Various interesting dynamical phenomena such as Raman central peak, Rashba effect, formation of indirect tail states, coupling of organic-inorganic moieties and ferroelectricity are reported to be observed in these OIHPs~\cite{etienne2016dynamical,beecher2016direct,guo2017polar,yaffe2017local,wu2019indirect}. These effects are believed to influence the electronic structure leading to enhancement of PCE at room temperature. Thus, understanding the mechanism of these phenomena in OIHPs, particularly through their structural origins, is crucial to further tuning their interesting properties. From the perspective of phase transitions, OIHPs are quite interesting as evident by the presence of non-trivial features such as glassy dynamics and mixed displacive/order-disorder components.~\cite{fabini2016dielectric,simenas2020suppression,letoublon2016elastic}

Despite its lower power conversion efficiency (10.4\%)~\cite{heo2014planar} than the prototypical MAPbI$_3$, MAPbBr$_3$ is a desirable candidate for solar cells given its higher thermal and structural stability~\cite{mali2015highly,mcgovern2020understanding} as well as suitability for tandem architecture~\cite{jacobsson2016exploration} along with other applications.~\cite{li2016carbon,dou2014solution} At ambient pressure, MAPbBr\textsubscript{3} transitions from cubic ($\alpha$) to tetragonal-I ($\beta$) to tetragonal-II ($\gamma$) to orthorhombic ($\delta$) phase \cite{poglitsch1987dynamic, swainson2003phase, chen2015under} as the temperature is reduced. The transitions are associated with increasing disorder of MA \cite{swainson2003phase, mashiyama2007disordered}. The exact nature of the orientational order in the $\delta$ ($<$149.5 K) and $\beta$ (155.1-236.9 K) phase has been under debate with some studies assigning non-polar space groups to these structures~\cite{swainson2003phase, niesner2016giant, lopez2017elucidating}, while others claiming them to be polar~\cite{poglitsch1987dynamic,gesi1997effect,gao2019ferroelectricity}. Notably, a recent study employing a combination of dielectric measurements, pyroelectric current and positive-up-negative-down measurements claimed the existence of polar ferroelectric behavior along $\langle001\rangle$ in the $\beta$ phase, ruling out the assignment of non-polar \textit{I4/mcm} space group~\cite{gao2019ferroelectricity}. However, this claim has been subsequently challenged~\cite{lehmann2021long}, and the observation of pyroelectric current was attributed to the presence of metastable polar states. 
On the other hand, although the high temperature $\alpha$ phase ($>$236.9 K) is described as cubic \textit{Pm-3m} \cite{swainson2003phase,rakita2016ch3nh3pbbr3,lopez2017elucidating,brown2017molecular}, recent studies~\cite{page2016short,bernasconi2017direct} 
identified presence of short-range octahedral distortion at room temperature leading to breaking of cubic symmetry. Thus, the exact nature of the structural phases of MAPbBr$_3$ is as yet unclear, and efforts to reconcile seemingly conflicting experimental evidence are on-going. In this context, while simple models have had limited success~\cite{lahnsteiner2019long} in describing the phase transitions, \textit{ab initio} molecular dynamics (AIMD) based approaches have proven crucial by allowing access to structural and electronic details of OIHPs at high spatial and temporal resolution accurately~\cite{mosconi2014structural,quarti2015structural,quarti2014interplay,carignano2015thermal,lahnsteiner2016room,jinnouchi2019phase}.


\begin{figure*}[tbh!]
 \centering
 \includegraphics[width=1.0\textwidth]{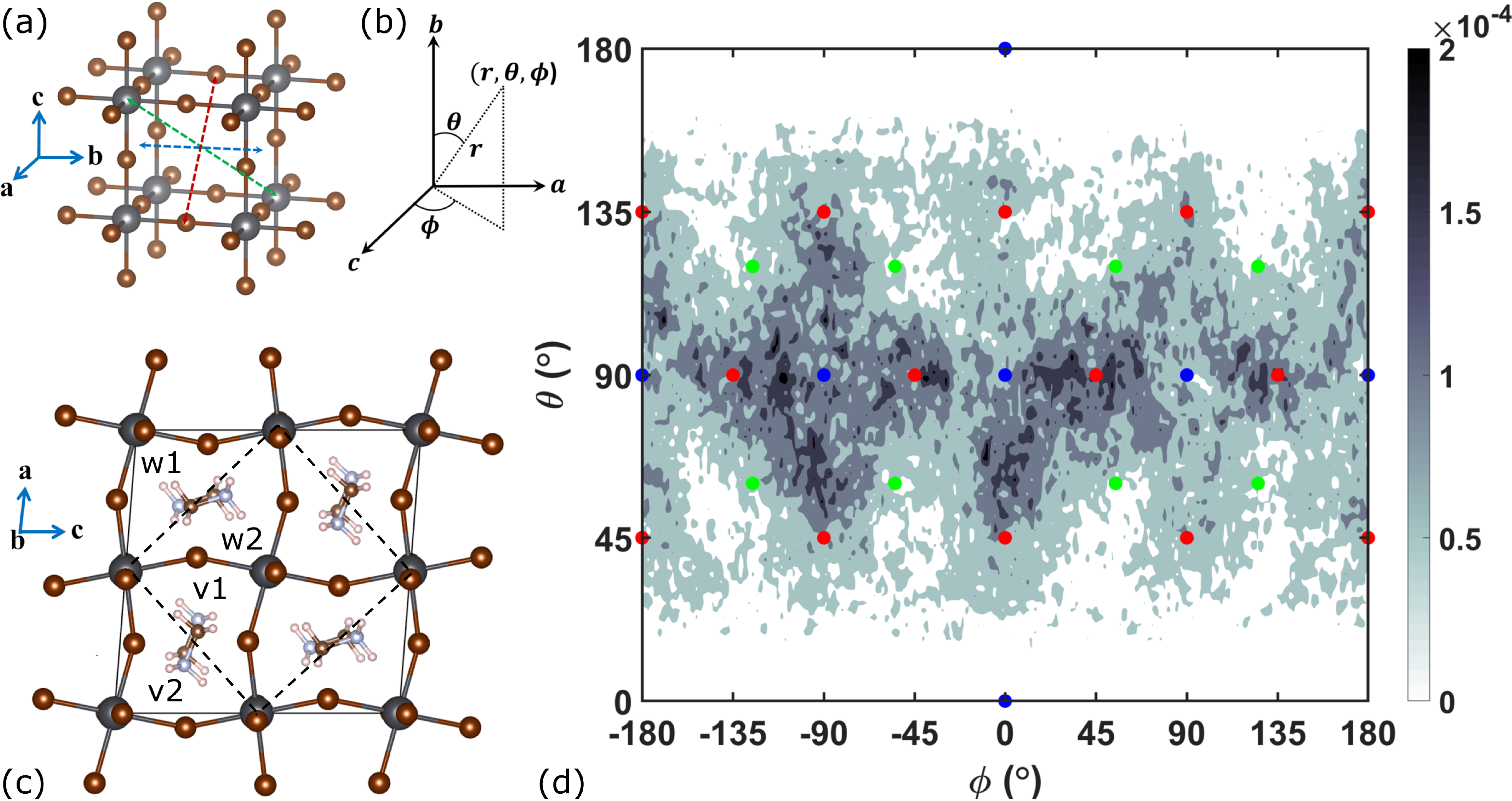}
 \caption{(a) Structure of the cubic cell where different types of high-symmetry MA orientations are shown. Blue, green and red arrow correspond to \textit{ff}, $bd$ and $ed$ orientations as described in the text; (b) Choice of order parameter $\theta$ and $\phi$ in polar coordinate system with respect to local axes; (c) Structure of the 48-atom orthorhombic $Pnma$ unit-cell (marked by the dashed boundary) inside (a portion of) the simulation cell. Grey, brown, blue, small-brown and light-pink balls indicate lead, bromine, nitrogen, carbon and hydrogen atoms, respectively; 
 (d) Orientational ($\theta-\phi$) probability distribution plot of all MA at 300 K. Blue, green and red circles correspond to \textit{ff}, $bd$ and $ed$ orientations} 
 \label{distribution}
\end{figure*}

In this study, we employ AIMD  in the isothermal-isobaric ensemble (NPT)~\cite{car1985unified,bernasconi1995first} to investigate the nature of the temperature-induced phases in MAPbBr$_3$. 
We show that, the low-temperature phase is not ferroelectric, and is correctly described by the orthorhombic space group \textit{Pnma}. We confirm that, while the motion of MA or the inorganic sub-lattice lead to local polar fluctuations at all temperatures, they do not result in an overall polarization. Our simulations at room temperature resulted in a cubic \textit{Pm-3m} phase with no  discernible orthorhombic distortions, in contrast to expectations from recent experiments~\cite{page2016short,bernasconi2017direct}. 
Our study reveals that the scissoring motion of bromide octahedra leads to a multiple peak feature in the Br-Br pair distribution functions at low temperatures which, however, evolves into a single peak at room temperature. Displacements of MA ions couple strongly to octahedral distortions through X-H$\cdots$Br (X=C, N) hydrogen-bonds (H-bonds) evidenced by significant correlation between the relevant order parameters at all temperatures. The timescales of  MA  motion extracted from our simulations are found to be in good agreement with experiments, thereby validating our model, and supporting a dynamic disordering of the molecular dipoles via a combination of local angular fluctuations and reorientational jumps. The latter is only significant at 300 K where N-H$\cdots$Br bonds are frequently broken due to fast rotations about the molecular axis, emphasizing the key role of H-bonds in the dynamics of OIHPs. We also show that, in addition to disordering of MA orientations, the transition to the cubic phase is also associated with displacive characteristics arising from both MA as well as Br ions in the lattice.

\section{Results and discussion}

\subsection{Is MAPbBr$_3$ Ferroelectric?}
The  structure of the MAPbBr$_3$ in any phase is made up of corner-sharing PbBr$_6$ octahedra with an MA ion associated with every (pseudo-)cubic unit. Each MA ion is expected to align along one of 13 high symmetry directions in the cube. These are classified into 3 groups (see Figure~\ref{distribution}(a)) -- {\it face-to-face} ({\it ff}), {\it body diagonal} ({\it bd}) and {\it edge-diagonal} ({\it ed}) -- corresponding to the 3 $C_4$, 4 $C_3$ and 6 $C_2$ rotation axes of the cube, respectively. Along with the relative orientation of the C-N bond in MA, these account for 26 possible orientations~\cite{onoda1990calorimetric}. In the low temperature $\delta$ phase, two groups of MA are oriented approximately along the $\langle100\rangle_c$ and $\langle001\rangle_c$ pseudo-cubic directions, respectively ($\langle101\rangle$ and $\langle10\overline{1}\rangle$ with respect to the orthorhombic axes). In the {\it Pnma} structure, the MA are stacked along the {\it b}-axis in an anti-polar fashion giving rise to a net zero dipole moment in the cell. Our DFT-based optimizations employing the van der Waals (vdW) interaction corrected PBE+D2 exchange-correlation (xc) functional~\cite{John21996,grimme2006semiempirical}  yielded the  {\it Pnma} structure for MAPbBr$_3$ at 0 K in agreement with previous theoretical findings~\cite{swainson2007pressure, sarkar2017role, lehmann2021long}. The non-ferroelectric ground state thus predicted is robust to variation of the vdW scheme employed or use of hybrid functionals (see Supporting Information-SI). Our conclusion agrees with several X-ray diffraction experiments~\cite{swainson2003phase, niesner2016giant, lopez2017elucidating} and with more recent reports of lack of ferroelectric switching~\cite{lehmann2021long} in the orthorhomic phase. It has been suggested that there may exist local minima slightly higher in energy than the {\it Pnma} structure differing from it by small relative rotations and displacements of the MA ions~\cite{lehmann2021long}. While these were ruled out in our 0 K optimization studies, we considered their possible occurrence at higher temperatures through our 0 GPa AIMD simulations at 40 K, 180 K and 300 K (see SI). To this end we identify the crucial symmetry elements characterizing the MA sub-lattice in the {\it Pnma} structure. We first define the polar angles $\theta$ and $\phi$ associated with each MA ion as depicted in Figure~\ref{distribution}(b). The orientational distribution plots at 40 K (Figure S1 in SI) and 180 K (Figure S2) indicate that MA rotations are restricted and centered around four sets of polar angles. Assuming a bivariate normal probability distribution function about each set centered at $\mu = (\mu_\theta,\mu_\phi)$ given by:
\begin{equation}\label{eq:a1}
 \rho(x)=\frac{1}{2\pi}|\zeta|^{-1/2}exp\left[-\frac{1}{2}(x-\mu)^{T}\zeta^{-1}(x-\mu)\right]
\end{equation}
we use Gaussian Mixture Modelling (GMM) to calculate the equilibrium orientations listed in Table S5 (see SI). In the anti-polar ground state (as seen at 0 K), the MA are clustered into two groups, $v$ (along $\langle100\rangle_c$) and $w$ (along $\langle001\rangle_c$), based on the orientation of the MA molecular axis, with an additional index 1 (parallel) or 2 (antiparallel) indicating the alignment of the C$\rightarrow$N vector with this axis. 

Local centrosymmetry between both types ($LCM_{12}$), in every group described above,  exists when $\mu_\theta(1)+\mu_\theta(2)=180^o$~and $\left\vert\mu_\phi(1) - \mu_\phi(2)\right\vert = 180^o$. Similar criteria can also be set for preservation of the four $2_1$ screw axes connecting the $v$ and $w$ groups of MA ($SM_{vw}$) as well as the mirror symmetry elements ($MSM$) of {\it Pnma} (see SI).

At 40 K the $LCM_{12}$ criteria are satisfied for each of the two groups, $v$ and $w$, to within 1$^o$ indicating MA dipole moments will be canceled due to local centrosymmetry. Along with the preservation of $SM_{vw}$ and $MSM$  criteria this result confirms the assignment of  \textit{Pnma} to the $\delta$ phase and, hence, a non-ferroelectric arrangement of MA. We note here that a 768-atom ($4\times4\times4$) supercell was required to ensure sufficient sampling of the electronic Brillouin zone. In particular, 96-atom supercells used previously along with the $\Gamma$-point only approximation led to a spurious polar minimum in MAPbBr$_3$ (see SI). A unimodal distribution of the dihedral angle $\theta_{12}$ between MA 1 and 2 under $v$ or $w$ group confirms absence of any local (intermediate-polar) energy minima close to the non-ferroelectric/anti-polar arrangement ($\theta_{12}$=180$^o$) (see Figure S4) and is in agreement with observations made in a recent MD study by Bokdam \textit{et al.}~\cite{bokdam2021exploring}. The occurrence of such intermediate-polar minima at 0 K in previous studies~\cite{lehmann2021long} is an artifact of fixed cell optimizations, a constraint not imposed by our variable-cell simulations (see SI). While a polar arrangement of MA ($\theta_{12}=$0$^o$) with slightly higher energy ($\approx$~2 meV/atom) as the anti-polar one, is also possible~\cite{sarkar2017role} (see SI), 
it was not accessed in the 40 K simulations. Since both these states have the same octahedral tilting pattern, they are essentially separated by MA rotational barrier which is not accessible at low temperatures.

\begin{figure}[t!]
 \centering
 \includegraphics[width=0.48\textwidth]{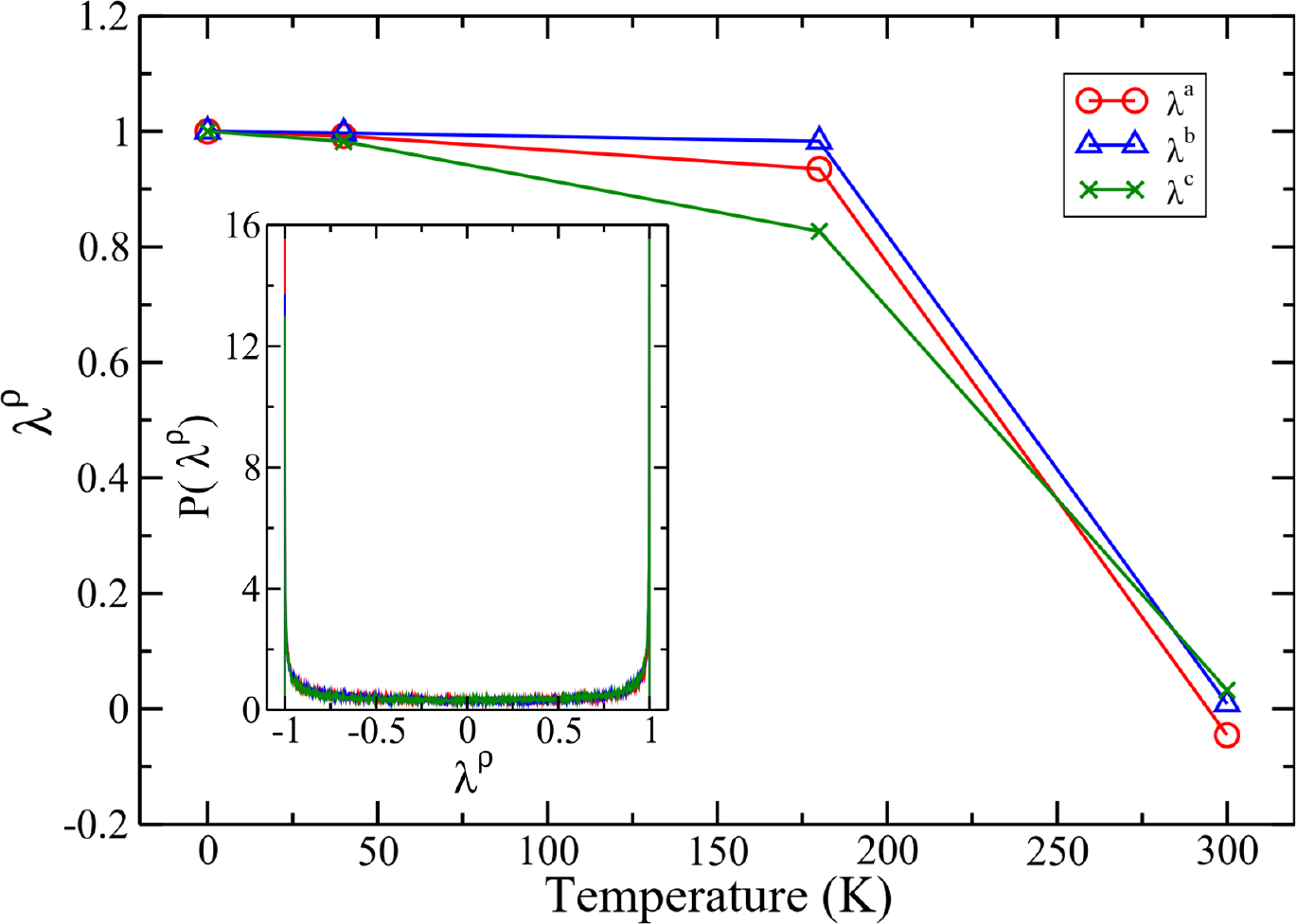}
 \caption{Evolution of order parameter $\lambda$ (time-averaged) with temperature; In inset distributions of $\lambda$ along the three different cubic axes at 300 K are shown.}  
 \label{lambda_all}
\end{figure}

At 180 K, both the $LCM_{12}$ and $MSM$ criteria are satisfied suggesting an overall non-polar arrangement of MAs, similar to MAPbI$_3$~\cite{breternitz2020role}. It should be noted that, to capture the transformation of lattice constants and the ``partial disorder" of the $\beta$ phase, longer simulations are required than presently undertaken. 

At 300 K, the simulations were started on the cubic lattice with MA orientations chosen randomly from the aforementioned 26 possibilities. The orientations subsequently sampled by the AIMD span the whole $\theta$-$\phi$ space (see Figure~\ref{distribution}(d)) with $ed$ orientations being more likely than others, in line with previous experimental expectations~\cite{lopez2017elucidating}. Among all \textit{ff} orientations, those along (pseudo-)cubic $c$ axis ($\theta=\pm 180^o$) are extremely less sampled, thus creating one strip along the $\theta$ axis and four strips along the $\phi$ axis. At every time step, the MA are categorized into groups based on the least angle they make with the aforementioned 26 orientations, thus obtaining a ratio of occurrence of $ed:\textit{ff}: bd = $1:0.55:0.40 indicating the trend in the relative potential depths of these minima. Indeed, a comparison of DFT energies of these orientations in a cubic unit cell revealed that \textit{ed} is the most stable while \textit{ff} and \textit{bd} are higher in energy by 24 meV/f.u. or more (Figure S3). This is in contrast to MAPbI$_3$ where $bd$ orientation is most stable~\cite{motta2015revealing,carignano2015thermal}, indicating that  the mixed halide MAPbI$_x$Br$_{3-x}$ would be statically disordered~\cite{selig2017organic}.

Thermal fluctuations about the anti-polar stacked MA orientations can generate short-lived local polar configurations in some OIHPs~\cite{beecher2016direct,laurita2017chemical,guo2017polar,yaffe2017local}. To identify such configurations, we define the order parameters $\lambda^\rho$ as:
\begin{equation}
\lambda^{\rho} = \frac{1}{T}\sum\limits_{t}\frac{1}{N_u}\sum\limits_{n}-{\rm cos}\left(\Delta_\rho~\phi_{\vec{R}_n}^{\mu\nu}(t)\right)\ \\
\end{equation} 
where, $\phi_{\vec{R}_n}^{\mu\nu}(t)$ is the orientation angle of the n$^{th}$ MA unit in the $\mu\nu$-plane at time $t$, $N_u$ is the number of cubic unit cells, $\rho$-$\mu$-$\nu$ indicate crystal directions and $\Delta_\rho$ is the forward-difference operator along the $\rho$ direction. $-$1 and $+$1 values of $\lambda$ indicate polar and anti-polar configurations for subunits consisting two neighboring MAs. Figure~\ref{lambda_all} shows that anti-polar MA ordering persists in MAPbBr$_3$ in the low temperature phases while at 300 K, both local anti-polar and polar subunits are equally sampled through nearly free MA rotation. Thus, such local configurations may be stabilized at 300 K under external conditions such as electric-field/pressure/strain/surface-reconstruction leading to permanent ferroelectric and anti-ferroelectric domains~\cite{ohmann2015real}. In MAPbBr$_3$, polar fluctuations have been shown experimentally by softening of acoustic phonon in cubic phase in pump-probe reflection spectroscopy~\cite{guo2017polar} and emergence of central peak (CP) in both Raman scattering~\cite{yaffe2017local} and inelastic neutron scattering (INS)~\cite{letoublon2016elastic,hehlen2022pseudospin}. The INS-CP has been attributed to MA relaxation modes (either rotation or center-of-mass rattling of A-site cation) while the Raman CP is shown to occur primarily because of coupled motion of the cation (head-to-head) and Br. Through the analysis presented above, we show that transient polar and anti-polar states also arise due to MA rotational fluctuations. 
As the polar sub-units are short-lived, we rule out formation of polar nano-regions~\cite{bari2021ferroelastic} unlike in MAPbI$_3$~\cite{garten2019existence}. Additionally, off-center displacements of different kinds of MA and Br ions occur in pairs at 40 and 180 K leading to no overall polarization (see SI, Figure S7-S9). Interestingly, these displacements completely vanish at 300 K, indicating that the phase transition has not only MA orientational disordering (pseudospin~\cite{lynden1994translation,yamada1974dynamical}) which appear as secondary order parameters in Landau theory of phase transition~\cite{parlinski1985secondary,marais1991phenomena}, but also MA/Br displacive components (see below). 

\subsection{Static vs. dynamic MA disorder?}
\begin{figure}[t!]
 \centering
 \includegraphics[width=0.48\textwidth]{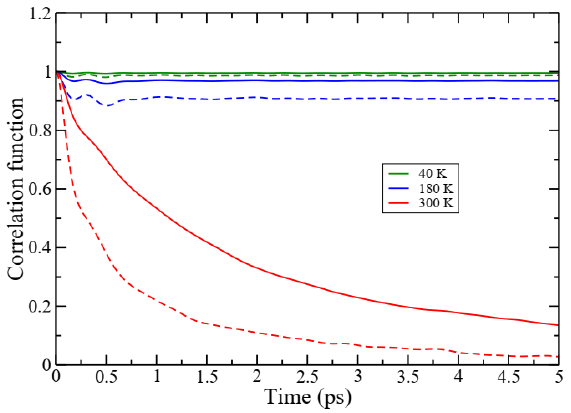}
 \caption{Correlation functions of C-N motion at 40, 180, and 300 K indicated as green, blue, and red curves. The continuous and dashed curves are for $C_1$(t) and $C_2$(t), respectively.}  
 \label{corr_rdf}
\end{figure}

While the MA have been largely believed to show dynamic disorder~\cite{wasylishen1985cation,poglitsch1987dynamic,leguy2016dynamic} at 300 K, Page \textit{et al.}~\cite{page2016short} suggested the possibility of static disorder where cations can show large fluctuations around their local orientations. Dynamic orientational disorder in MA is confirmed here by the diffusive nature of the rotational autocorrelation function (ACF) $C_1(t)$ at 300 K shown in Figure~\ref{corr_rdf}. 
The relevant time-scales can be extracted using a general diffusive model proposed by Mattoni \textit{et al.}~\cite{mattoni2015methylammonium}
\begin{equation}\label{eq:b3}
 \begin{aligned}
 C_1^{M}(t)=&e^{-t/\tau_{0}} \left[ A_{1} + \frac{t}{\tau_{0}} e^{-(t/\tau_{1})^n} + (1-A_{1}) \right. \\ 
 &\left. cos\left (\frac{2\pi}{\tau_{2}}t \right) \frac{1}{(t/\tau_{3}+1)^m} \right] 
 \end{aligned}
\end{equation}
where $\tau_{0}$ controls the overall exponential decay (particularly that observed at 300 K), $\tau_1$ is the decay time for nearly free rotations occurring at short times, and  $\tau_{2}$ describes the time-scale for harmonic motion with a power-law damping governed by $\tau_{3}$. At 40 and 180K, the ACF show almost no long time decay ($\tau_{0}\to\infty$) confirming their ordered behavior. Note that in the latter phase, reorientation of MA ions leading to partial disorder is expected but could involve larger (but finite) $\tau_{0}$~\cite{mattoni2015methylammonium} and hence not seen in our 20 ps long simulations. 
At 300 K, the timescales extracted for the fastest ($\tau_1$=341 fs) 
and slowest motions ($\tau_0$=2.035 ps) 
match well with the ssNMR ($\tau_1$=355 fs)~\cite{wasylishen1985cation} and TR-OKE ($\sim$2 ps) experiments~\cite{zhu2016screening}. 

\begin{figure*}[t!]
 \centering
 \subfigure[]{\includegraphics[height=0.38\textwidth]{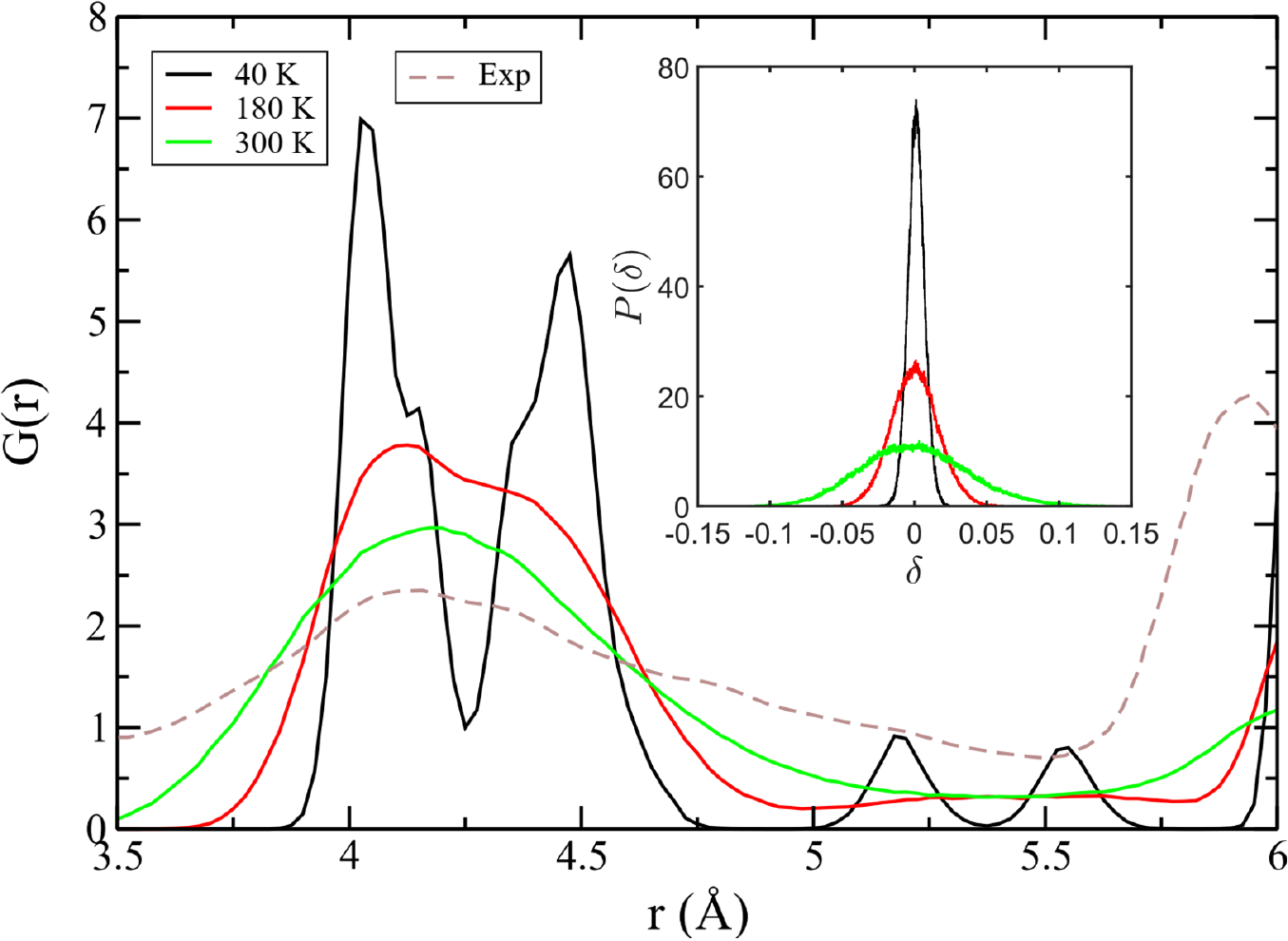}}\hspace{0.05\textwidth}
 \subfigure[]{\includegraphics[height=0.38\textwidth]{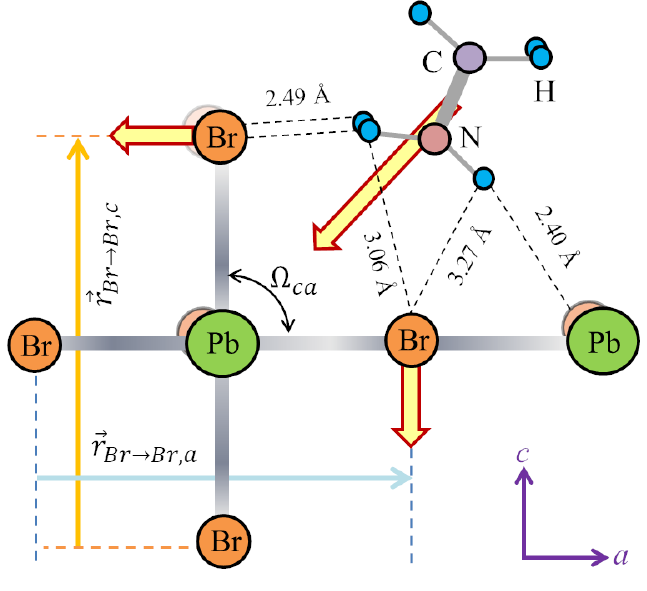}} 
 \caption{(a) Evolution of the calculated radial distribution functions of Br-Br pair (G(r)\{Br-Br\}) with temperature in 3.5-5~${\rm \AA}$~region (continuous lines) along with the total RDF at room temperature taken from experiment~\cite{bernasconi2017direct} (broken line). Inset depicts the evolution of distributions of $\delta$ as a measure of tetragonal distortion of octahedra with temperature; (b) Schematic description of the parameters used to calculate $\delta$ and $\Omega$ distortions. The coupling motion between $\Omega_{ca}$ and translation of MA in \textit{ca} plane is also shown. The distances between H$^{...}$Br, shown here for reference, are taken from the 0 K optimized structure.}  
 \label{rdf_tetragonal_distortion}
\end{figure*}

It has been shown in MAPbI$\textsubscript{3}$ that the dynamics of MAs are governed by a fast `wobbling-in-a-cone' (corresponding to local angular fluctuations) and a slow `jump-like' reorientation motion~\cite{bakulin2015real}. In our 300 K simulation of MAPbBr$\textsubscript{3}$, the dynamical disordering of MA orientations over multiple minima (FIG.~\ref{distribution}(d)) indicate similar behavior captured by the wobbling-in-a-cone/jump model~\cite{ji2011orientational,lipari1982model} for which the second order ACF $C_2(t)$ is given by
\begin{equation}\label{eq:b4}
C_2^{WCM}(t)=S^2e^{-t/\tau_0}+(1-S^2)e^{-(\frac{1}{\tau_0}+\frac{1}{\tau_1})t}
\end{equation}  
where $\tau_0$ and $\tau_1$ are relaxation times corresponding to the re-orientational jump and initial fast motion around a minimum, respectively. $S$ is a generalized order parameter measuring the degree of spatial restriction of motion. The computed $C_2(t)$ can be fit to Eq.~\ref{eq:b4} to extract $\tau_0=1.86$ ps and $\tau_1=320$ fs in excellent agreement with corresponding time-scales (1.5$\pm$0.3 ps,  300$\pm$100 fs) extracted from polarization resolved 2-D IR spectroscopic measurements~\cite{selig2017organic}, where anisotropy decay is directly proportional to $C_2(t)$~\cite{lin2010calculation}. Furthermore, $\tau_0$ and $\tau_1$ predicted by both the models are in agreement in line with their expected equivalence.~\cite{mattoni2016modeling}. We also reproduce the experimentally suggested Glazer notations~\cite{glazer1972classification} for different phases (see SI), thus confirming the validity of our model within the simulation timescales.

\subsection{Is the $\alpha$ phase distorted?}
Although the room temperature phase is largely understood to be cubic, some recent studies have suggested the persistence of orthorhombic distortions in this phase manifesting either as orthorhombic lattice parameters~\cite{page2016short} or as distorted PbBr\textsubscript{6} octahedra~\cite{bernasconi2017direct}. In either case, the primary evidence is obtained by fitting the radial distribution functions (RDF) from X-ray diffraction data to either a pseudo-cubic (PC) or an orthorhombic (OR) lattice models. In order to verify the possibility of such distortions at 300 K, we estimated the lattice parameters in two ways: by measuring the nearest neighbor Pb-Pb distances (PC) and by using a $\sqrt2\times1\times\sqrt2$ transformed cell tensor (OR) (see SI). Both the time-averaged PC ($a=$5.978$\pm$0.008 \AA, $b=$5.987$\pm$0.011 \AA, $c=$5.989$\pm$0.016 \AA) and OR ($a=$8.492$\pm$0.060 \AA, $b=$11.973$\pm$0.022 \AA, $c=$8.455$\pm$0.057 \AA)
lattice parameters confirm the absence of any significant deviation from cubic symmetry. Thus, cubic models should yield a better fit with experimental data since the presence of local distortions do not manifest through the lattice constant values. 

Orthorhombic distortions in the high temperature $\alpha$-phase were also inferred in experiments from the persistence of a doublet feature in the Br-Br RDF in the 3.5-5~${\rm \AA}$ region~\cite{bernasconi2017direct}. The evolution of G(r)\{Br-Br\} with temperature is shown in Figure~\ref{rdf_tetragonal_distortion}(a) where this doublet is clearly observed up to 180 K. This feature evolves into a broad peak around 4.2~${\rm\AA}$ at 300 K where the resolution of a doublet, if present, is difficult warranting further investigation into its origin. In any phase, the Br-Br doublet indicates the presence of two distinct Br-Br distances in PbBr\textsubscript{6} octahedra which can arise due to two primary reasons - a tetragonal distortion defined as 
\begin{equation}
\delta(j,t)=\frac{2\times|\overrightarrow{r}_{Br \to Br, b}(j,t)|}{|\overrightarrow{r}_{Br \to Br, a}(j,t)|+|\overrightarrow{r}_{Br \to Br, c}(j,t)|}-1 
\end{equation}
or a scissoring distortion defined as 
\begin{equation}
\Omega_{\mu\nu}(j,t)=\hat{r}_{Br \to Br, \mu}(j,t) \cdot \hat{r}_{Br \to Br, \nu}(j,t)
\end{equation}
Here, $\overrightarrow{r}_{Br \to Br, \mu}(j,t)$ is the vector connecting two Br at opposite vertices of the octahedra situated approximately along the crystal axis $\mu$ (see Figure~\ref{rdf_tetragonal_distortion}(b)). Such scissoring distortions would also influence electronic properties through large displacement of Br~\cite{bird2021large,gehrmann2022transversal}.

\begin{figure*}[t!]
 \centering
 \includegraphics[width=1.0\textwidth]{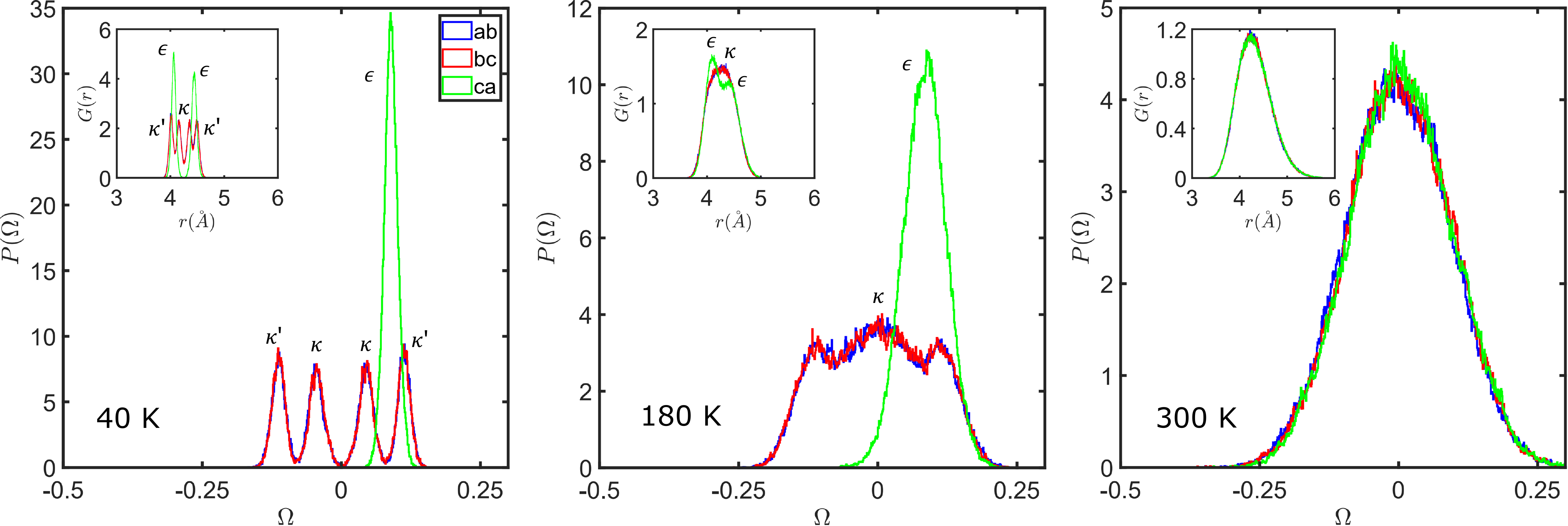}
 \caption{Distribution of $\Omega_{ab}$, $\Omega_{bc}$ and $\Omega_{ca}$ as measure of scissoring distortion of octahedra at 40, 180 and 300 K. Plane-wise radial distribution functions arising due to $\Omega$ are shown in insets.}  
 \label{scissoring_distortion}
\end{figure*}

Distributions of $\delta$ at all temperatures are unimodal and centered at 0 (inset in Figure~\ref{rdf_tetragonal_distortion}(a)), indicating the absence of any tetragonal distortion at all temperatures and thus ruling it out as a cause of Br-Br doublet. On the other hand, distributions of the scissoring distortion parameters clearly reveal the origin of doublet-like features. Figure~\ref{scissoring_distortion} shows the distribution of $\Omega_{\mu\nu}(j,t)$ at various temperatures with each peak at a non-zero value corresponding to a Br-Br doublet. At 40 K, there are 5 such peaks whose contributions to the RDF (shown in in the corresponding inset) results in a doublet-like feature emerging primarily from the $\epsilon$ peak and partially from $\kappa'$.  A small splitting of every peak in the doublet 
occurs due to $\kappa$. Along $a$ axis, $\kappa$ and $\kappa'$ sit alternately, while their supplementary counterparts (with $\Omega_{ab} < 0$) are situated along $b$ axis. At 180 K, all the supplementary $\kappa$ peaks merge at $\Omega =$ 0 and only the $\epsilon$ peak at $\Omega \ne$ 0 yields a Br-Br doublet (second panel in Figure~\ref{scissoring_distortion}). However at 300 K, all the distributions of $\Omega$ are unimodal having most probable values at 0 (third panel in Figure~\ref{scissoring_distortion}). Assignment to a single peak was confirmed by a 1-D GMM analysis which found that each of the distributions at 300 K fits best with only a single gaussian component with its mean at zero (see Figure S12). Thus, we rule out presence of any multiplet in these distributions thereby confirming the absence of persistent distortions at 300 K.

\subsection{Nature of the MA/lattice coupling?}
\begin{figure*}[t!]
 \centering
 \subfigure[]{\includegraphics[height=0.34\textwidth]{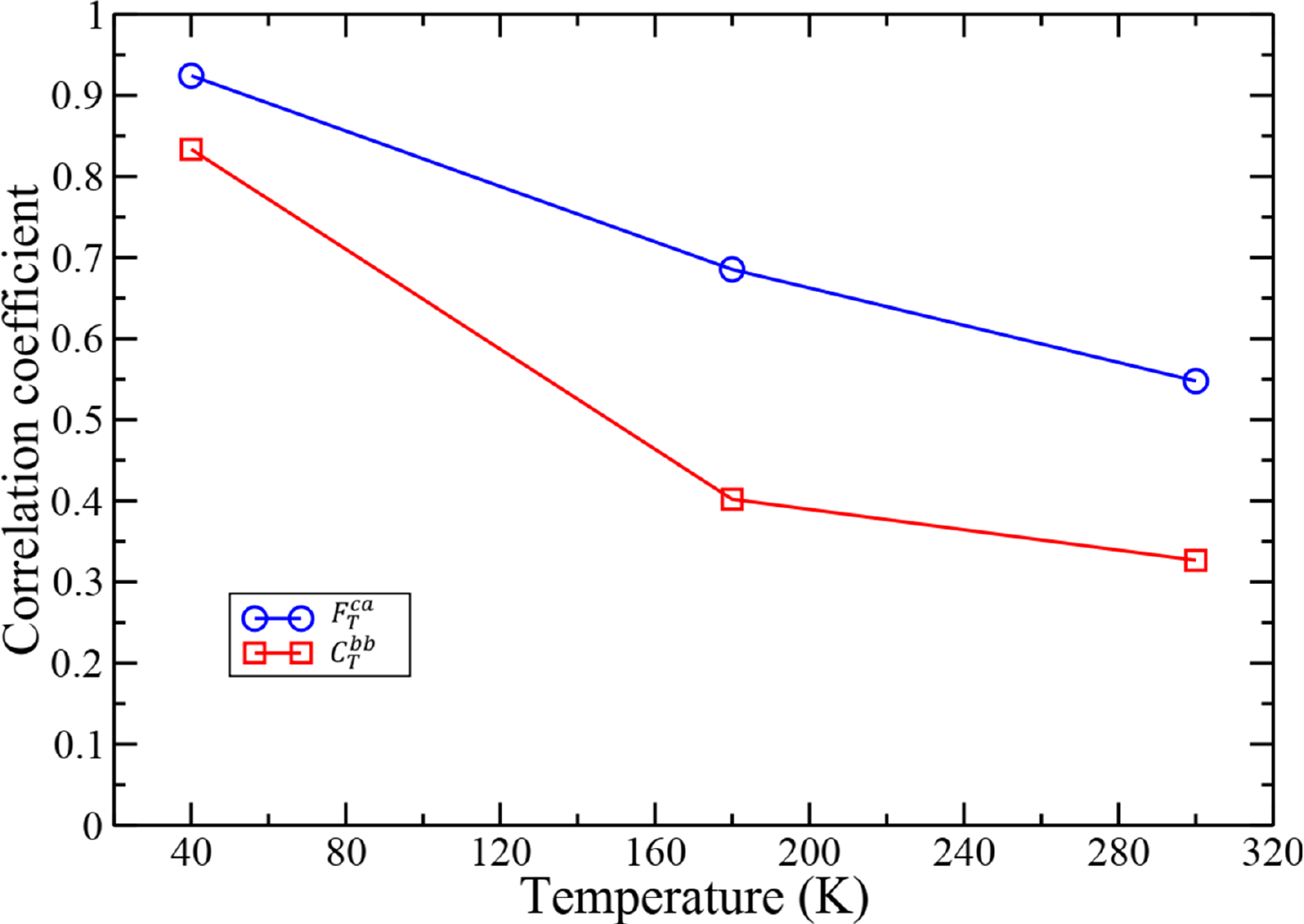}}\hspace{0.02\textwidth}
 \subfigure[]{\includegraphics[height=0.34\textwidth]{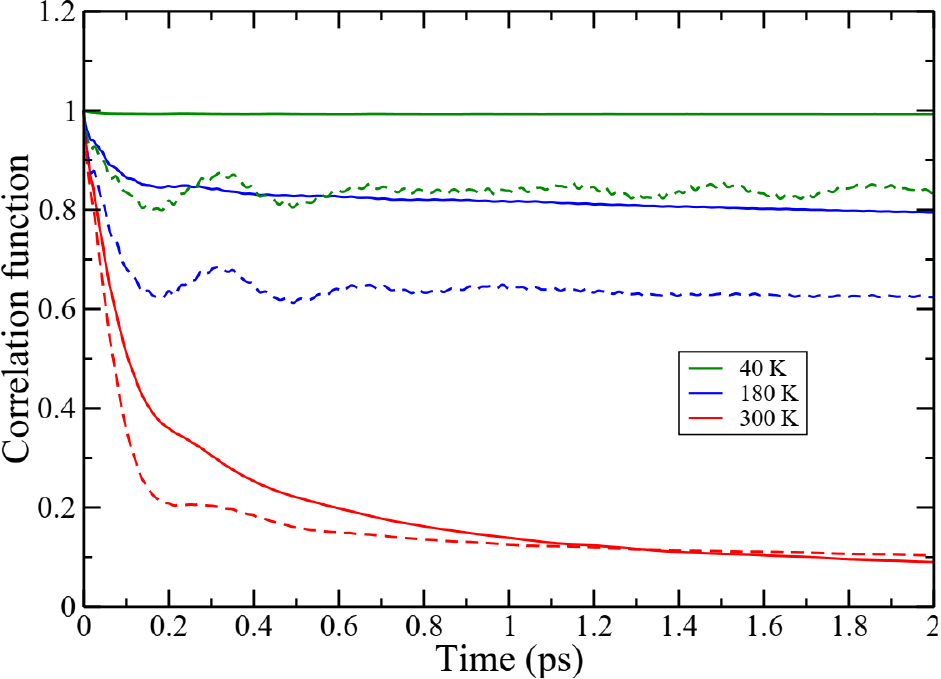}} 
 \caption{(a) $C^{bb}_T$ as a measure of coupling of MA translation and Pb-Br bond length along \textit{b} axis; and $F^{ca}_T$ as a measure of coupling of MA scissoring and MA translation in the \textit{ac} plane. (b) Time correlation functions of H-bond framework at 40, 180, and 300 K indicated as green, blue, and red curves. The continuous and dashed curves are for N-H$^{...}$Br and C-H$^{...}$Br, respectively.} 
 \label{coupling}
\end{figure*}

The coupling between the MA and lattice motion was investigated next. To this end, we computed the correlation coefficients C$^{\mu\nu}_T$ between the length of a Pb-Br bond lying along the $\mu$ direction and  the displacement (from average) of the MA in the $\nu$ direction within the same cubic unit cell. Figure~\ref{coupling}(a) shows that the coupling of Pb-Br distances and MA translations along the $b$ axis ($C^{bb}_{T}$) is reduced with temperature but remains significant even at 300 K. In contrast, we found very little correlation between Pb-Br distances and MA rotations (see Table S10). We also computed the correlation coefficients $F^{\mu\nu}_{T}$ between scissoring distortions $\Omega_{\mu\nu}$ and MA translations in the $\mu\nu$ plane. Interestingly, the scissoring motion in \textit{ac} plane (perpendicular to the MA stacking axis) and MA translation along the \textit{ac} edge-diagonal direction $\left[202\right]$ are strongly coupled ($F^{ca}_{T}$) at 40 K (see Figure~\ref{coupling}(a)) through the H-bonding interactions between the organic and inorganic sub-lattices. As illustrated in Figure~\ref{rdf_tetragonal_distortion}(b), if $\Omega_{ac}$ decreases, the associated scissoring angle increases, thus forcing the MA to translate along the \textit{ac} edge-diagonal direction towards the origin to maintain the H$^{...}$Br bonds and vice versa. While significant coupling behavior is also seen at 180 and 300 K, this correlation becomes lesser in the cubic phase due to frequent H-bond breaking effected via both MA reorientational jumps as well as rotation about the molecular axis. Note that at 300 K, as all axes are equivalent, all $F^{\mu\nu}_{T}$ become similar (Table S11). Thus, MA translations couple with the lattice scissoring in the \textit{ac} plane, whereas translations perpendicular to the plane couple with the corresponding Pb-Br bond stretch mode. Moreover, the MA reorientation is facilitated by the scissoring distortion as suggested by similar timescales (Figure S16(a)) of both motions at 300 K. To check this further, a model calculation was performed using a cubic 2$\times$2$\times$2 CsPbBr$_3$ supercell in which one Cs was replaced by MA. The MA rotational energy barrier was then calculated without ($\Omega=0$) and with ($\Omega\neq0$) scissoring distortion. The barrier from a \textit{ff} to \textit{ed} orientation increases when $\Omega > 0$ and decreases when $\Omega < 0$ compared to the undistorted lattice (see Figure S16(b)). This indicates that MA reorientation can be facilitated or delayed depending on the local distortion dynamics. Thus, the decrease in MA reorientation timescales in the series MAPbX$_3$ (X=I, Br, Cl) is related to the blue-shift of low-frequency lattice modes~\cite{niemann2016halogen,leguy2016dynamic} and not due to H-bond energetics (see SI for justification) as previously proposed~\cite{selig2017organic,gallop2018rotational}. We also note that, at 300 K, Pb-Br bond-length fluctuations on neighboring unit cells are strongly positively correlated (see Figure S17(b), Tables S13) strengthening the validity of the PC model for lattice parameters.

Since H-bonds enable the coupling of the two sub-lattices~\cite{bernasconi2018ubiquitous,singh2020origin}, we also investigated their dynamics. At 40 K N-H$^{...}$Br and C-H$^{...}$Br~\cite{bernasconi2018ubiquitous,capitani2017locking} type H-bonds form with H$^{...}$Br distances up to 2.75 and 3.17~${\rm \AA}$, respectively (See Figure S18). At 300 K, where dynamic disordering causes H-bonds to break and reform, a peak-like feature for N-H$^{...}$Br RDF at $\sim$2.46~${\rm \AA}$ confirms the persistence of the H-bond, while the absence of any structure in the C-H$^{...}$Br RDF suggests that the latter interaction weakens significantly~\cite{yin2017hydrogen}. Notably, the equilibrium length of the N-H$^{...}$Br H-bond does not change significantly with temperature~\cite{bernasconi2018ubiquitous}. To probe the dynamical behavior, we construct the following correlation function
\begin{equation}
 C_{X-H\cdots Br}(t)=\frac{1}{N}\sum_{i=1}^{N} \langle \mathbf{H}_i(0) \cdot \mathbf{H}_i(t) \rangle
\end{equation}
where $\mathbf{H}(t)=[\delta_{X-H\cdots Br_1},....,\delta_{X-H\cdots Br_{12}}]$, and $\delta_{X-H\cdots Br}$ equals 1 if the $H\cdots Br$ distance is within the aforementioned distance cutoffs for each type of H-bond or 0 otherwise. Figure~\ref{coupling}(b) shows that at all temperatures, the timescale of rotation about MA axis, indicated by the initial rapid decay of $C_{X-H\cdots Br}(t)$, is 0.16 - 0.18 ps in good agreement with experimentally measured value of 0.2 ps~\cite{bernard2018methylammonium}. There is no long time decay component in the correlations at 40 and 180 K confirming an ordered H-bonded framework. At 300 K the MA reorientational jumps significantly reduce the H-bond lifetimes thereby adding a slower component ($\sim$2 ps). It should be noted that, at all temperatures, the $C_{C-H^{...}Br}$ functions relax more than $C_{N-H^{...}Br}$ at the fast timescale indicating weaker H-bonding interactions for C than N.

\subsection{Nature of phase transition?}
Phase transitions in halide perovskites have been interpreted primarily through either order/disorder or displacive mechanisms~\cite{fontana1990displacive,stern2004character,bussmann2009precursor,yashima2009structural}. In MAPbBr$_3$, their (co-)existence is debated~\cite{swainson2015soft,letoublon2016elastic,hehlen2022pseudospin}. One way to distinguish between these two characters is through the distributions of order parameters derived from the atomic positions. For instance, in a simple double-well model~\cite{dove1997theory} an order-disorder transition would be associated with the conversion of a unimodal, off-centered distribution of the order parameters at low temperatures to a bimodal distribution at high temperatures. In contrast, displacive transitions would result in the centering of the low temperature off-centered unimodal distributions across the transition.

Having established the presence of MA orientational order-disorder component, we next briefly discuss the existence of displacive character in the transitions. In MAPbBr$_3$, the evolution of MA positions (Figure S7) from off-centered (40 and 180 K)  to centered distributions (300 K) indicates the existence of MA-displacive feature during the high-temperature transition. Similarly, centered distributions of Br positions (Figure S9) at 300 K indicate Br-displacive character. Interestingly, the x-component displacement of Br situated along $a$ axis goes from non-zero at 40 K to zero at 180 K, indicating a possible Br-displacive feature at the mid-temperature transition too. Additionally, displacive transitions are also associated with softening of a low-energy phonon mode in a pretransitional state~\cite{milesi2020archetypal}. In MAPbBr$_3$, some low-energy octahedral bending modes~\cite{ferreira2020direct} at zone boundary $M$ (1/2, 1/2, 0) and $R$ (1/2, 1/2, 1/2) points are expected to soften during the transition to the high-temperature cubic phase to justify a displacive behavior~\cite{letoublon2016elastic,hehlen2022pseudospin}. Although, such mode-softening was seen in INS experiments~\cite{swainson2015soft}, subsequent experiments ruled it out~\cite{letoublon2016elastic,hehlen2022pseudospin}. To this end, we calculated amplitudes of octahedral bending modes such as scissoring-
\begin{equation}
\Omega^{\prime}_{\mu\nu}(j,t)=D^{\mu}_{\nu}(j,t)+D^{\nu}_{\mu}(j,t)
\end{equation}
and rocking-
\begin{equation}
\xi_{\mu\nu}(j,t)=D^{\mu}_{\nu}({j,t})-D^{\nu}_{\mu}(j,t)
\end{equation}
where $D^{\mu}_{\nu}(j,t)$ is the $\nu$-component of the displacement of the Br situated along $\mu$ in the $j^{\rm th}$ unit cell at time $t$. It should be noted that both $\Omega_{\mu\nu}$ and $\Omega^{\prime}_{\mu\nu}$ are equivalent as evidenced by their similar power spectra (see Figure S19). 
Interestingly, all three rocking modes $\xi_{\mu\nu}$ show clear softening in the cubic phase (see Figure S20) confirming the associated phase transition has octahedral displacive components as well.

It is interesting to compare the phase transition features of MAPbBr$_3$ and CsPbBr$_3$, where the latter lacks orientational degrees of freedom at the A-site. Similar to the case of MA, Cs displacement shows displacive behavior : the off-center positions of Cs in orthorhombic phase ($<$361 K) become cage-centered at tetragonal (361-403 K) and cubic ($>$ 403 K) phases~\cite{boziki2021molecular}. Interestingly, in CsPbBr$_3$ a recent study~\cite{zhu2022probing} has shown that Br transverse displacement (e.g. displacement along $y$/$z$ direction for Br situated along $a$ direction) in cubic phase is disordered in a shallow double-well confirming order-disorder transition with respect to the orthorhombic phase. However, as discussed above, we do not see any disordering in Br displacements in MAPbBr$_3$.

\section{Conclusions}
In conclusion, our AIMD simulations in the isothermal-isobaric ensemble clearly establish that the low-temperature phase of MAPbBr$_3$ is not ferroelectric and instead conforms to a centrosymmetric space group \textit{Pnma}, while the room temperature phase is cubic (\textit{Pm-3m}) down to the local scale. Using time-correlation functions we show that the MA disordering leading to the cubic phase is best described as dynamic with the MA ions thermally diffusing between various orientational minima. The predicted timescales of MA motion agree well with experiments validating this perspective. By designing appropriate order parameters we trace the origins of the experimentally observed doublet feature in Br-Br RDF~\cite{bernasconi2017direct} to a scissoring distortion of the PbBr$_6$ octahedra. However, this vanishes in our room temperature simulations indicating that no orthorhobmic distortions persist in cubic MAPbBr$_3$. While this conclusion is in disagreement with the results of Bernasconi \textit{et al.}~\cite{bernasconi2017direct} we note that the Br-Br doublet feature identified by these authors at 300 K is rather weak and unlikely to have originated from the same doublet-like features seen at low temperatures in the neutron diffraction studies by Page \textit{et al.}~\cite{page2016short} or in our simulations at 40 K. While our simulations indicate the presence of local polar fluctuations arising from the motion of MA, Pb and Br separately, we confirm that they do not lead to any overall polarization at all temperatures. Both MA and Br ions show off-center displacements at low temperatures that vanish at 300 K, accompanied by softening of Br sublattice mode, suggesting a coexistence of displacive and order-disorder features for the structural transition. We find that MA translations correlate with the lattice scissoring in \textit{ac} plane but with Pb-Br bond lengths in the \textit{b} direction, thus indicating an anisotropic coupling enabled by H-bonding. This coupling also facilitates the MA reorientation and is likely a common feature among all the members of the MAPbX$_3$ series. An analysis of H-bond lifetimes yielded timescales similar to those involved in MA rotational motion emphasising the importance of the H-bonding in controlling the order-disorder transition of the MA sub-system. At 180 K, the MA reorientation dynamics possibly occurs at a much slower timescale and would require longer simulations.

In addition to confirming the non-ferroelectric nature of the low temperature phases of MAPbBr$_3$, a recent study~\cite{bari2021ferroelastic} has also pointed out the existence of ferroelastic domains in various phases making the system even more intriguing. Also, it is interesting to explore how the nearly degenerate polar state found at 0 K, can be accessed in experiments. Recently, an orthorhombic polymorph was found at 150 K, instead of the previously proposed tetragonal structure showing the complexity of the system in the mid-temperature region~\cite{wiedemann2021hybrid}. AIMD based methods would prove valuable for further exploration of these new properties and phases.

\begin{acknowledgements}
The authors thank DAE-BRNS for providing funding for this work and gratefully acknowledge the HPC facilities at IISER Bhopal as well as National Supercomputing Mission (NSM) for providing computing resources of ‘PARAM Shivay’ at Indian Institute of Technology (BHU), Varanasi, which is implemented by C-DAC and supported by the Ministry of Electronics and Information Technology (MeitY) and Department of Science and Technology (DST), Government of India. S.M. acknowledges funding through the Integrated Ph.D. program at IISER Bhopal.
\end{acknowledgements}

\bibliography{reference}

\end{document}


\preprint{APS/123-QED}

\title{\Large{Supporting Information : Deciphering the nature of temperature-induced phases in \ce{MAPbBr3} by {\it ab initio} molecular dynamics}}

\author
{Sayan Maity$^{1}$, Suraj Verma$^{1}$, Lavanya M. Ramaniah$^{2}$, Varadharajan Srinivasan$^{1}$}

\affiliation{$^{1}$Department of Chemistry, Indian Institute of Science Education and Research Bhopal, Bhopal 462 066, India}
\affiliation{$^{2}$High Pressure and Synchrotron Radiation Physics Division, Bhabha Atomic Research Centre, Trombay, Mumbai 400085, India}
\maketitle

\section{Computational methods} 
Car–Parrinello molecular dynamics \cite{car1985unified} simulations are performed using the QUANTUM ESPRESSO (QE) suite of electronic-structure codes~\cite{giannozzi2009quantum}. To accurately model the van der Waals (vdW) interactions in the crystals, we employed the DFT-D2 \cite{Stefan2006} scheme with a generalized gradient approximation (PBE) \cite{John21996} to the exchange-correlation functional. Ionic cores were modeled by ultrasoft pseudopotentials\cite{Vanderbilt1990}. The valence electrons, including the $1s^1$ electron of H, the $2s^2p^2$ electrons of C, the $2s^2p^3$ electrons of N, the $4s^4p^5$ electrons of Br, and the $5d^{10}6s^26p^2$ electrons of Pb were treated explicitly. All simulations were performed using 768 atom cell (4$\times$4$\times$4 with respect to the 12 atom pseudo-cubic unit-cell) in the isothermal-isobaric ensemble (constant NPT) by using the variable-cell Parrinello-Rahman barostat \cite{parrinello1980crystal} and a single Nose-Hoover thermostat \cite{martyna1992nose} with a frequency of 60 THz to maintain a constant pressure (P) and temperature (T), respectively. All of the plane waves {G} with kinetic energies below 65 Ry were included and to maintain a constant plane wave kinetic energy cutoff of $E_0$ = 55 Ry for a fluctuating cell, a smooth step function with height $A$ = 150 Ry and width $\sigma$ = 5 Ry to the plane wave kinetic factor is added as proposed by Bernasconi \textit{et al.} \cite{bernasconi1995first}: $G^2\rightarrow G^2+A[1+\textit{erf}(\frac{G^2/2-E_0}{\sigma})]$, where \textit{erf} is the error function. The augmented charge was represented by the kinetic energy cutoff of 400 Ry. The fictitious mass of the electrons was set to 500 atomic units, and the corresponding mass preconditioning with a kinetic energy cutoff of 2.5 Ry was used to all Fourier components of wavefunctions \cite{tassone1994acceleration}. Convergence criteria for electron minimization were chosen as $10^{-6}$ Ry. A time-step of 10 atomic units (0.24189 \textit{fs}) and real masses for all atoms are used for ionic propagation. For ionic minimization, $10^{-4}$ Ry and $10^{-3}$ Ry/Bohr of convergence thresholds on total energy and forces are used, respectively. At 0 GPa and 40, 180, and 300 K, molecular dynamics were performed, where at each temperature, the initial 2 ps trajectory was discarded due to equilibration. At 40 K $\sim$13 ps and 180, 300 K $\sim$20 ps trajectories after equilibration were used for all analysis. 
 
\section{Is ${\rm \ce{MAPbBr3}}$ Ferroelectric?}
\subsection{Possibility of polar structures at 0 K}
All geometry optimizations at 0 K were done by using the BFGS algorithm \cite{yuan1991modified}, with a force convergence criterion of 10$^{-5}$ Ry/Bohr. Monkhorst-Pack (MP) k-points mesh of $2\times2\times2$ and $4\times4\times4$ were used to sample the Brillouin Zone (BZ) of the bulk crystal \cite{monkhorst1976special}, for the 96 atom orthorhombic $\delta$-phase and 12 atom cubic $\alpha$-phase, respectively. For the $\delta$-phase, the parameters of the orthorhomic cell (48 atom) existing inside the 96 atom supercell are reported. For the purpose of benchmarking, van der Waals corrections (vdW) due to Grimme's DFT-D2 \cite{grimme2006semiempirical} (D2), Grimme's DFT-D3 \cite{grimme2010consistent}, Tkatchenko-Scheffler~\cite{tkatchenko2009accurate} (TS) and the exchange-correlation functional vdW-DF2 (DF2) \cite{lee2010higher}, vdW-DF3-opt3 (DF3) \cite{chakraborty2020next} are compared, wherever required. Phonon calculations were carried out on the optimized structures using density functional perturbation theory (DFPT) as implemented in QE at $\Gamma$ point~\cite{Baroni2001}.

In TABLE~\ref{ortho_vdw}, the results of optimizations (ions+cell) of orthorhombic $\delta$ phase, using various vdW corrected functionals, are shown. Here $\theta_{12}$ is the dihedral angle between MA 1 and 2, which are stacked along crystal $b$ axis. If the value of $\theta_{12}$ is 180$^o$, the MAs are arranged in anti-polar fashion and the zero value represents polar arrangement. Any value of $\theta_{12}$ between these extremes, can be indicated as intermediate-polar arrangement. In TABLE ~\ref{ortho_vdw}, optimizations with each functional are started from anti-polar ($\theta_{12}$=180$^o$) and intermediate-polar ($\theta_{12}$=145$^o$) structures and in both cases exactly same final structure was obtained where MA arrangements become anti-polar ($\theta_{12}$=180$^o$). We also test the existence of this anti-polar minimum with hybrid functional, shown in TABLE~\ref{ortho_vdw}. In all these cases, the calculated lattice constants at 0 K can be compared with the experimentally measured values of the same anti-polar $\delta$ phase at 115 K ($a$=8.565 \AA, $b$=11.841 \AA and $c$=7.976 \AA)~\cite{mashiyama2007anti}. It can be observed that our calculated values are reasonably close to the experimental data. 

The results of other optimizations of the orthorhombic phase are shown in TABLE~\ref{ortho_geom},~\ref{ortho_pbesol}. 
and discussed in the next section with proper context.

IN TABLE~\ref{cubic_vdw}, optimization data of the $\alpha$ phase are shown. Starting from any orientation of MA (face-to-face=\textit{ff},edge-diagonal=\textit{ed} and body-diagonal=\textit{bd}), in the optimized structure the orientation becomes \textit{ed} with distortion in cell, thus breaking the cubic symmetry. 

\begin{table*}[h!]
 \centering
 \begin{tabular}{|c|c|c|c|c|c|c|}
 \hline
 \textbf{functional} & \textbf{PBE+D2} \cite{grimme2006semiempirical} & \textbf{PBE+D3} \cite{grimme2010consistent} & \textbf{PBE+TS} \cite{tkatchenko2009accurate} & \textbf{DF2} \cite{lee2010higher} & \textbf{DF3} \cite{chakraborty2020next} & \textbf{PBE0} \cite{perdew1996rationale,adamo1999toward}\\
 \hline
 a (\AA)               & 8.537 & 8.699 & 8.749 & 8.872 & 8.628 & 8.105    \\
 \hline
 b (\AA)               & 11.930 & 11.983 & 12.000 & 12.342 & 11.935 & 11.481 \\
 \hline
 c (\AA)               & 7.903 & 7.912 & 7.884 & 8.144 & 7.898 & 7.649     \\
 \hline
 $\alpha$ ($^\circ$)   & 90.0 & 90.0 & 90.0 & 90.0 & 90.0  & 90.0     \\
 \hline
 $\beta$ ($^\circ$)    & 90.0 & 90.0 & 90.0 & 90.0 & 90.0  & 90.0     \\
 \hline
 $\gamma$ ($^\circ$)   & 90.0 & 90.0 & 90.0  & 90.0 & 90.0  & 90.0    \\
 \hline
 V (\AA$^3$)           & 804.915 & 824.793 & 827.761 & 891.732 & 813.403 &  711.730    \\
 \hline
 $\theta_{12}$ ($^\circ$)   & 180.0 & 180.0 & 180.0 & 180.0 & 180.0      & 180.0 \\
 \hline
 \end{tabular}
 \caption{\footnotesize{Optimization (ions+cell) results of orthorhombic $\delta$-phase using various vdW corrected and a hybrid exchange-correlation functionals. $\theta_{12}$ is the dihedral angle (CN $\to$ CN) between MA 1 and 2 under $v$ or $w$ group. Starting from anti-polar ($\theta_{12}$=180$^o$) and intermediate-polar arrangement ($\theta_{12}$=145$^o$), same structures were obtained, which are reported here. Note that for the PBE0 case, the optimization was only performed around the anti-polar arrangement and the force optimization was done with a higher threshold of $10^{-3}$ Ry/Bohr due to the computational cost involved.}}
 \label{ortho_vdw}
\end{table*}

\begin{table*}[h!]
 \centering
 \begin{tabular}{|c|c|}
 \hline
 \textbf{functional} & {\textbf{PBE+D2}~\cite{grimme2006semiempirical}} \\
 \hline
 a (\AA)                  &  8.575                             \\
 \hline
 b (\AA)                  &  12.515                           \\
 \hline
 c (\AA)                  &   7.658                         \\
 \hline
 $\alpha$ ($^\circ$)      &   90.448                                \\
 \hline
 $\beta$ ($^\circ$)       &   89.739                                \\
 \hline
 $\gamma$ ($^\circ$)      &   85.397                                \\
 \hline
 V (\AA$^3$)              &   819.174                           \\
 \hline
 $\theta_{12}$ ($^\circ$) & $\sim$174                                \\
 \hline
 E (meV/atom)              &   6.8     \\                    
 \hline
 \end{tabular}
 \caption{\footnotesize{Optimization (ions) results of orthorhombic $\delta$-phase starting from intermediate-polar ($\theta_{12}$=145$^o$) arrangement. The largest component of the stress tensor is 0.8 GPa, resulting as a consequence of not allowing cell relaxation. This structure is $\sim$81.6 meV/f.u. higher in energy than the cell-optimized structure.}}
 \label{ortho_geom}
\end{table*}

\begin{table*}[h!]
 \centering
 \begin{tabular}{|c|c|}
 \hline
 \textbf{functional} & {\textbf{PBEsol}~\cite{perdew2008restoring}} \\
 \hline
 a (\AA)                  &      8.168                  \\
 \hline
 b (\AA)                  &      7.606                   \\
 \hline
 c (\AA)                  &     11.506                  \\
 \hline
 $\alpha$ ($^\circ$)      &      90.0                       \\
 \hline
 $\beta$ ($^\circ$)       &      90.0                       \\
 \hline
 $\gamma$ ($^\circ$)      &      90.0                     \\
 \hline
 V (\AA$^3$)              &     714.782                 \\
 \hline
 $\theta_{12}$ ($^\circ$) &     180.0                      \\
 \hline
 E (meV/atom)              &    0.0 \\ 
 \hline
 \end{tabular}
 \caption{\footnotesize{Optimization (ions+cell) results of orthorhombic $\delta$-phase starting from intermediate-polar ($\theta_{12}$=145$^o$) arrangement using PBEsol functional as used in the work by Lehmann \textit{et al.}~\cite{lehmann2021long}.}}
 \label{ortho_pbesol}
\end{table*}

\begin{table*}[h!]
 \centering
 \begin{tabular}{|c|c|c|c|}
 \hline
 \textbf{functional} & \textbf{PBE+D2} \cite{grimme2006semiempirical} & \textbf{PBE+D3} \cite{grimme2010consistent} & \textbf{PBE+TS} \cite{tkatchenko2009accurate} \\
 \hline
 a (\AA)               & 5.918 & 5.914 & 5.929     \\
 \hline
 b (\AA)               & 5.964 & 6.016 & 6.061     \\
 \hline
 c (\AA)               & 5.918 & 5.974 & 5.974     \\
 \hline
 $\alpha$ ($^\circ$)   & 93.557 & 93.155 & 92.464    \\
 \hline
 $\beta$ ($^\circ$)    & 93.526 & 92.122 & 90.147      \\
 \hline
 $\gamma$ ($^\circ$)   & 93.529 & 92.617 & 90.172      \\
 \hline
 V (\AA$^3$)           & 207.624 & 211.829 & 214.524 \\
 \hline
 E (meV/f.u.)          & 0.0 & 0.0 & 0.0    \\
 \hline
 \end{tabular}
 \caption{\footnotesize{Optimization (ions+cell) results of cubic $\alpha$-phase starting from face-to-face (\textit{ff}),  body-diagonal (\textit{bd}) and edge-diagonal (\textit{ed}) arrangement using various vdW corrected exchange-correlation functionals. In all cases, the optimized arrangement is \textit{ed}, which is reported here.}}
 \label{cubic_vdw}
\end{table*}

\subsection{Possibility of (persistent) polar structures in the $\delta$ (40 K), $\beta$ (180 K) and $\alpha$ (300 K) phases}
To describe the orientations of C-N vectors, we have defined two orientational order parameter $\theta(t)$ (0 to $180^o$) and $\phi(t)$ ($-180^o$ to $180^o$) similar to the orthogonal coordinate system, where $\theta$ is the angle between the vector and \textit{b} axis, and $\phi$ is the angle between \textit{c} and the projection of the vector on \textit{ca} plane. However the cell parameter vectors (\textit{a(t)}, \textit{b(t)} and \textit{c(t)}) changes during the NPT simulation, indicating the necessity to define a three-dimensional coordinate where this change is accounted for. In this regard, for every MA, a local coordinate system corresponding to the smallest cube is defined, where local \textit{a(t)}, \textit{b(t)} and \textit{c(t)} are taken as three time-dependent Pb-Pb distances, lying approximately along to their corresponding crystallographic axes of the (pseudo)-cube. It is to be noted that, as the orthogonality of these local axes might be lost due to the allowed fluctuations of the cell-parameters, $\theta(t)$ and $\phi(t)$ are calculated after the basis transformation of those axes into mutually orthogonal axes. After calculating these throughout the simulation at any temperature ($\theta(t)$, $\phi(t)$) probability distribution plots are generated. These give the idea of average orientations of MAs which are connected to the nature of the potential energy surface described by these orientational order parameters.

\begin{figure*}[tbh!]
 \centering
 \includegraphics[width=1\textwidth]{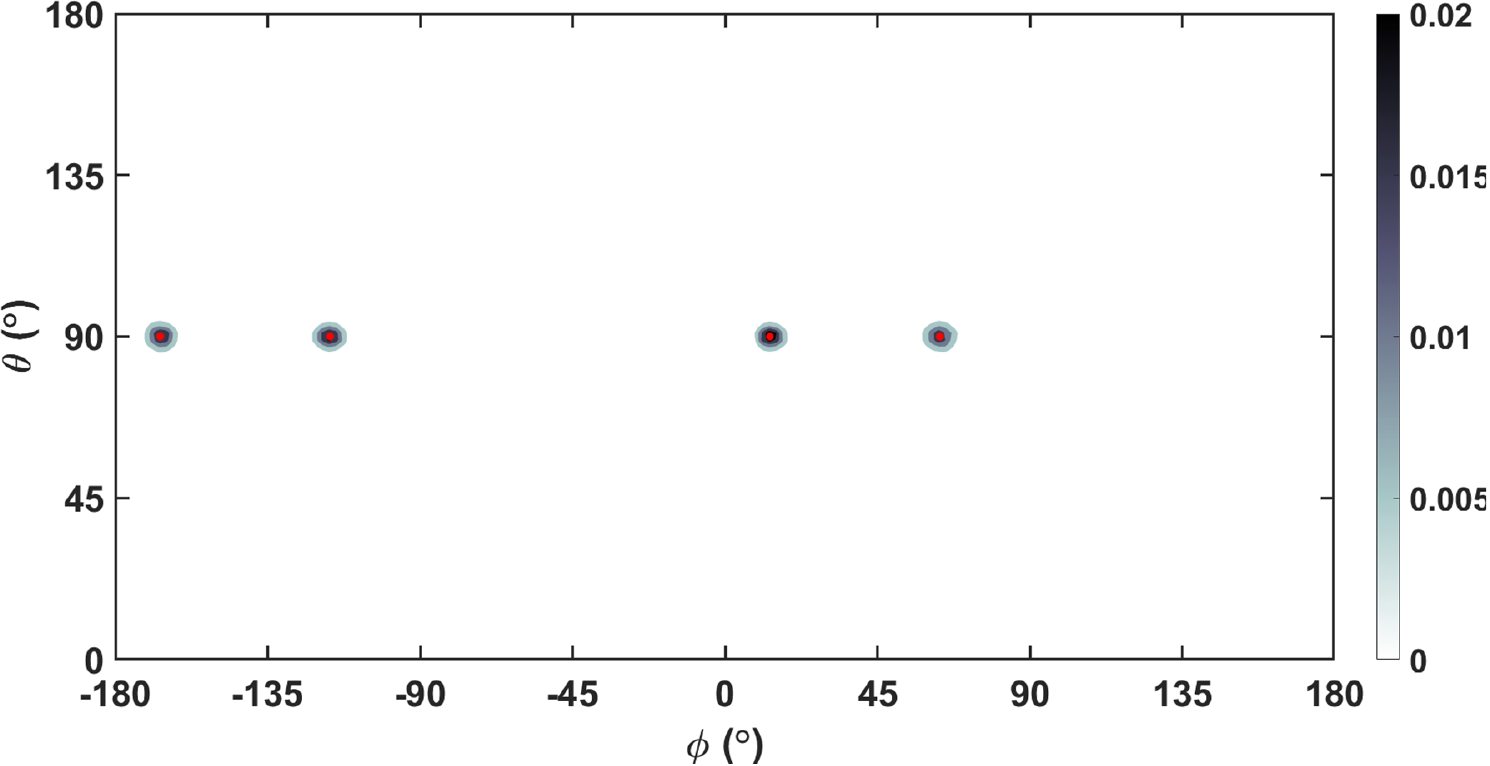}
 \caption{\footnotesize{Orientational ($\theta-\phi$) probability distribution plot of all MAs at 40 K. The red circles are orientations at 0 K, shown as references.}} 
 \label{40_dis}
\end{figure*}

\begin{figure*}[tbh!]
 \centering
 \includegraphics[width=1\textwidth]{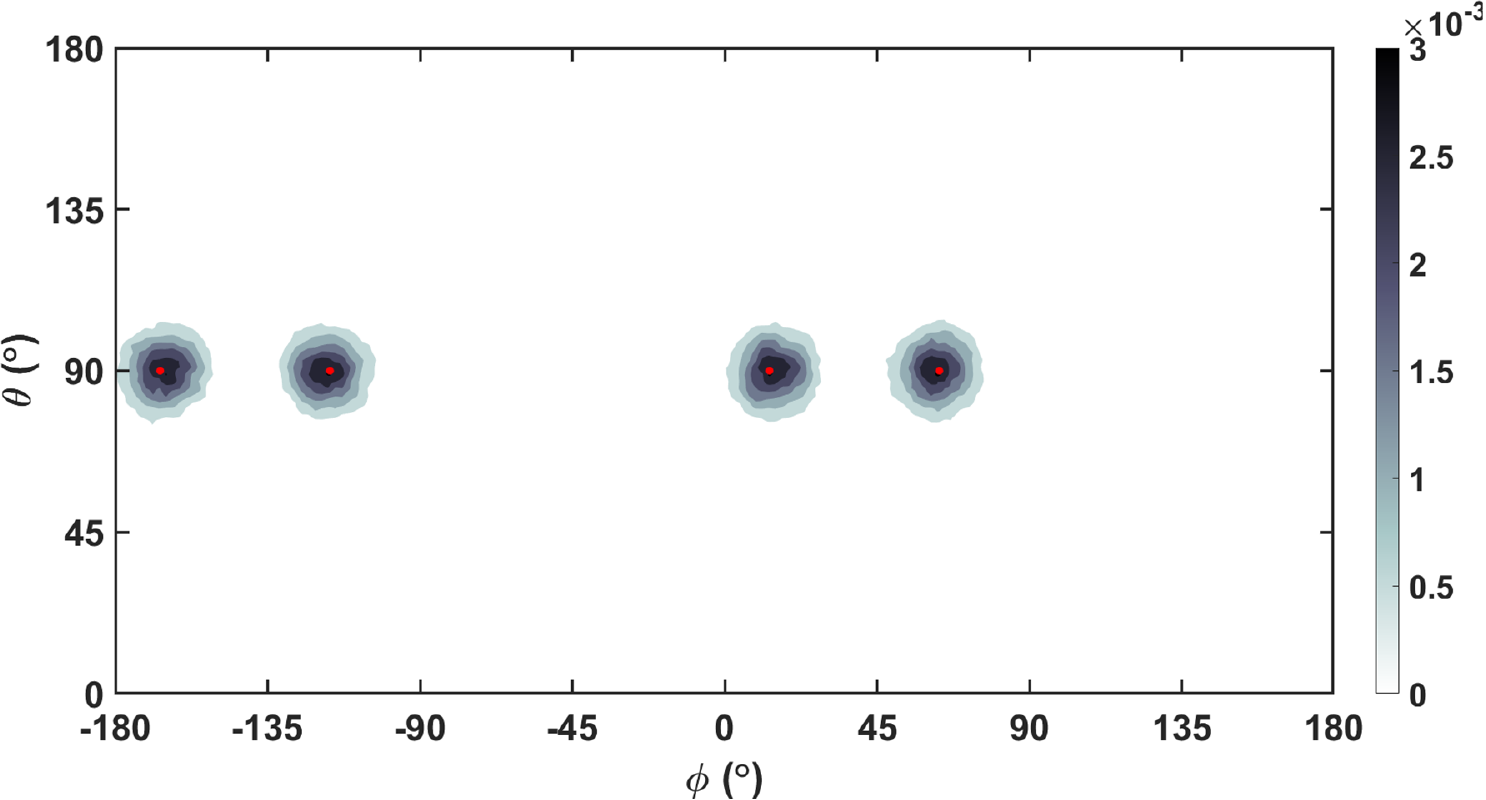}
 \caption{\footnotesize{Orientational ($\theta-\phi$) probability distribution plot of all MAs at 180 K. The red circles are orientations at 0 K, shown as references.}} 
 \label{180_dis}
\end{figure*}

By using the trajectory generated by the 768 atoms cell, at 40 K, it is observed that the values of $\theta(t)$, $\phi(t)$ (and the independent projections of C-N vectors on local \textit{a(t)}, \textit{b(t)} and \textit{c(t)} axes) for every MA remains constant on an average and their fluctuations are least compared to other temperature, indicating ordered nature of these cations. This can also be seen from the orientational distribution plot at 40 K (see Figure~\ref{40_dis}). We can assume, this overall distribution is a summation of several components, each described by a bivariate normal probability distribution function as follows:
\begin{equation}\label{eq:a1}
 \rho(x)=\frac{1}{2\pi}|\zeta|^{-1/2}exp\left[-\frac{1}{2}(x-\mu)^{T}\zeta^{-1}(x-\mu)\right]
\end{equation}
Here, $x$ is the vector \{$\theta$,$\phi$\}, $\mu$ is the population mean vector \{$\mu_\theta$,$\mu_\phi$\} and $\zeta$ is the population variance-covariance matrix, where $\rho$ is the correlation between $\theta$ and $\phi$:
\[
\zeta =
\begin{vmatrix}
\sigma_{\theta}^2 & \rho\sigma_{\theta}\sigma_{\phi} \\ 
\rho\sigma_{\theta}\sigma_{\phi} & \sigma_{\phi}^2  \\
\end{vmatrix}
\]
We use Gaussian Mixture Modelling (GMM), a clustering machine learning algorithm, to generate all the Gaussian components for the orientational distribution after accounting for periodicity in $\phi$ at $\pm180^o$. The presence of several Gaussians indicates the presence of different local energy minima in potential energy surfaces. The covariance matrix for every Gaussian component ($\sigma_{\theta}$ and $\sigma_{\phi}$) gives the idea of the shape or spread of potential energy surface around the corresponding energy minima. After accounting for all MAs the total distributions at 40 and 180 K are clustered using GMM and for both of them, four minima are found as listed in Table \ref{positions and spreads}. 
\begin{table*}[tbh!]
   \centering
    \begin{tabular}{|c|c|c|c|c|c|c|c|} 
    \hline
    \multicolumn{1}{|c|}{\textbf{Temperature}} & \multicolumn{1}{|c|}{\textbf{index}} & \multicolumn{1}{|c|}{\textbf{$\mu_\theta (^o)$}} & \multicolumn{1}{|c|}{\textbf{$\mu_\phi (^o)$}} &
    \multicolumn{1}{|c|}{\textbf{$\sigma_{\theta} (^o)$}} &
    \multicolumn{1}{|c|}{\textbf{$\sigma_{\phi} (^o)$}} &
    \multicolumn{1}{|c|}{\textbf{$p$}} \\
    \hline
    & $v1$ & 82.730 & 67.075 & 6.215 & 5.279 & 0.2503 \\
    40 K & $v2$ & 108.967 & -146.913 & 4.650 & 4.655 & 0.2376 \\
    (96 atom) & $w1$ & 100.098 & -171.864 & 6.987 & 8.229 &0.2624 \\
    & $w2$ & 69.848 & 41.746 & 4.758 & 4.499 & 0.2497 \\
    \hline
    & $v1$ & 90.028 & 63.275 & 2.834 & 3.273 & 0.2500 \\
    40 K & $v2$ &  89.980 & -116.789 & 2.384 & 3.036 & 0.2500 \\    
    (768 atom) & $w1$ &  90.039 & -166.456 & 2.490 & 3.031 & 0.2500 \\
    & $w2$ &  89.958 & 13.547 & 2.329 & 2.771 & 0.2500 \\ 
    \hline
    & $v1$ & 89.880 &	62.568 & 7.626 & 7.462 & 0.2500 \\
    180 K & $v2$ & 89.9096	& -117.634 & 7.206 &	7.568 & 0.2500 \\
    (768 atom) & $w1$ & 90.082	& -165.532 & 7.449 & 7.443 & 0.2500 \\
    & $w2$ & 89.8478	& 14.430 & 7.391 & 7.386 & 0.2500 \\ 
    \hline
    \end{tabular}
  \caption{\footnotesize{Positions ($\mu$) and shapes ($\sigma$) of total ($\theta$,$\phi$) orientational distribution of all groups of MAs at 40 and 180 K. $p$ indicates the contribution of every component to the overall distribution. $v$ and $w$ are two groups of MAs based on their orientations. Every group is further subdivided into two types (indicated as $1$ ($C\to N$) and $2$ ($N\to C$)) which are arranged anti-ferroelectrically in the starting centrosymmetric simulation cell.}}
  \label{positions and spreads}
\end{table*} 
\begin{table*}[]
\begin{tabular}{|c|c|c|c|}
\hline
Symmetry & 40 K & 180 K & 300 K \\
\hline
$LCM_{12}$    &  $\checkmark$    &   $\checkmark$    &   $\times$    \\
\hline
$SM_{vw}$     &  $\checkmark$    &   $\checkmark$        &   $\times$    \\
\hline
$MSM$         &  $\checkmark$    &   $\checkmark$        &   $\times$    \\
\hline
\end{tabular}
  \caption{\footnotesize{Evolution of some defined symmetry elements with temperature.}}
\label{symmetry}
\end{table*}
At these temperatures, MAs are clustered into two groups, $v$ and $w$, based on their orientations. Every group is further subdivided into two types (indicated as $1$ and $2$ ($C\to N$ and $N\to C$) which were arranged anti-ferroelectrically to each other (molecular axes lying on \textit{ac} plane) in the starting centrosymmetric simulation cell of \textit{Pnma} symmetry. Considering only the MA orientations, criteria for the local centrosymmetry between both types ($LCM_{12}$), in every group described above, is defined as: 
\begin{align}\label{eq:a2}
 \mu_\theta(1)+\mu_\theta(2)=180^o \\
 \left\vert\mu_\phi(1) - \mu_\phi(2)\right\vert = 180^o
\end{align}
This local centrosymmetry arises due to combination of both of the two $2_1$ screw axis parallel to $b'$ (one along $b'$ and the other at $a'/2$) in the $\delta$-orthorhombic $Pnma$ unit cell, which will be preserved upon satisfying $LCM_{12}$. With respect to this unit-cell, the $v$ and $w$ groups, which are situated alternatingly along $a'$ axis, are connected to each other by four $2_1$ screw axes parallel to $c'$ (two at $a'/4$, $3a'/4$ and other two not only at the same but also at $b'/2$) and by two $2_1$ screw axes parallel to $a'$ (at $b'/4$, $3b'/4$ and simultaneously at $1/4$ above of $a'b'$ plane). Preserving all these screw symmetry between $v1$,$v2$,$w1$ and $w2$ ensures the following condition, indicated by $SM_{vw}$:
\begin{align}\label{eq:a3}
 \mu_{\theta_v}(j) = \mu_{\theta_w}(j), j = 1,2 \\
 \mu_{\phi_v}(1) + \mu_{\phi_w}(1) =  \mu_{\phi_v}(2) + \mu_{\phi_w}(2)
\end{align} 
In the $Pnma$ unit cell, the two mirror plane parallel to the $a'c'$ plane (at $b'/4$ and $3b'/4$) ensures the angles between all the cations and the  $b'$ axis is $90^o$. This will also be true if local (pseudo-)cubic axes $a$,$b$ and $c$ is considered and criteria for the mirror symmetry is defined as indicated by $MSM$:
\begin{align}\label{eq:a4}
 \mu_{\theta_v}(j) = \mu_{\theta_w}(j) = 90^0, j = 1,2 \\
 \left(\vec{r}^v_{MA} - \vec{r}^w_{MA} \right)\cdot\vec{b} = 0
\end{align} 
where $\vec{r}^{v/w}_{MA}$ is the center of mass coordinates of a MA ion in group v/w. The absence of the $MSM$ due to the absence of these mirror symmetry could allow movement of atomic positions (due to rotation of MA indicated by $LCM_{12}$, and translation of MA or lattice) along the $b'$ axis which could break overall or global centrosymmetry of the crystal, may be responsible for ferroelectricity. It is to be noted that the symmetry elements described above, are with respect to $Pnma$ unit cell whose $a',b'$ and $c'$ axes are $\langle101\rangle$, $\langle010\rangle$ and $\langle10\overline{1}\rangle$ respectively ($i.e$ $a,b=b',c$) of the pseudocubic cell. 

\begin{figure*}[tbh!]
 \centering
 \subfigure[]{\includegraphics[width=0.48\textwidth]{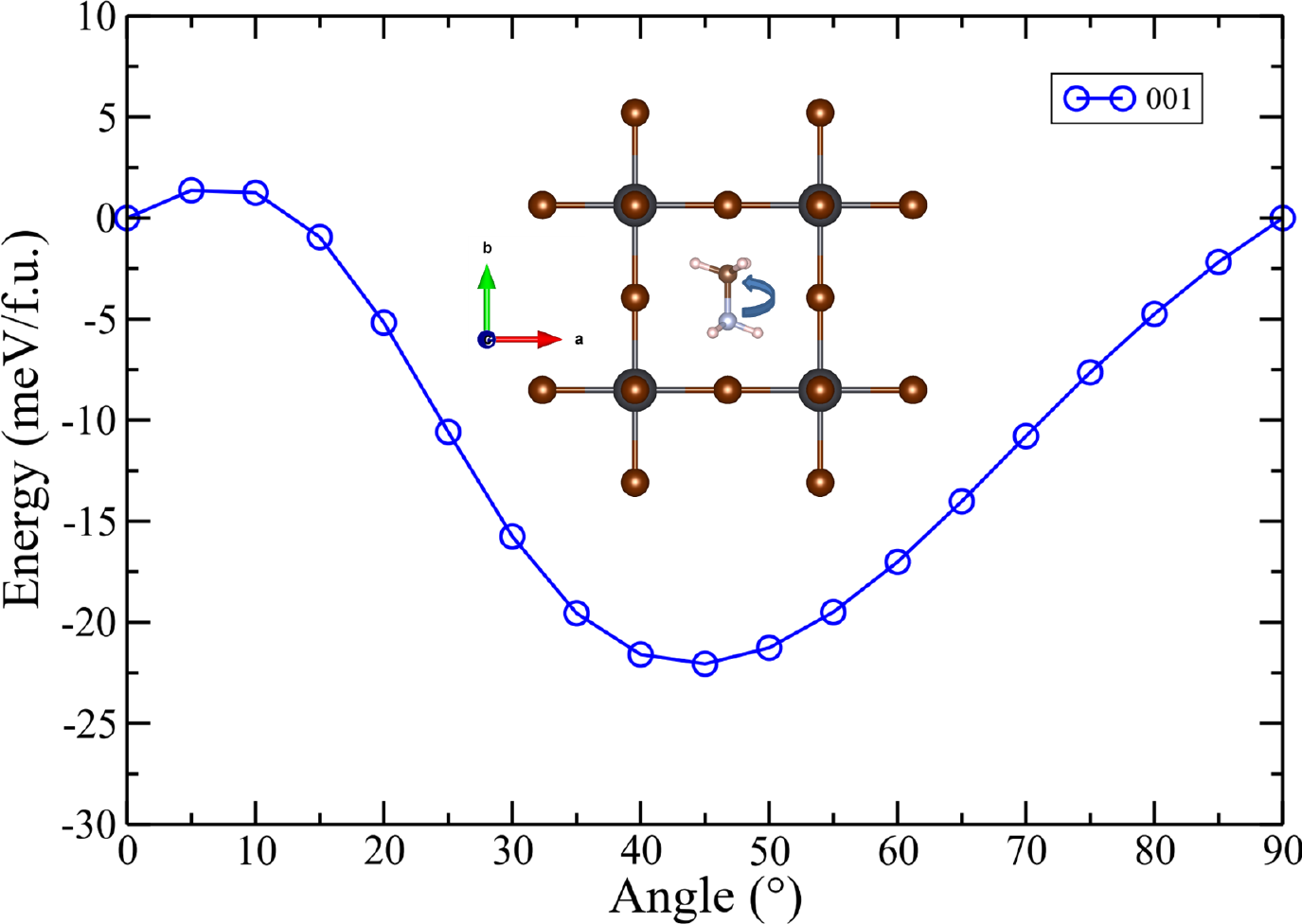}} 
 \subfigure[]{\includegraphics[width=0.48\textwidth]{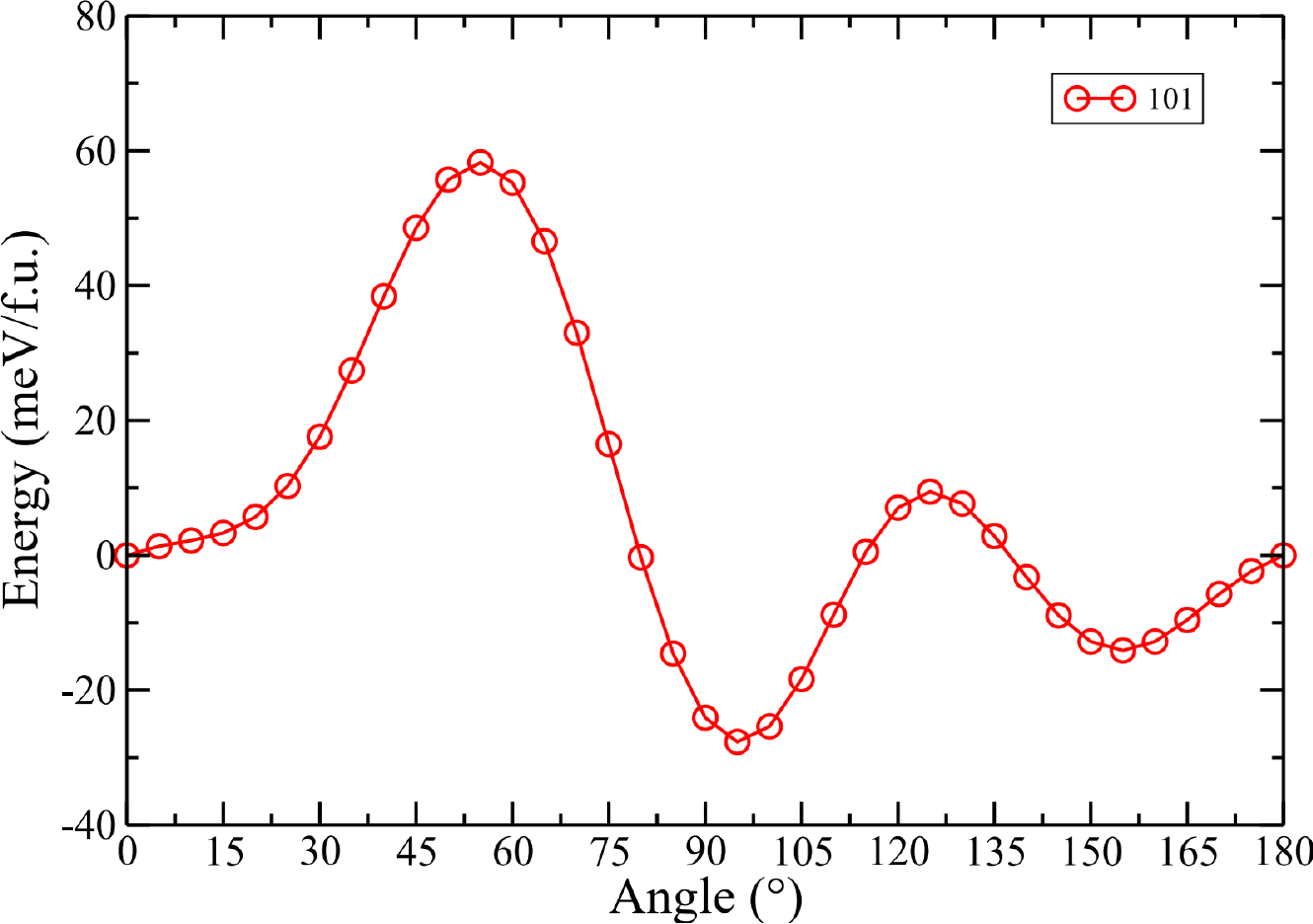}} 
 \caption{\footnotesize{Energy as function of angle for rotation in (a) [001] and, (b) [101] planes in the cubic phase, where cubic lattice constant is taken as $a$=5.926 $\AA$. Starting orientation is taken along $b$ axis as shown in inset of (a) whose corresponding angle is 0$^o$. For rotation in [001] and [101] plane the corresponding \textit{ed} oriention occur at 45$^o$ and 90$^o$, respectively.}} 
 \label{pes}
\end{figure*}

\begin{figure*}[tbh!]
 \centering
 \subfigure[]{\includegraphics[height=0.3\textheight]{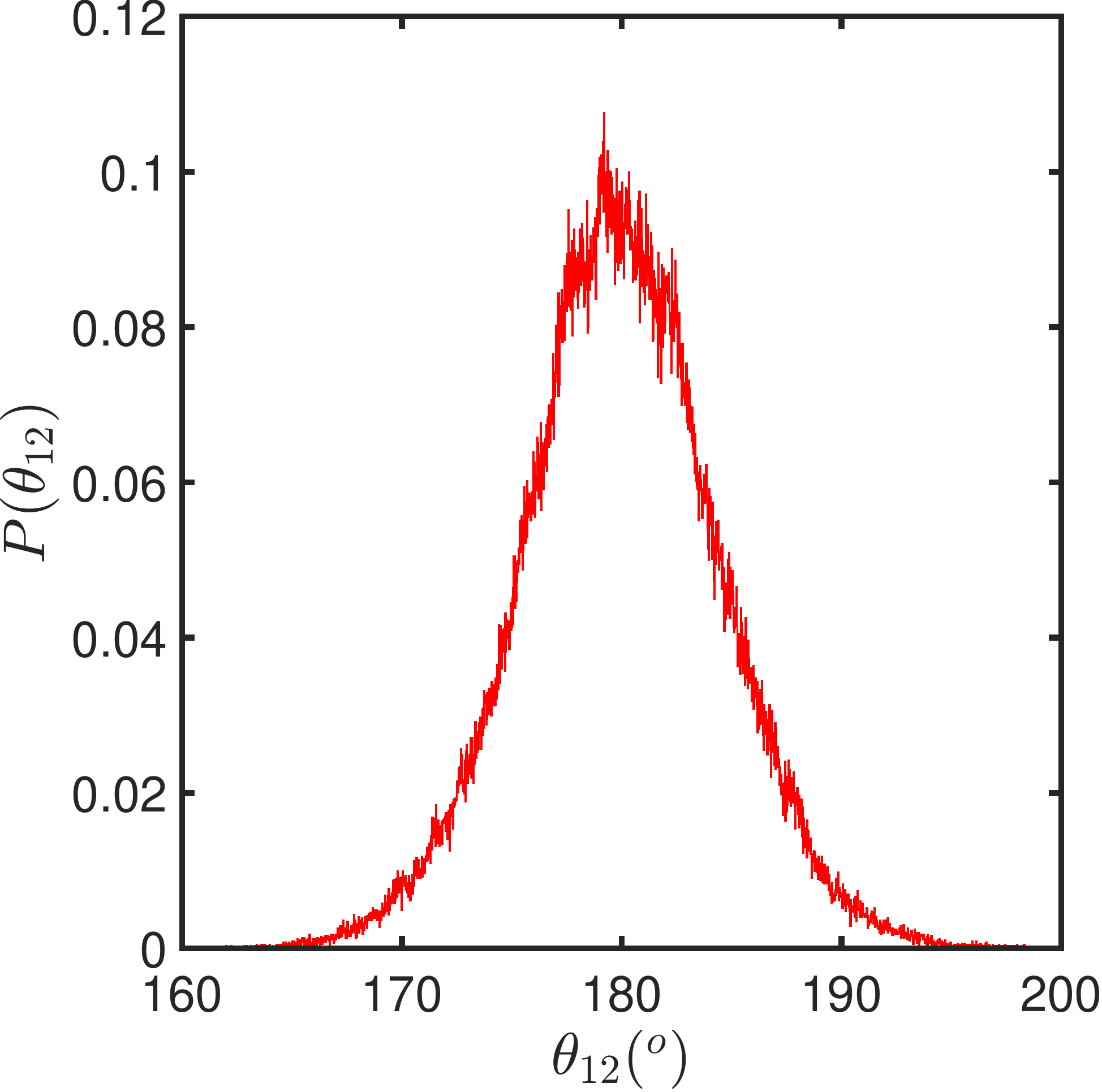}} 
 \subfigure[]{\includegraphics[height=0.3\textheight]{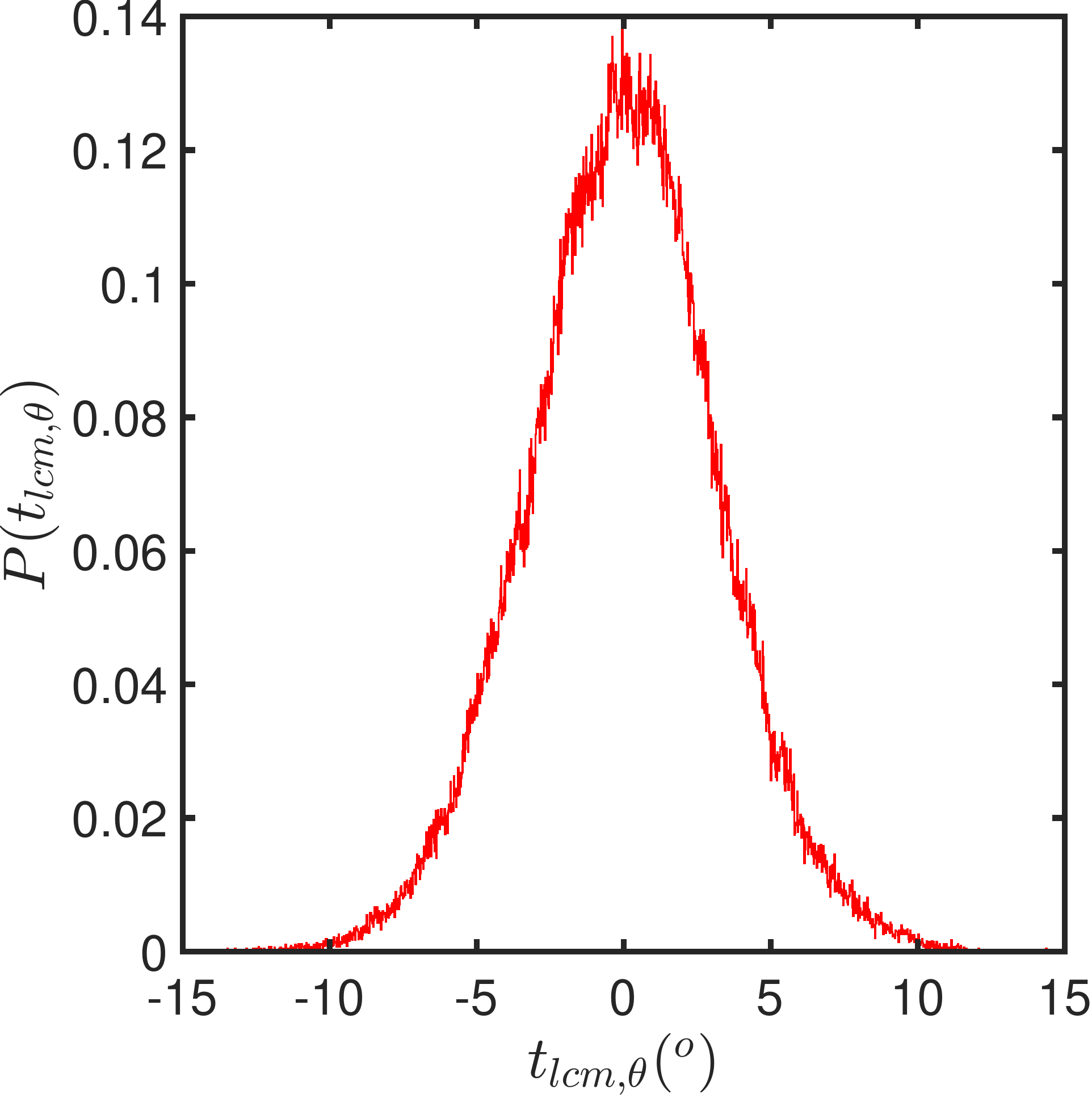}} 
 \subfigure[]{\includegraphics[height=0.3\textheight]{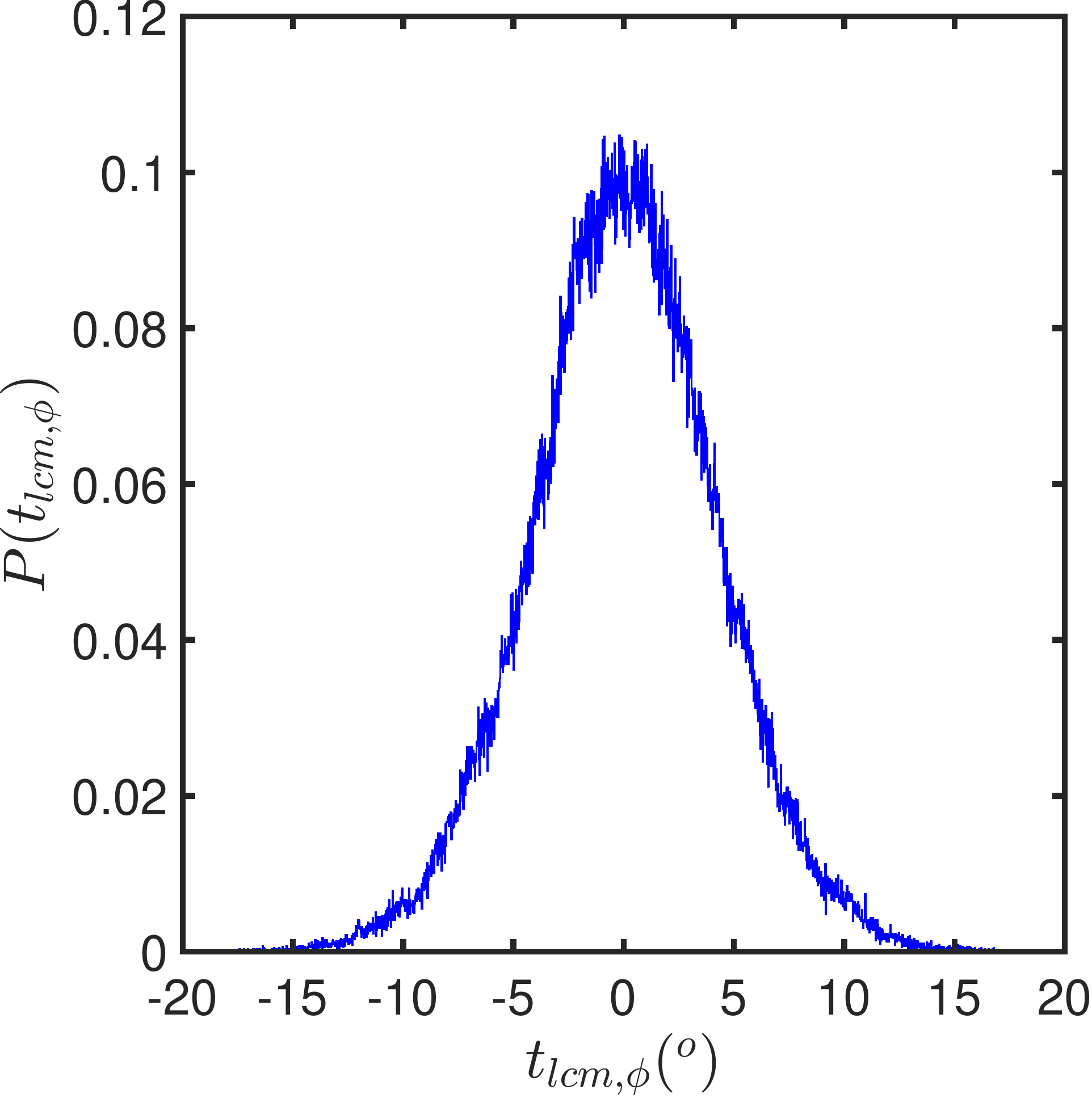}} 
 \caption{\footnotesize{Distribution of (a) $\theta_{12}$, (b) $t_{lcm,\theta}$, and (c) $t_{lcm,\phi}$ at 40 K.}} 
 \label{t_lcm}
\end{figure*}

Using the $LCM_{12}$ criteria we plot the distribution of $t_{lcm,\theta}=180^o-(\mu_\theta(1)+\mu_\theta(2))$ and $t_{lcm,\phi}=180^o-(|\mu_\phi(1)-\mu_\phi(2)|)$ along with $\theta_{12}$ at 40 K (see Figure~\ref{t_lcm}). The unimodal nature of these distributions confirms the absence of any local energy minima configuration which is close to the non-ferroelectric or anti-polar arrangement. Due to the large spread of the distribution, we suggest that the presence of configurations that deviated from the 0 K equilibrium arrangement are the consequence of thermal fluctuations. Thus we rule out the presence of similar intermediate-polar minima, however, suggested by Lehmann \textit{et al.}~\cite{lehmann2021long}. We suggest that these minima, obtained as results of geometry optimizations, will fall back to the anti-polar ground state if cell relaxations are also allowed. Using PBE+D2, a ionic optimization of an intermediate-polar structure resulted into a polar state ($\theta_{12}$=174$^o$) (see TABLE~\ref{ortho_geom}), which, upon cell optimization reaches anti-polar ground state ($\theta_{12}$=180$^o$) by lowering the energy by $\sim$82 meV/f.u. (see TABLE~\ref{ortho_vdw}). Also, using the PBEsol exchange-correlation~\cite{perdew2008restoring} as used in the work, we performed complete cell relaxation of the intermediate-polar structure ($\theta_{12}$=145$^o$) which reached the anti-polar ground state ($\theta_{12}$=180$^o$) (see TABLE~\ref{ortho_pbesol}). However, it should be noted that, apart from the experimentally suggested anti-polar structure, a polar structure with the same Pb-Br lattice but different MA arrangement ($\theta_{12}=$0$^o$) was also proposed by Sarkar \textit{et al.}~\cite{sarkar2017role}, which we do not observe in our molecular dynamics simulation at 40 K. Our optimization starting from the same structure also led to the polar ground state, 
which is only 2 meV/atom higher in energy (can be considered within standard DFT error) than the anti-polar ground state. As by X-ray experiment, differentiation between C and N is difficult~\cite{mashiyama2007anti}, thus the existence of this polar minimum cannot be ruled out.   


\begin{figure*}[tbh!]
 \centering
  \subfigure[]{\includegraphics[height=0.3\textheight]{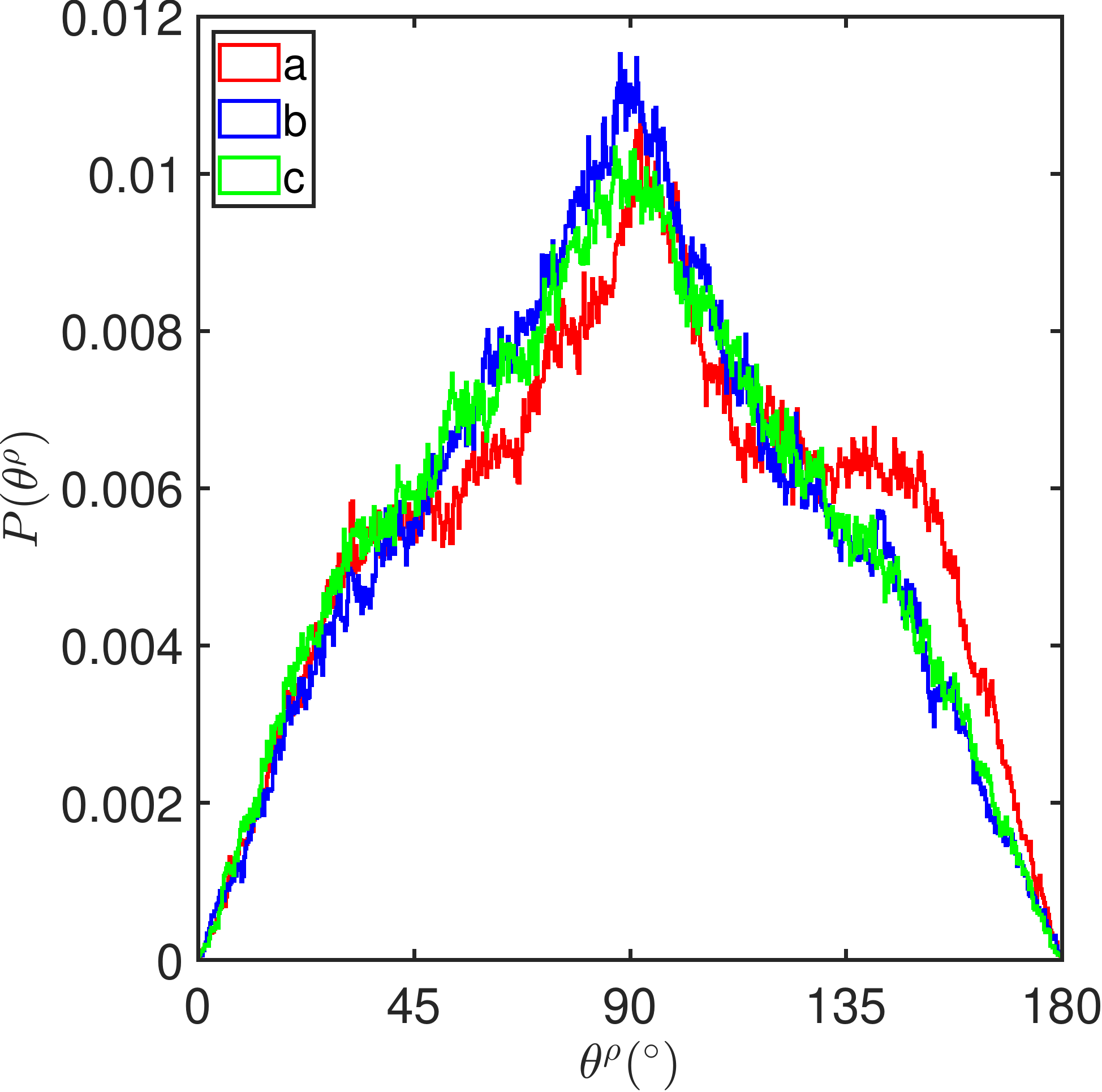}} \hspace{0.02\textwidth} 
  \subfigure[]{\includegraphics[height=0.314\textheight]{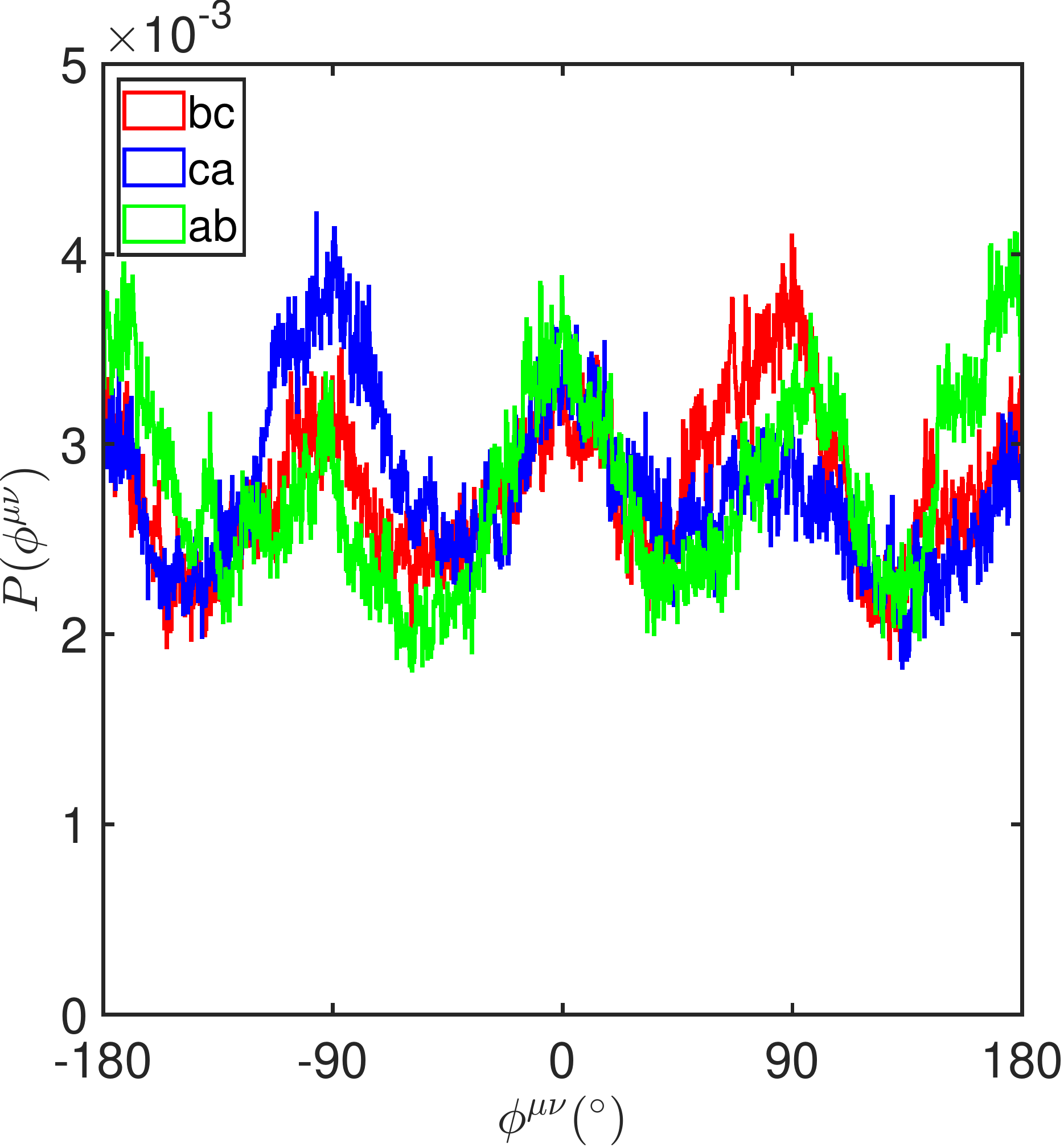}}
 \caption{\footnotesize{
 (a) distribution of $\theta$ along the three different cubic axes (b) distribution of $\phi$ in the three planes.}} 
 \label{op_lambda}
\end{figure*}

At 40 K, MAs are stacked along \textit{b} axis. Change in their orientations due to temperature effect, can generate short-lived local polar or ferroelectric type
fluctuations. In order to probe this at any temperature the following order parameters ($\lambda^a$, $\lambda^b$, $\lambda^c$ for \textit{a}, \textit{b} and \textit{c} stacking respectively) are defined:
\begin{equation}
\lambda^{a} = \frac{1}{T}\sum_{t}\frac{1}{N_u}\sum_{i,j,k}-cos(\phi_{i+1,j,k}^{bc}(t)-\phi_{i,j,k}^{bc}(t))\ \\
\end{equation} 
\begin{equation}
\lambda^{b} = \frac{1}{T}\sum_{t}\frac{1}{N_u}\sum_{i,j,k}-cos(\phi_{i,j+1,k}^{ca}(t)-\phi_{i,j,k}^{ca}(t))\ \\
\end{equation} 
\begin{equation}
\lambda^{c} = \frac{1}{T}\sum_{t}\frac{1}{N_u}\sum_{i,j,k}-cos(\phi_{i,j,k+1}^{ab}(t)-\phi_{i,j,k}^{ab}(t))\ \\
\end{equation} 
where, $i$,$j$,$k$ are lattice indexes of every methylamine, $N_u$ is the number of cubic unit cells in the supercell, and $T$ is the total time. At 0 K, along any crystal direction, two consecutive MAs are stacked in a way that the difference between their $\phi$ values is $\pm180^o$, thus resulting in $\lambda=1$. From Figure~\ref{op_lambda}(a) it can be seen that at low temperatures the same stacking is obtained. At 300 K, however, $\lambda\approx 0$, indicating that the anti-ferroelectric ordering of consecutive MA ions does not persist. It can be seen from the distribution of the order parameter $\lambda$ (Figure 3) and orientation angle $\theta$ and $\phi$ (Figure~\ref{op_lambda}(a) and (b)) that, both polar and anti-polar arrangements amongst neighboring MA ions are equally sampled. The existence of such transient polar sub-units can influence the electronic properties of the crystal at room temperature and need to be explored further.

\begin{figure*}[tbh!]
 \centering
 \includegraphics[width=0.6\textwidth]{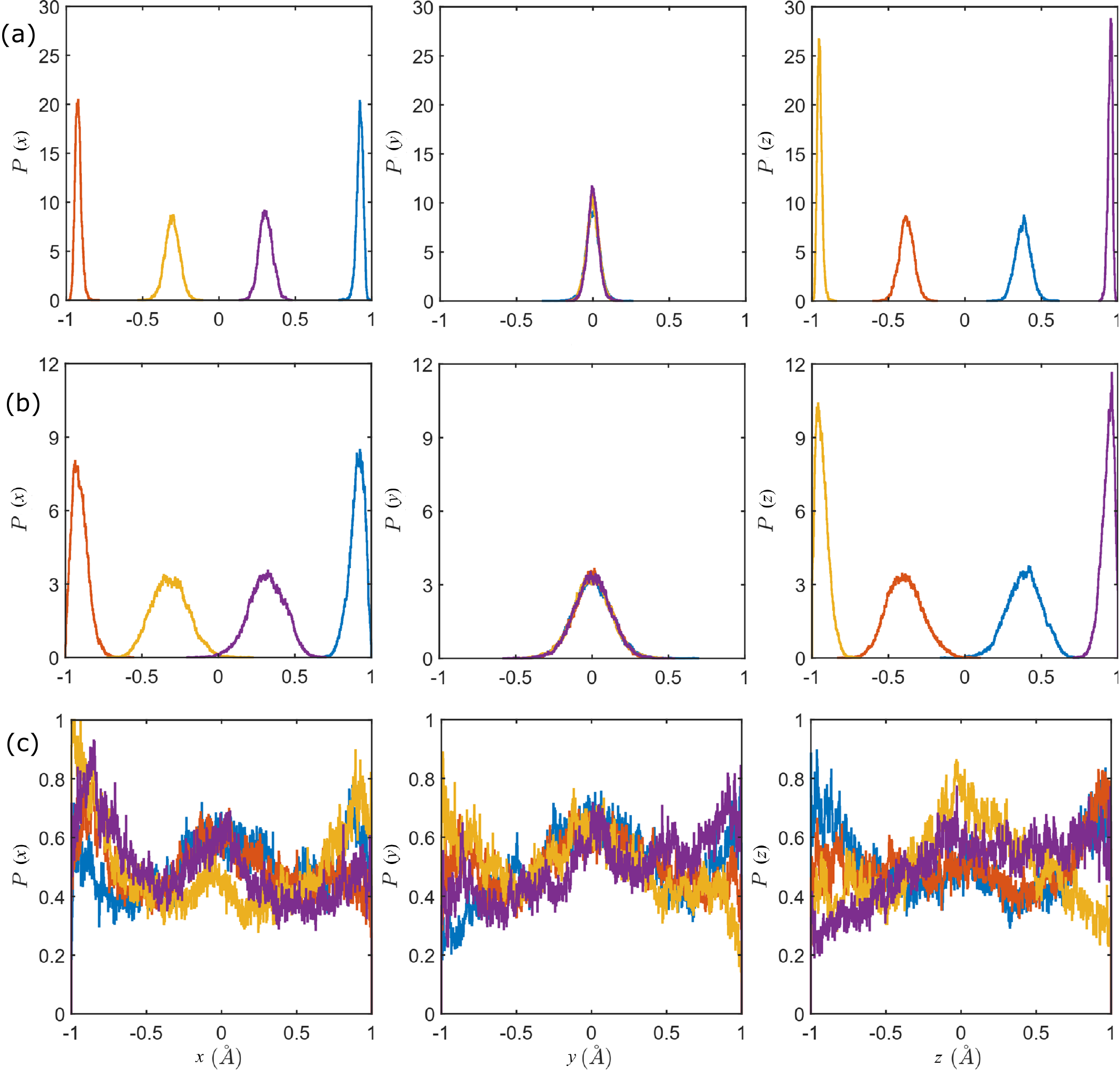}
 \caption{\footnotesize{Distribution function for x, y and z components for MA rotation $p^{MA}_{rot}$ at (a) 40 (b) 180 and (c) 300 K. Contributions are decomposed in blue, orange, yellow and indigo representing $v1$, $v2$, $w1$ and $w2$ type MAs, respectively.}} 
 \label{MA_rot_dis}
\end{figure*}

\begin{figure*}[tbh!]
 \centering
 \includegraphics[width=0.6\textwidth]{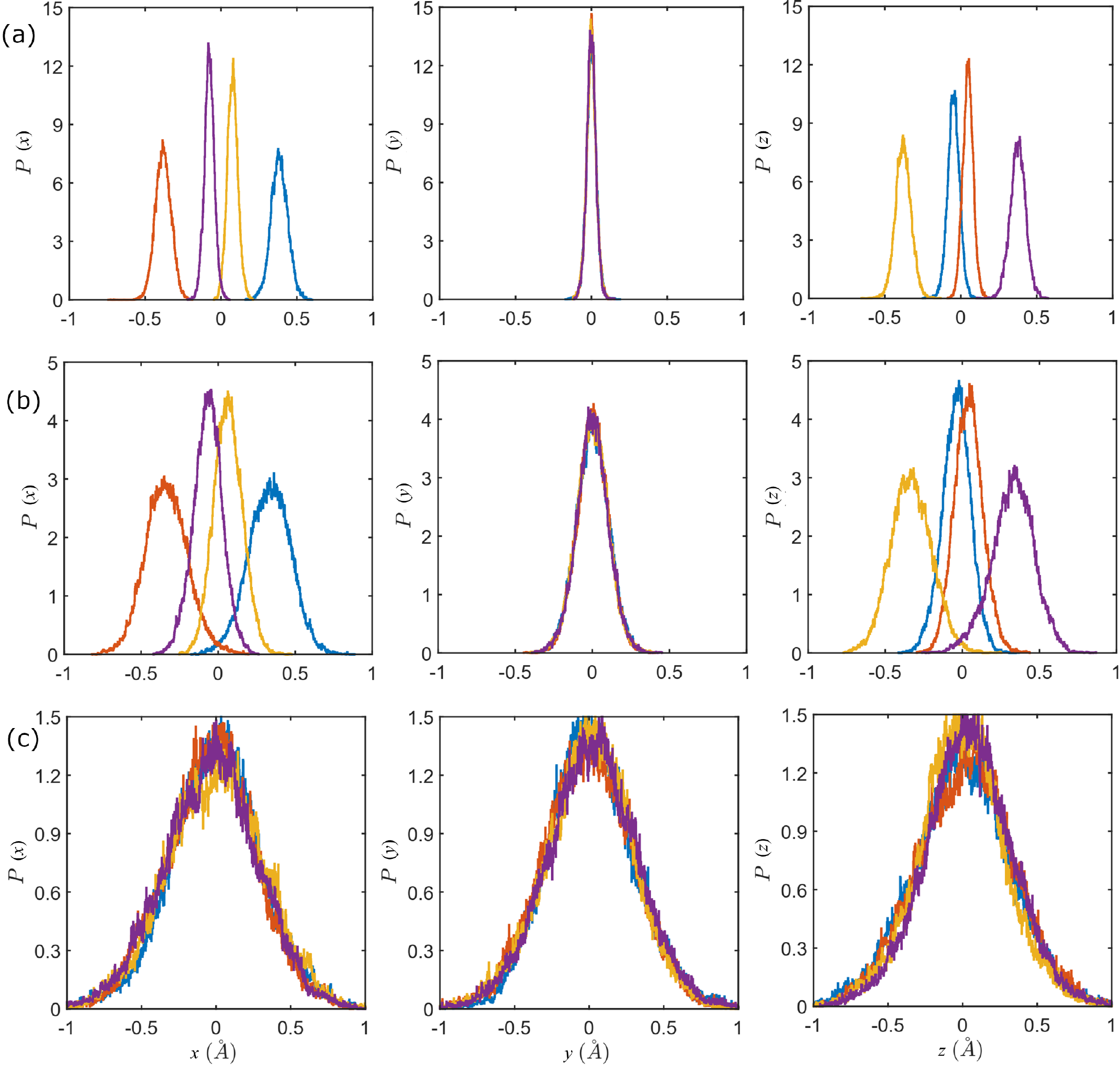}
 \caption{\footnotesize{Distribution function for x, y and z components for MA displacements $p^{MA}_{dis}$ at (a) 40 (b) 180 and (c) 300 K. Contributions are decomposed in blue, orange, yellow and indigo representing $v1$, $v2$, $w1$ and $w2$ type MAs, respectively.}} 
 \label{MA_trans_dis}
\end{figure*}

\begin{figure*}[tbh!]
 \centering
 \includegraphics[width=0.6\textwidth]{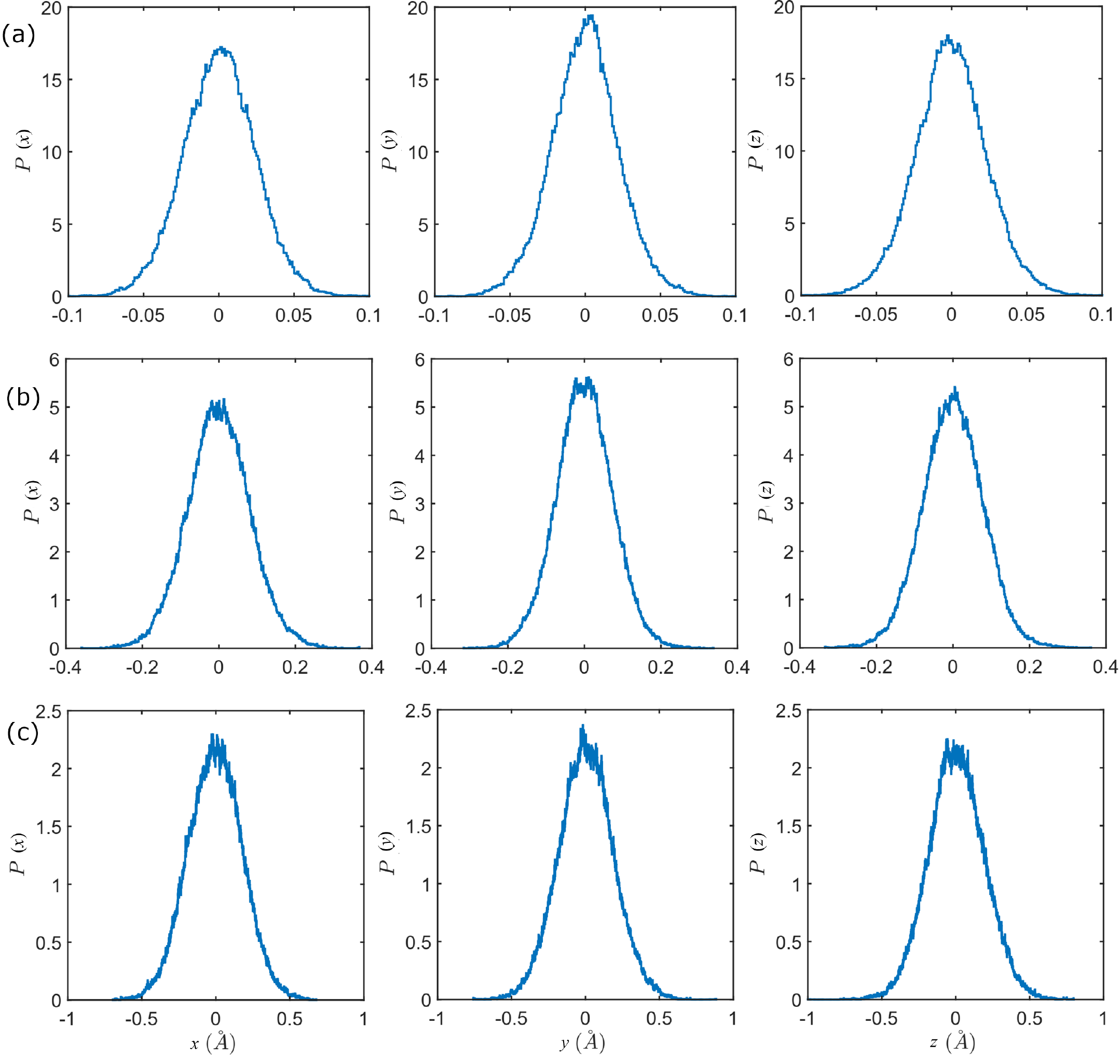}
 \caption{\footnotesize{Distribution function for x, y and z components for lead displacements $p^{Pb}_{dis}$ at (a) 40 (b) 180 and (c) 300 K.}} 
 \label{Pb_trans_dis}
\end{figure*}

\begin{figure*}[tbh!]
 \centering
 \includegraphics[width=0.6\textwidth]{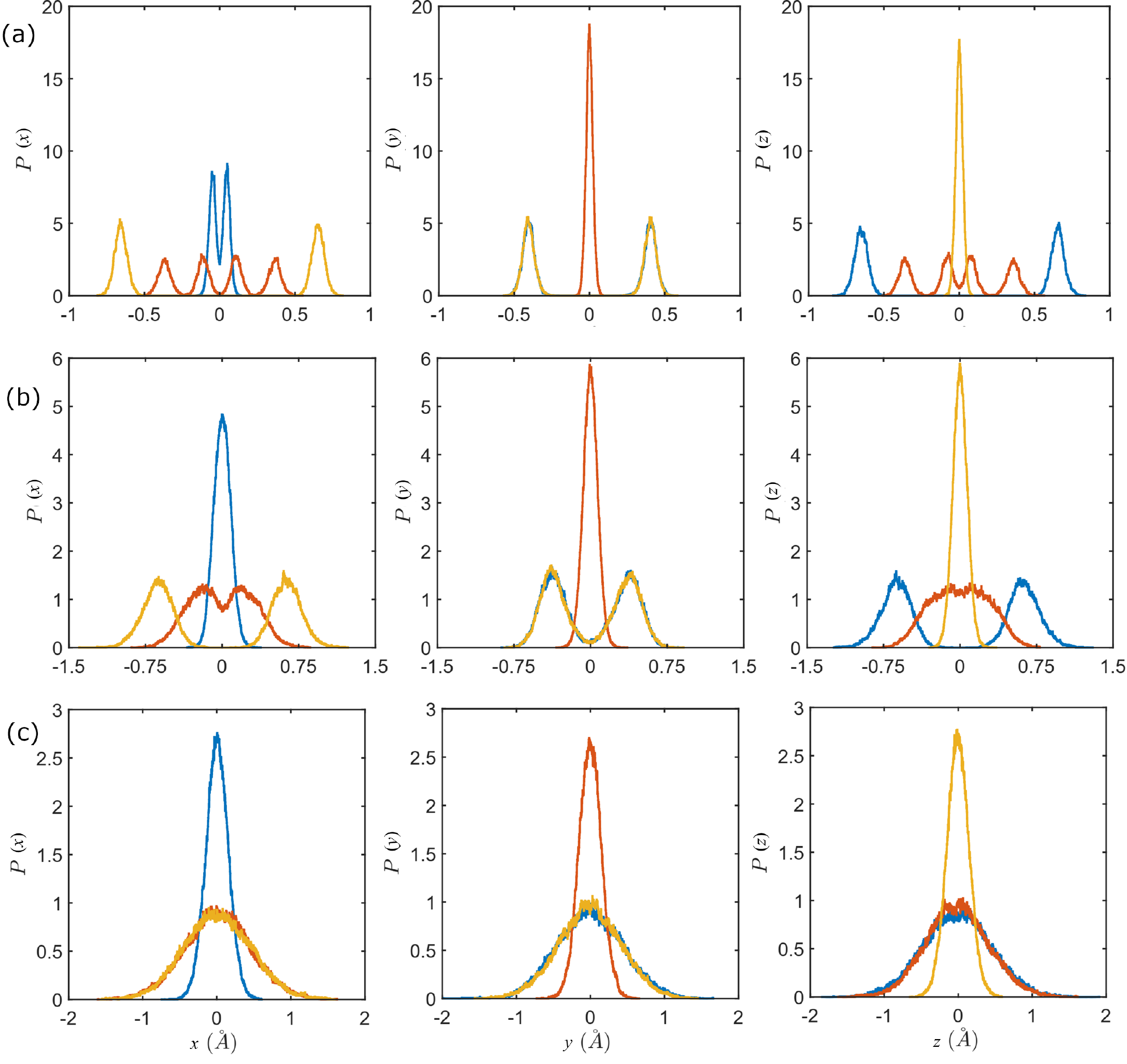}
 \caption{\footnotesize{Distribution function for x, y and z components for bromine displacements $p^{Br}_{dis}$ at (a) 40 (b) 180 and (c) 300 K. Contributions are decomposed in blue, orange and yellow representing $p$, $q$ and $r$ type bromines, respectively.}} 
 \label{Br_trans_dis}
\end{figure*}

The instantaneous symmetry breaking can happen due to four possibilities such as MA rotation, MA displacement, lead displacement, and bromine displacement. Each of them is quantified as:  $P^{MA}_{rot}(t)=\frac{1}{64}\sum_{k=1}^{64}p^{MA}_{rot,k}(t)$, $P^{MA}_{dis}(t)=\frac{1}{64}\sum_{k=1}^{64}p^{MA}_{dis,k}(t)$, $P^{Pb}_{dis}(t)=\frac{1}{64}\sum_{k=1}^{64}2p^{Pb}_{dis,k}(t)$ and $P^{Br}_{dis}(t)=\frac{3}{192}\sum_{k=1}^{192}p^{Br}_{dis,k}(t)$, respectively, where:
\begin{equation}\label{eq:e1}
 p^{MA}_{rot}(t)=\hat{r}_{k}^{C \to N}(t) 
\end{equation}
\begin{equation}\label{eq:e2}
 p^{MA}_{dis}(t)=\overrightarrow{r}_{k}^{MA}-\frac{1}{12}\sum_{j=1}^{12}\overrightarrow{r}_{Br,j}^k
\end{equation}
\begin{equation}\label{eq:e3}
 p^{Pb}_{dis}(t)=\overrightarrow{r}_{k}^{Pb}-\frac{1}{6}\sum_{j=1}^{6}\overrightarrow{r}_{Br,j}^k
\end{equation}
\begin{equation}\label{eq:e4}
 p^{Br}_{dis}(t)=\overrightarrow{r}_{k}^{Br}-\frac{1}{8}\sum_{j=1}^{4}\overrightarrow{r}_{MA,j}^k-\frac{1}{4}\sum_{j=1}^{2}\overrightarrow{r}_{Pb,j}^k 
\end{equation}
$\hat{r}_{k}^{C \to N}$ is the unit-vector connecting C and N; $\overrightarrow{r}_{k}^{MA}$, $\overrightarrow{r}_{k}^{Pb}$, $\overrightarrow{r}_{k}^{Br}$ are the center-of-mass of MA, position of Pb and Br; and $\overrightarrow{r}_{Br,j}$, $\overrightarrow{r}_{MA,j}$, $\overrightarrow{r}_{Pb,j}$ are the nearest neighbors. The nominal charges (+1 for MA, +2 for Pb, and -1 for Br) are used to calculate polarization values. The zero value of them indicates the absence of any polarization. The global polarization is calculated as the mean of distributions of all the parameters. For MA, Pb, and Br displacements, the means of the distributions for each component are close to zero indicating that they do not contribute to polarization. However, at different temperatures, the origin of zero polarization can differ. For example, at 40K, in the case of MA and Br displacements, there are different types of atoms each contributing to non-zero polarization peaks which are distributed symmetrically about zero, thus resulting in overall zero values on average. On the other hand, at 300 K, these atoms become indistinguishable due to disorder and they all contribute to peaks at zero. The effect of MA rotations shows interesting behavior. At all temperature, the average polarization is close to zero (the maximum value for MA rotational component is at 300 K: 0.02-0.03 ${\rm \AA}\times$charge/768 atom cell). 

\section{Static vs. dynamic MA disorder?}
Characterization of motion of the organic cation MA (CH\textsubscript{3}NH\textsubscript{3}$^+$) inside the Pb-Br lattice is crucial to obtain a clear understanding of the effect of these molecular ions on various structural, vibrational and electronic features of the hybrid perovskite. The \textit{ab initio} MD simulations at finite temperature (and ambient pressure) allow us to probe the dynamics and various correlations of individual or ensemble of dipoles at microscopic scale as well as a shorter timescale. Considering the CH\textsubscript{3}-NH\textsubscript{3}$^+$ axis unit vector ($\hat{n}_i$) as the dipole corresponding to every MA molecule we have constructed first and second-order auto (self)-correlation functions (ACFs) $C_1(t)$ and $C_2(t)$ at time $t$, respectively, as follows:
\begin{equation}\label{eq:b1}
 C_1(t)=\frac{1}{N}\sum_{i=1}^{N} \langle P_1(\hat{n}_i(0)\cdot\hat{n}_i(t)) \rangle
\end{equation}
\begin{equation}\label{eq:b2}
 C_2(t)=\frac{1}{N}\sum_{i=1}^{N} \langle P_2(\hat{n}_i(0)\cdot\hat{n}_i(t)) \rangle
\end{equation}
Where N is the total number of MAs considered in simulation; $P_1(x)=x$ and $P_2(x)=\frac{1}{2}(3x^2-1)$ are the first and second-order Legendre polynomial, respectively; $\hat{n}_i(0)$ is the initial orientation of the $i$-th dipole; and $\langle...\rangle$ indicates time average. These ACFs are calculated using configurations at every 1.20945 \textit{\textit{fs}} time-interval for total 5 \textit{\textit{ps}}, using only the equilibrated parts of all trajectories. This small enough time-interval, as well as usage of sufficient statistics obtained from longer MD trajectories ($\gg5$ \textit{\textit{ps}}) ensure capturing of fast dynamics of MAs and accurate estimation of decay times at any temperature. In general, the decay of these functions over time gives the idea of the amount of time taken to lose the memory of any arbitrary starting orientation of the MA.

\begin{table*}[t!]
   \centering
    \begin{tabular}{|c|c|c|c|c|} 
    \hline
    \multicolumn{1}{|c|}{\textbf{Temperature}} & \multicolumn{1}{|c|}{\textbf{Fitting type}} & \multicolumn{3}{c|}{\textbf{Fitted timescales}} \\
    \cline{3-5}
    \textbf{} & \textbf{} & \textbf{$\tau_{1}$} & \textbf{$\tau_{2}$} & \textbf{$\tau_{0}$} \\
    \hline
    40 K & $C_1^M(t)$ & 99.909 fs & 338.646 fs & $\infty$ \\
    \hline
    180 K & $C_1^M(t)$ & 100.227 fs & 1.085 ps & $\infty$ \\
    \hline
    300 K & $C_1^M(t)$ & 341 fs & 545 fs & 2.035 ps \\
    \cline{2-5}
    & $C_2^{WCM}(t)$ & 320 fs & - & 1.863 ps \\
    \hline
    300 K & Experiments & 355 fs~\cite{wasylishen1985cation}, & &    \\
    & & 140$\pm$10 fs~\cite{zhu2016screening}, & 400-600 fs~\cite{zhu2016screening} & $\sim$2 ps~\cite{zhu2016screening}, \\
    & & 300$\pm$100 fs~\cite{selig2017organic} & & 1.5$\pm$0.3~\cite{selig2017organic} \\
    \cline{1-5}
   \end{tabular}
  \caption{\footnotesize{Summary of MA rotational timescales fitted with $C_1^M(t)$ and $C_2^{WCM}(t)$ model obtained from \textit{ab initio} MD in MAPbBr\textsubscript{3} at 40, 180 and 300 K, along with experimentally extracted data. For $C_1^M(t)$ and $C_2^{WCM}(t)$ type fitting, 2 and 5 $ps$ of time correlation data are used, respectively. Due to lack of reorientational jump the $\tau_0$ values at 40 and 180 K, are very large, indicated as infinity.}}
  \label{correlation_timescales}
\end{table*}

The $C_{1}(t)$ ACFs at 40, 180, and 300 K are shown in main text in Figure 2. It is evident that at 40 K it remains constant on average, however, they show oscillatory behavior which gets damped over time. The constant behavior on average confirms that, at equilibrium, the rotations of MAs are severely restricted and they get stuck in their corresponding minima in potential energy surface. At $t\to0$, a very weak decay is observed which is not simple exponential in nature. The damped oscillations at 40 K are due to the coupled motion of the molecule and lattice. The qualitative nature of $C_{1}(t)$ obtained at 180 K is the same as all of the above-mentioned features are present. However, at 180 K, the decay at $t\to0$ is more than 40 K but doesn't go to zero over time and overall it is of lesser magnitude than 40 K, suggesting less ordered motion of MAs at 180 K, which could indicate a partial disorder. It is worth mentioning that in the simulation time of $\sim$20 ps at 180 K, orthorhombic to tetragonal transition does not happen, due to which characterization of disorder requires more simulation time. 
On the other hand, at 300 K, the $C_{1}(t)$ ACF decay to zero exponentially over a certain time indicating motion of the cations become completely uncorrelated which is indicative of a complete disordered phase. However, at $t\to0$, a fast decay governed by powered exponential law is observed. In this faster timescale, MAs fluctuate around one of the local potential energy minima before hopping to another adjacent equilibrium at a slower timescale, suggesting ``wobbling in a cone" and ``reorientation jump" respectively~\cite{gallop2018rotational}. MA dynamics particularly in the high temperature (pseudo-)cubic phase of OHIPs have recently generated extreme interest and many experimental efforts by means of dielectric relaxation~\cite{poglitsch1987dynamic}, solid-state NMR (ssNMR)~\cite{bernard2018methylammonium,wasylishen1985cation,kubicki2017cation,fabini2017universal}, quasi-elastic neutron scattering (QENS)~\cite{leguy2015dynamics,chen2015rotational}, time-resolved optical Kerr effect spectroscopy (TR-OKE)~\cite{zhu2016screening}, and polarization-resolved 2-dimensional IR (2DIR) spectroscopy~\cite{taylor2018investigating,bakulin2015real,selig2017organic} have been made. Here we make connections with some of these experimental findings done in MAPbBr\textsubscript{3}~\cite{selig2017organic,wasylishen1985cation,zhu2016screening}. 

To extract the relavent timescales of MA motion, a general diffusive model proposed by Mattoni \textit{et al.}~\cite{mattoni2015methylammonium} ($C_1^{M}(t)$) and a wobbling-in-a-cone/jump model~\cite{ji2011orientational,lipari1982model} ($C_2^{WCM}(t)$) have been employed as described in the main text. 
\begin{figure*}[tbh!]
 \centering
 \includegraphics[width=1.00\textwidth]{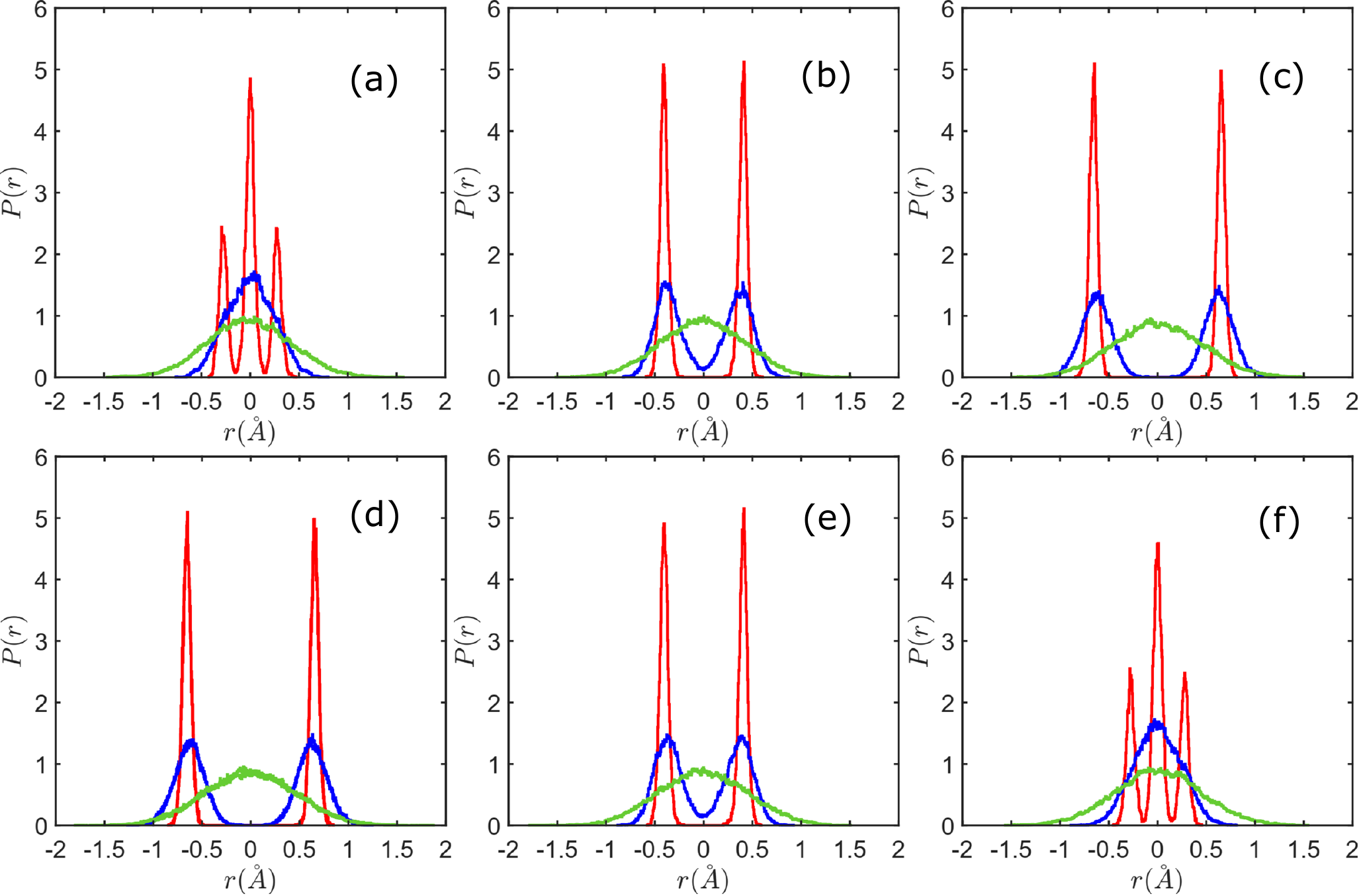}
 \caption{\footnotesize{Distribution of various $D^{\mu}_{\nu}$ ($\nu$ component displacement of Br situated along $\mu$) as (a) $D_{c}^q$, (b)$D_{b}^r$, (c)$D_{a}^r$, (d)$D_{c}^p$, (e)$D_{b}^p$, (f) $D_{a}^q$ at 40 K (red), 180 K (blue) and 300 K (green). The three $\mu$ are p (along a axis), q (along b axis) and r (along c axis).}}  
 \label{br_glazer}
\end{figure*}
Upon fitting, all the extracted timescales at various temperatures are shown in Table~\ref{correlation_timescales} along with experimental findings.
Time-resolved optical Kerr effect (TR-OKE) spectroscopy experiments \cite{zhu2016screening} in MAPBBr\textsubscript{3} identified three regimes of motion (140$\pm$10 \textit{fs}, 400-600 \textit{fs} and $\sim$2 $ps$). Also in the MD correltion plots $C_1(t)$ and $C_2(t)$, right after the rapid decay, an oscillatory region is observed before the longer exponential decay starts. These three timescales can be extracted from $C_1^M(t)$ as $\tau_1$, $\tau_2$ and $\tau_0$, approximately ($\sim$341 \textit{fs}, $\sim$545 $ps$ and $\sim$2.035 $ps$ ). The ultrafast response is caused by initial rapid motion inside wobbling-cone which is like the motion of almost free-rotor. However, immediately the local interaction of molecule with lattice restricts the motion up to maximum cone-angle resulting in a intermediate response. After that the slow response is caused by the diffusive reorientation.  On the other hand, the calculated timecales from $C_2^{WCM}(t)$ ($\tau_0=1.86$ ps and $\tau_1=320$ fs) are in excellent agreement with corresponding time-scales (1.5$\pm$0.3 ps,  300$\pm$100 fs) extracted from polarization resolved 2-D IR spectroscopic measurements~\cite{selig2017organic,lin2010calculation}, as described in the main text. 





\begin{figure*}[t!]
 \centering
 \subfigure[]{\includegraphics[width=0.46\textwidth]{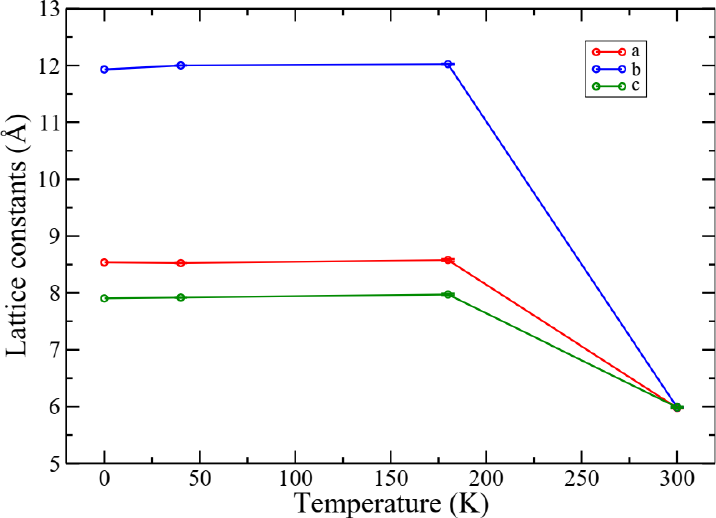}}\hspace{0.02\textwidth} 
 \subfigure[]{\includegraphics[width=0.46\textwidth]{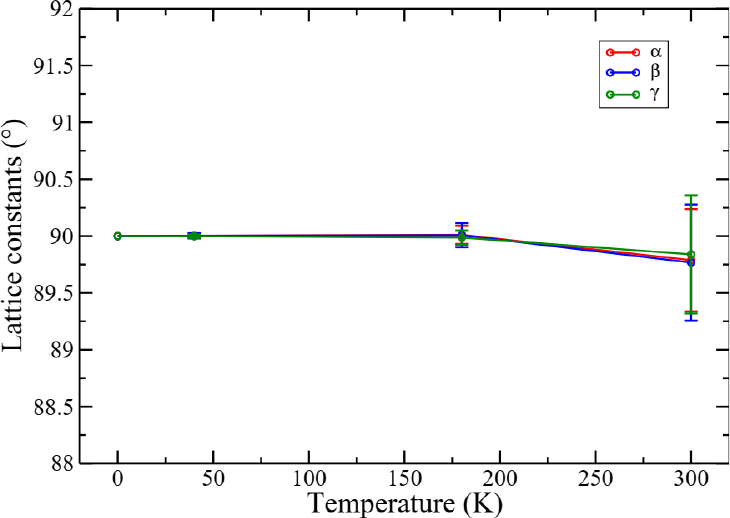}} 
 \caption{\footnotesize{Evolution of lattice parameters with temperature. In (a) crystal axes lengths and, (b) crystal angles are shown. In some cases, the associated errors of means as mentioned in Table~\ref{lattice_parameter} are small compared to the y-axis scales in the plots.}}
 \label{latpam_T}
\end{figure*}


\begin{table*}[tbh!]
   \centering
    \begin{tabular}{|c|c|c|c|c|c|c|c|} 
    \hline
    \multicolumn{1}{|c|}{\textbf{Temperature}} & \multicolumn{1}{|c|}{\textbf{lattice}} & \multicolumn{1}{|c|}{\textbf{$a$ (\AA)}} &
    \multicolumn{1}{|c|}{\textbf{$b$ (\AA)}} &
    \multicolumn{1}{|c|}{\textbf{$c$ (\AA)}} &
    \multicolumn{1}{|c|}{\textbf{$\alpha$ ($^o$)}} &
    \multicolumn{1}{|c|}{\textbf{$\beta$ ($^o$)}} &
    \multicolumn{1}{|c|}{\textbf{$\gamma$ ($^o$)}} \\
    \hline
    & PC-l & 5.818 & 6.001 & 5.818 & 89.999 & 85.822 & 89.998 \\
    & (err) & ($\pm$0.005) & ($\pm$0.004) & ($\pm$0.006) & ($\pm$0.090) & ($\pm$0.113) & ($\pm$0.085) \\
    \cline{2-8}
    40 K & PC-g & 5.818 & 6.001 & 5.818 & 90.000 & 85.823 & 89.999 \\
    & (err) & ($\pm$0.001) & ($\pm$0.001) & ($\pm$0.001) & ($\pm$0.007) & ($\pm$0.020) & ($\pm$0.006) \\
    \cline{2-8}
    & OR-g & 8.525 & 12.001 & 7.919 & 90.002 & 90.003 & 89.999 \\
    & (err) & ($\pm$0.004) & ($\pm$0.003) & ($\pm$0.004) & ($\pm$0.021) & ($\pm$0.023) & ($\pm$0.015) \\
    \cline{2-8}    
    \hline
    & PC-l & 5.858 & 6.015 & 5.858 & 89.988 & 85.925 & 89.970 \\
    & (err) & ($\pm$0.024) & ($\pm$0.014) & ($\pm$0.022) & ($\pm$0.320) & ($\pm$0.414) & ($\pm$0.324) \\
    \cline{2-8}
    180 K & PC-g & 5.856 & 6.012 & 5.855 & 90.003 & 85.931 & 89.982 \\
    & (err) & ($\pm$0.006) & ($\pm$0.004) & ($\pm$0.006) & ($\pm$0.053) & ($\pm$0.064) & ($\pm$0.040) \\
    \cline{2-8}
    & OR-g & 8.578 & 12.024 & 7.973 & 90.011 & 90.007 & 89.987 \\
    & (err) & ($\pm$0.020) & ($\pm$0.007) & ($\pm$0.014) & ($\pm$0.080) & ($\pm$0.106) & ($\pm$0.063) \\
    \cline{2-8}
    \hline
    & PC-l & 5.990 & 5.999 & 6.001 & 89.853 & 89.712 & 89.992 \\
    & (err) & ($\pm$0.070) & ($\pm$0.070) & ($\pm$0.070) & ($\pm$1.250) & ($\pm$1.290) & ($\pm$1.270) \\
    \cline{2-8}    
    300 K & PC-g & 5.978 & 5.987 & 5.989 & 89.906 & 89.757 & 90.038 \\
    & (err) & ($\pm$0.008) & ($\pm$0.011) & ($\pm$0.016) & ($\pm$0.260) & ($\pm$0.240) & ($\pm$0.280) \\
    \cline{2-8}    
    & OR-g & 8.492 & 11.973 & 8.455 & 89.788 & 89.767 & 89.839 \\
    & (err) & ($\pm$0.060) & ($\pm$0.022) & ($\pm$0.057) & ($\pm$0.450) & ($\pm$0.511) & ($\pm$0.520) \\
    \hline
    \end{tabular}
  \caption{\footnotesize{Global (g) and local (l) lattice parameters obtained from molecular dynamics simulation of 768 atom cell at various temperatures. PC refers to pseudo-cubic and OR refers to orthorhombic type lattice parameters. For comparison purposes, global lattice parameters are scaled to the corresponding smallest unit cell. The associated errors (err) are shown in bracket which is calculated as standard deviation of means of randomly chosen 500 blocks from the trajectory, each block having 2000 time-frames.}}
  \label{lattice_parameter}
\end{table*}

In MAPbBr$_3$ the inorganic lattice consists of lead and Bromine ions. Apart from the CH$_3$NH$_3^+$ motion it’s also interesting to look at motion of the PbBr$_3^+$ lattice, particularly of nature of positional fluctuations of bromine ions as they indicate the rotation of PbBr$_6$ octahedra. With respect to the local (pseudo-)cubic axes, three unique bromines ($p$,$q$ and $r$, each type having 64 Br among 768 atoms) are identified which are along local $a$, $b$ and $c$ axes, respectively. Now for every types of bromine we can define three unique displacements (Total 9 types of displacements) as displacements of $\nu$ component (a, b and c components). Among them 6 following order parameters are used to describe the motion of bromines leading to the idea of Glazer symbol\cite{glazer1972classification}: (1) Looking along $a$ axis - $D_{c}^q$ and $D_{b}^r$; (2) Looking along $b$ axis - $D_{a}^r$ and $D_{c}^p$; (3) Looking along $c$ axis - $D_{b}^p$ and $D_{a}^q$.

At 40 K the expected Glazer notation of the $Pnma$ phase is $a^-b^+a^-$. Looking along $a$ axis, $D_{c}^q$ have trinodal distribution (0, $\sim\pm$0.3 \AA) where all of the peaks have contribution due to each group of bromines sitting alternately along $b$ and $D_{b}^r$ have bimodal distribution ($\sim\pm$0.4 \AA) where all of the peaks have contribution due to each group of bromines sitting alternately along $c$, thus confirming $a^-$ notation (see Figure~\ref{br_glazer}). Looking along the $b$ axis, $D_{a}^r$ have bimodal distribution ($\sim\pm$0.7 \AA) where one of the peaks have contribution only due to one group of bromines (other peaks have contribution only due to another group of bromines) sitting alternately along $c$ and same is true for $D_{c}^p$ for bromines sitting alternately along $a$, thus confirming $b^+$ notation (as one group results into one peak at non-zero value). Looking along the $c$ axis, $D_{b}^p$ have bimodal distribution ($\sim\pm$0.4 \AA) where all of the peaks have contribution due to each group of bromines sitting alternately along with $a$ and $D_{a}^q$ have trinodal distribution (0, $\sim\pm$0.3 \AA) where all of the peaks have contribution due to each group of bromines sitting alternately along $b$, thus confirming $a^-$ notation (as positions of peaks are also same when looking along $a$ axis). 

\begin{figure*}[t!]
 \centering
  \subfigure[]{\includegraphics[width=0.6\textwidth]{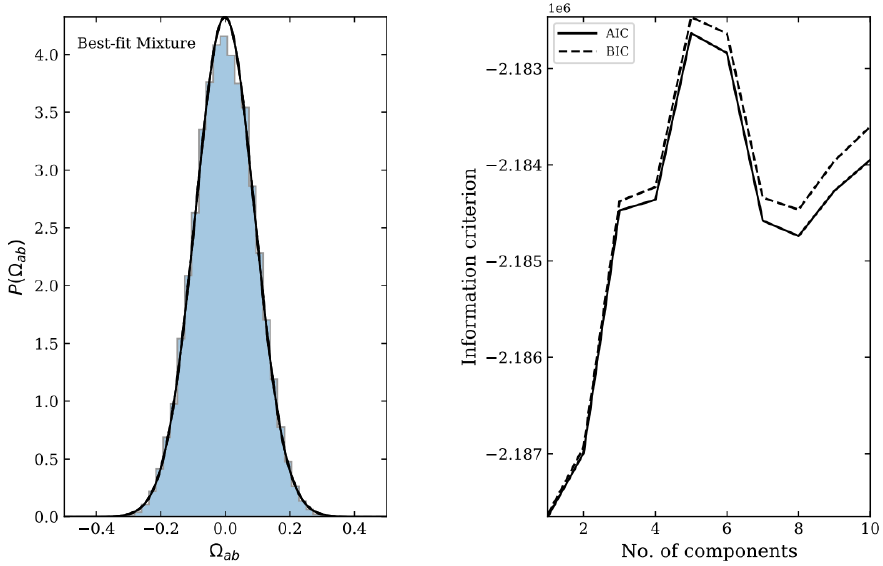}}
  \subfigure[]{\includegraphics[width=0.6\textwidth]{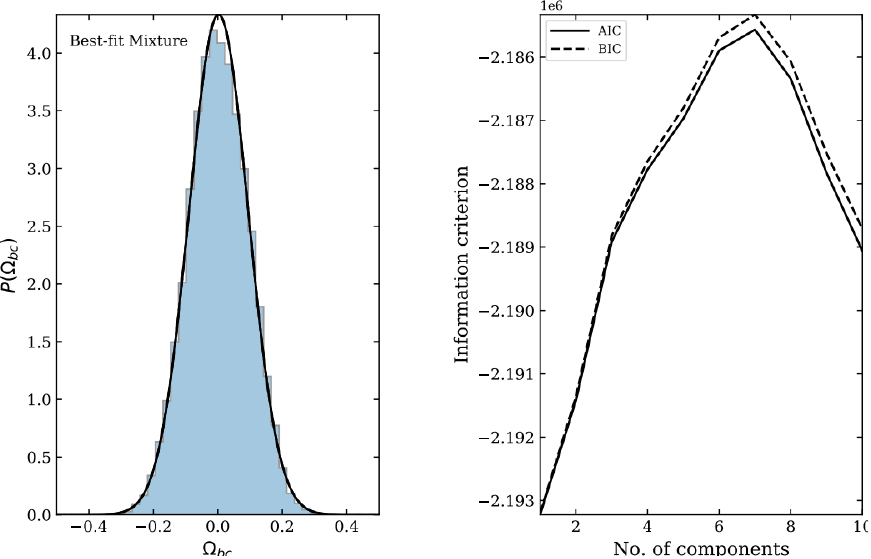}}
  \subfigure[]{\includegraphics[width=0.6\textwidth]{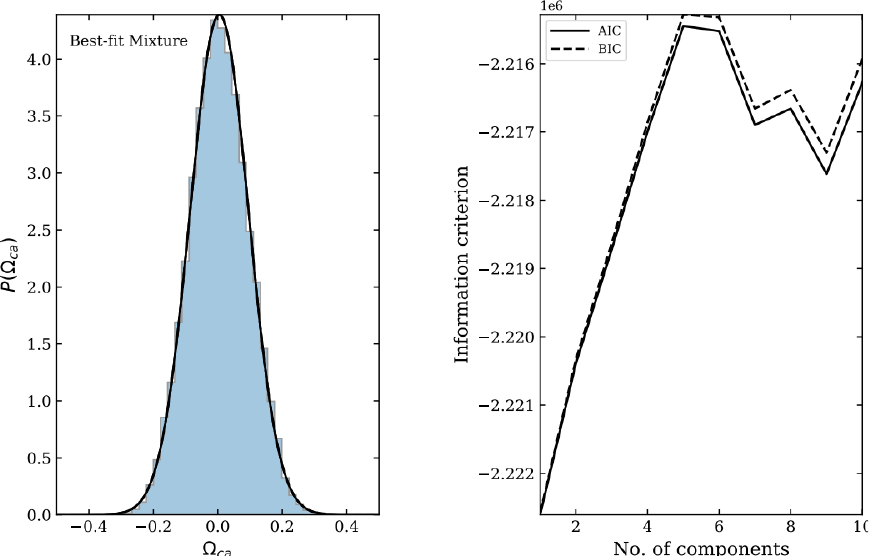}}
 \caption{\footnotesize{1D-GMM fit for $\Omega_{ab}$, $\Omega_{bc}$ and $\Omega_{ca}$ at 300 K with component analysis. AIC and BIC refer to Akaike's Information Criteria and Bayesian Information Criteria respectively, which are minimized for the correct number of components.}} 
 \label{gmm_ab_scissor}
\end{figure*}

At 180 K, very similar observations are made except for the $D_{c}^q$ and $D_{a}^q$, where the trinodal nature at 40 K is lost, instead they merge to one at 0 \AA (see Figure~\ref{br_glazer}). Thus when looking along (pseudo-)cubic $a$ and $c$ axes, one of the order parameter have zero fluctuation on average (e.g. $a^0$) but the other indicates opposite types of rotation of the octahedra on average (e.g. $a^-$). As in the tetragonal $I4/mcm$ phase, Glazer notation is $a^0c^-a^0$, this may indicate the transformation into tetragonal phase is not yet happened, indeed a mixed type Glazer notation is seen.

At 300 K, all of the order parameters are unimodal whose peaks are at 0 \AA. This indicates on average there exists no octahedral rotation confirming the Glazer symbol as $a^0a^0a^0$ which is the same as the cubic \textit{Pm-3m} phase (see Figure~\ref{br_glazer}).

\section{Is the $\alpha$ phase locally distorted?}
The calculated lattice constants are shown in TABLE~\ref{lattice_parameter}. At 40 K, OR model fits extremely well as described in the main text. Surprisingly at 180 K, a similar observation is made instead of overall tetragonality. It is important to note that, the simulation timescale of ~20 $ps$ at 180 K, is not enough to realize the transition into the tetragonal cell. At 300 K presence of (pseudo-)cubic cell is confirmed, however, the existence of orthorhombic or tetragonal distortion is also proven, as described in the main text. The evolution of the lattice constants with temperature is shown in Figure~\ref{latpam_T}.

The local structural information and its stability and dynamical change throughout the simulation can be investigated by pair distribution functions. This gives the idea about how, on average, atoms in a system are radially packed around each other, thus also providing an effective way of describing order or disorderliness in the system. Mathematically, it is defined by the density of atoms $g(r)=\rho(r)/\rho$, where $\rho(r)$ is the local density at a given radius $r$ and $\rho$ is the overall density of the system. At any time $t$ of the simulation, the radial distribution function can be calculated as follows:  
\begin{equation}\label{eq:d1}
 g_t(r)=\frac{1}{V_{t}/N}\frac{<N_{t}(r\pm \frac{\Delta r}{2})>}{<\Omega_{t}(r\pm \frac{\Delta r}{2})>}
\end{equation}
Where the first term represents the density $\rho_t$ as the total number of particle $N$ divided by volume of the system $V_t$ and the second term represents the local density $\rho_t(r)$ as the number of particles \textless$N_{t}$\textgreater\ at time $t$ in the distance interval $r\pm \frac{\Delta r}{2}$ divided by the volume of that shell \textless$\Omega_{t}$\textgreater\ of width $\Delta r$. 

\begin{figure*}[tbh!]
 \centering
 \subfigure[]{\includegraphics[width=0.48\textwidth]{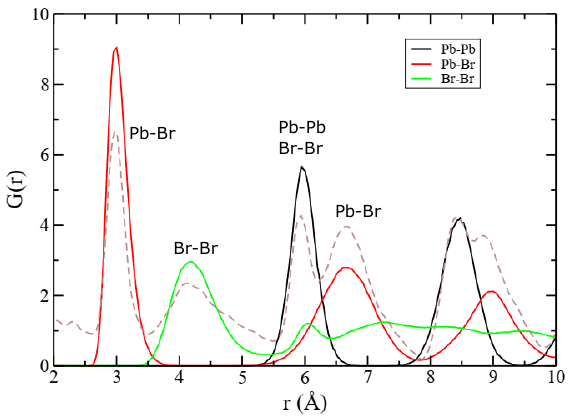}} 
 \subfigure[]{\includegraphics[width=0.48\textwidth]{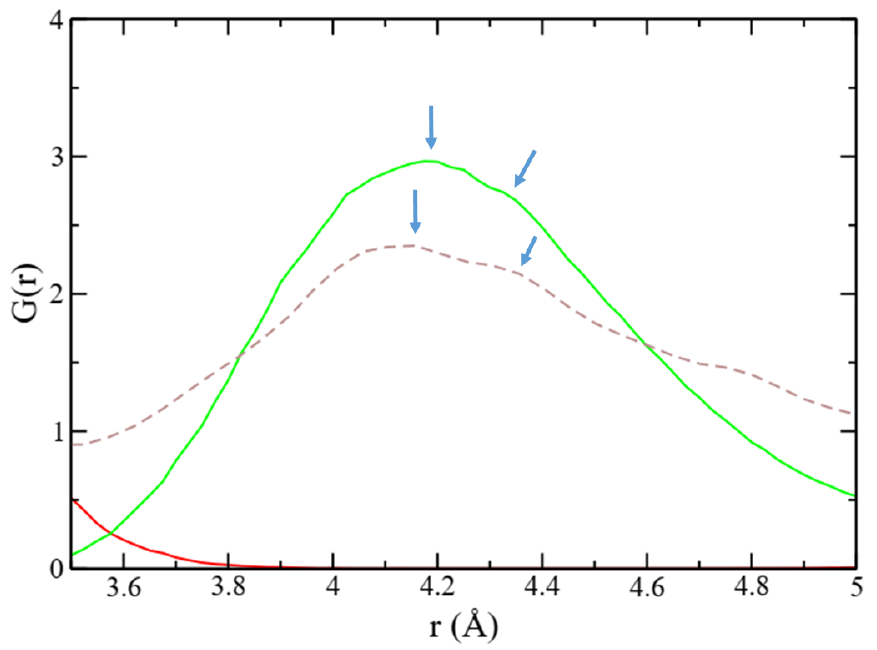}} 
 \caption{\footnotesize{(a) Comparison of radial distribution function at 300 K. Continuous lines are calculated and dashed line is from experiment~\cite{bernasconi2017direct}. The contributions of pairs suggested by the experiment are written near peaks. (b) Radial distribution function of Br-Br pair (G(r)\{Br-Br\}) at 300 K in 3.5-5 \AA~region. Arrows represent possible positions of doublet peaks.}} 
 \label{rdf_exp_Br}
\end{figure*}

\begin{figure*}[tbh!]
 \centering
 \subfigure[]{\includegraphics[width=0.48\textwidth]{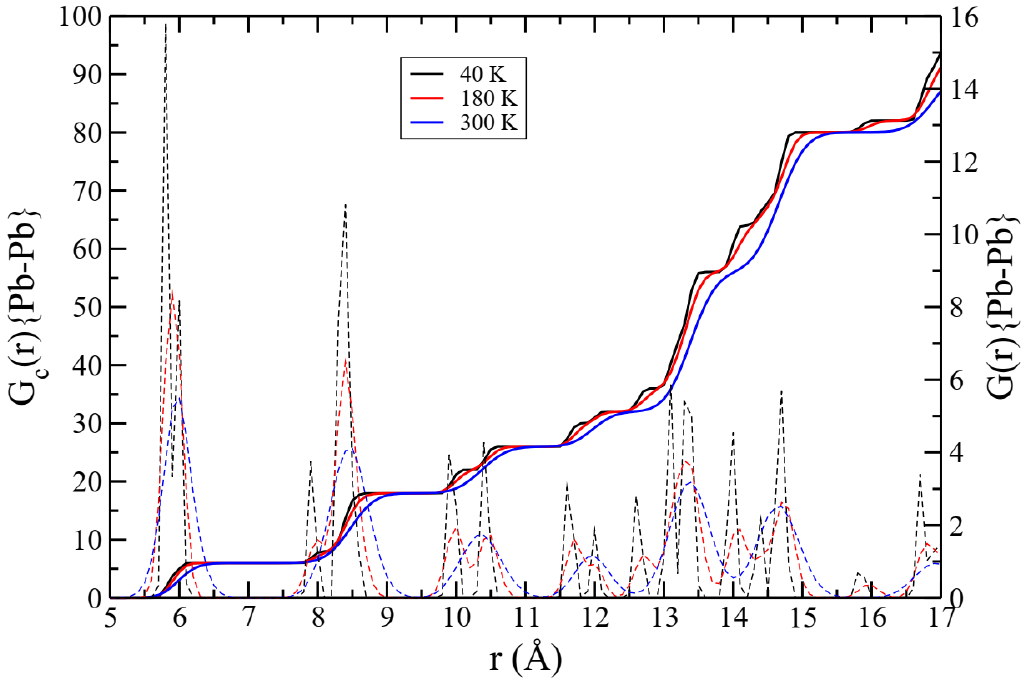}} 
 \subfigure[]{\includegraphics[width=0.48\textwidth]{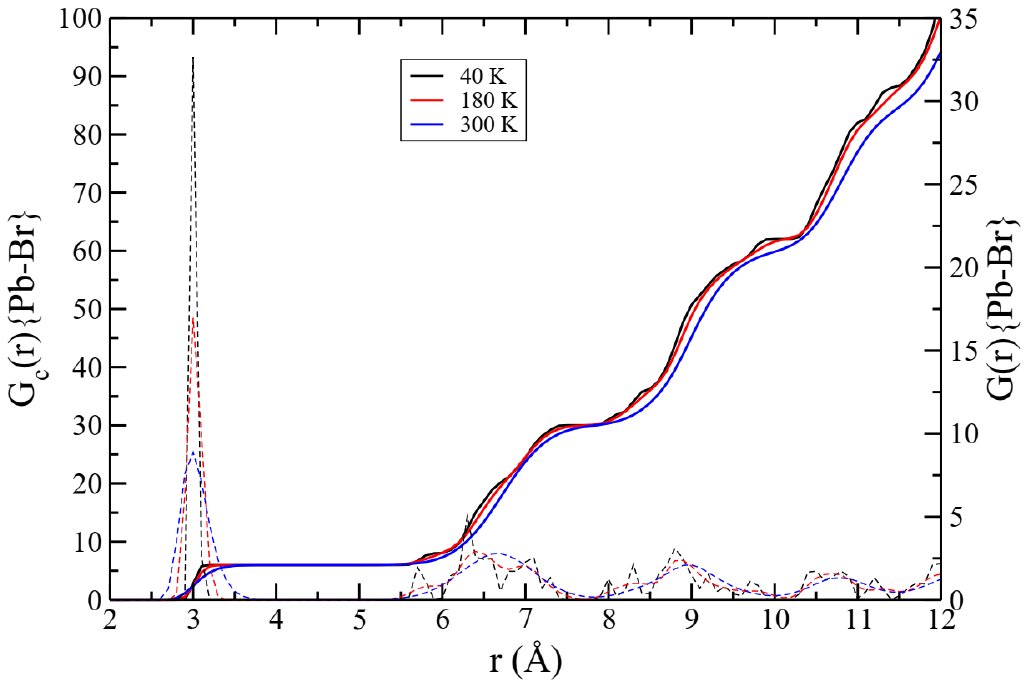}} 
 \caption{\footnotesize{Radial distribution functions $G(r)$ (dashed lines and right X axis) and cumulative radial distribution functions $G_c(r)$ (continuous lines and left X axis) at different temperatures between (a) Pb-Pb and (b) Pb-Br atoms.}} 
 \label{rdf_Pb_Br}
\end{figure*}


The coordination of different atoms are also calculated as $g_{t}^c(r)=\int 4\pi r^{2}\rho_t(r)g_t(r)dr$, displayed by cumulative radial distribution function. To get the overall idea of these distribution function accounting for many snapshots over all simulation time, we calculate the average over time as defined by $G(r)=\frac{1}{T}\sum_{k=1}^{T}g_k(r)$ and $G_{c}(r)=\frac{1}{T}\sum_{k=1}^{T}g_{k}^c(r)$. For averaging we have taken snapshots of structures at every 12.0945 \textit{fs} during simulation time.

\begin{figure*}[tbh!]
 \centering
 \includegraphics[width=0.8\textwidth]{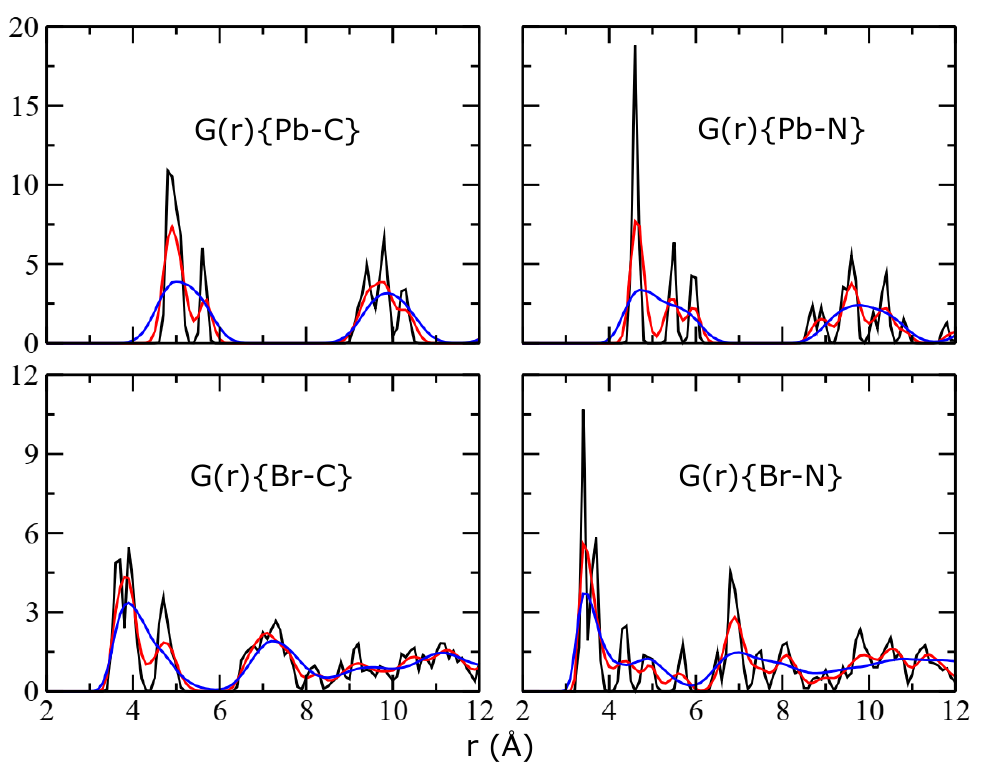}
 \caption{\footnotesize{Radial distribution functions $G(r)$ for (a) Pb-C, (b) Pb-N, (c) Br-C and (d) Br-N pairs. The black, red, and blue lines correspond to 40, 180, and 300 K, respectively.}} 
 \label{rdf_C_N}
\end{figure*}

The 768 atom cell spans 20-25 \AA\ in all three Cartesian directions which allows us to calculate pair distribution and cumulative functions properly up to $\sim$20 \AA, easily revealing the features and co-ordinations not only around first and second nearest neighbors but also around higher neighbors. These functions for the Pb-Pb pair have been shown in Figure~\ref{rdf_Pb_Br}(a). Radial distributions of $G(r)\{Pb-Pb\}$ consist of peaks whose sharpness decreases with temperature as at higher temperatures they become more spread out over distances. The second, third, and seventh peaks, at 7.9, 8.5, and 12 \AA\ at 40 K are due to the lead boundaries corresponding to the orthorhombic unit cell. Also, the regions around the first peaks (at $\sim$6 \AA) at different temperatures are of significant interest as the 12 atoms (pseudo-)cubic unit cell's periodicity lies there. The thermal motions of leads at 40 K are so restricted that the first peak splits into two at 5.8 and 6.0 \AA\ respectively, clearly indicating the existence of tetragonal distortion due to the choice of mentioned smallest cell. However, at 180 K, these two sub-peaks merge into one around their average position at 5.9 \AA\ due to thermal broadening, as the resolution of 0.1 \AA\ in the distance is not enough to distinguish the features for tetragonal distortion also confirmed from the $PC-l$ data shown in Table-\ref{lattice_parameter}. On the other hand, At 300 K, one broad peak at 6 \AA\ is seen due to the proper (pseudo-)cubic nature of the cell without the mentioned distortion. Both of the plots shown in Figure~\ref{rdf_Pb_Br} also indicate that the extent of fluctuations in the motion of Pb and Br increases with temperature as sharper peaks at lower temperatures become more spread out and also merge at the higher temperatures. For example, in the case of $G(r)\{Pb-Pb\}$, around ~9.5-11 \AA\, two separate peaks (at 9.9 and 10.4 \AA) observed at 40 K, shift towards right at 180 K due to thermal expansion and merge into one having two sub-peaks (at 10.0 and 10.5 \AA). At 300 K they completely become one single broad peak having its maximum at 10.4 \AA). These changes are also reflected as the changes in slope at those mentioned distances in the corresponding cumulative distribution $G_c(r)\{Pb-Pb\}$. In the $G_c(r)$ plots, the number of neighbors or the coordination number increases with distance following very smooth transition sequences due to the effect of temperature which will lead to overestimation of distance for a particular coordination number. Thus for any coordination number after any smooth transition, we have chosen its corresponding distance where the slope change of that transition occurs. With temperature, the number of sequential transitions decreases. Each Pb atom has 6 Pb atoms as first neighbors at 5.8, 5.9, and 5.9 \AA\ at 40, 180, and 300 K, respectively. At 40 K, 2-second neighbors and 10 third neighbors have seen at 7.9 and 8.3 \AA. Although the similar is observed at 180 K analyzed from slope change at 7.9 and 8.4 \AA, it's also important to note that these two transitions (first to second to third neighbors) are not distinct like 40 K, instead they almost merge into one transition towards 12-second neighbors. The 12-second neighbors at 300 K are seen at 8.4 \AA. In a perfect cubic condition, it should be at $\sqrt{2}$ of the first neighbor distance $i.e.$ at 8.34 \AA\ which is within the calculation error bar. From $G_c(r)\{Pb-Pb\}$ plot, it is observed that each Pb atom has 6 Br first neighbors at 3 \AA\ at all temperatures. This resembles a binary \textit{fcc} lattice where Pb and Br atoms sit at the corner and face-centers, respectively. For 40 K, 2-second, 10 third, and 12 fourth neighbors occur at 5.7 6.3 and 7 \AA, respectively. The same is seen at 180 K from the slope change at 5.8, 6.4, and 6.9 \AA, however, these three transitions are almost merged into one smooth transition to 24-second neighbors at 6.4 \AA. On the other hand, a clear transition to 24-second neighbors at 6.7 \AA\ is seen in the cubic phase at 300 K.

The radial distribution function corresponding to C or N atoms from Pb or Br is interesting to look at as those can reveal information about the extent of rotational degrees of freedom of methylamine cation in the Pb-Br cage. These are shown in Figure~\ref{rdf_C_N}. The sharp peaks at 40 K merge at higher temperatures. In fact, at 300 K, the first peak for $G(r){Pb-N}$ is non-symmetric due to this. The higher mobility of methylamine translates into the broadness of the peaks resulting in a disordered phase at 300 K. The first peak for Br-N pair distribution is sharper than Br-C, which is due to the hydrogen bonding between the N-Br donor-acceptor pair.

\section{Nature of guest/host coupling?}
\subsection{The octahedral scissoring mode and MA translations/rotations}
To calculate coupling strength between two parameters $x$ and $y$ varying with time $t$, Pearson correlation coefficients are calculated by the following equation
\begin{equation}
r=\frac{\sum_{t}\left(x_{t}-\bar{x}\right)\left(y_{t}-\bar{y}\right)}{\sqrt{\sum_{t}\left(x_{t}-\bar{x}\right)^{2} \sum_{t}\left(y_{t}-\bar{y}\right)^{2}}}
\end{equation}
where the averages are over time. The coupling strength varies from -1 (perfectly anti-correlated) to 1 (perfectly correlated). For example, the coupling between Br-Br scissoring in \textit{ac} plane ($x_t=\Omega_{ca}(t)$) and translation of MA in \textit{ac} plane ($y_t=d_{ca}^{MA}(t)$), indicated as $F_T^{ca}$, is calculated using the above described formula. 

\subsection{Low-frequency modes of host facilitate MA reorientation}
\begin{figure*}[tbh!]
 \centering
 \subfigure[]{\includegraphics[width=0.37\textheight]{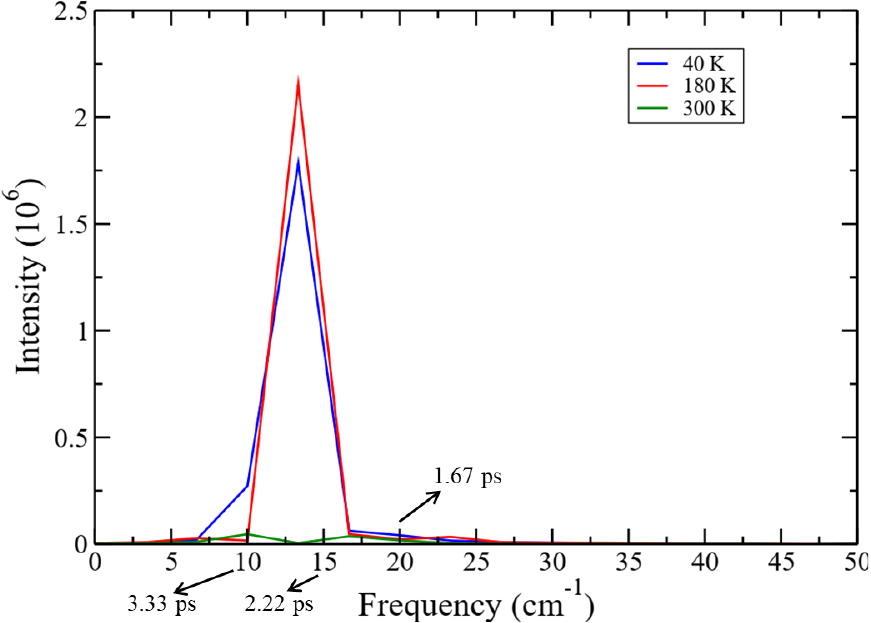}}\hspace{0.02\textwidth}  
 \subfigure[]{\includegraphics[width=0.37\textheight]{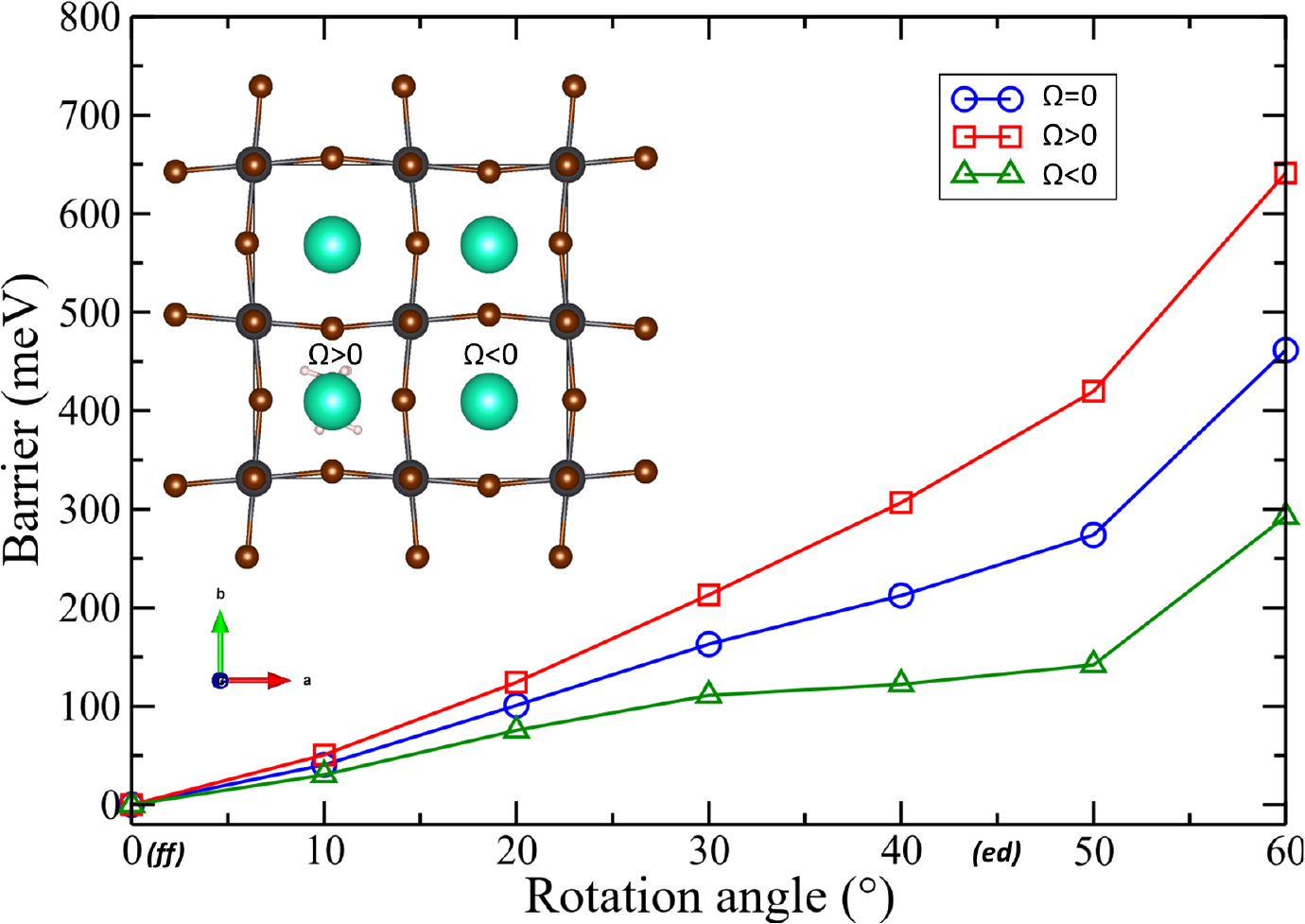}} 
 \caption{\footnotesize{(a) The power spectra of Br$\to$Br scissoring mode in ab plane ($\Omega_{ab}$) calculated at zone-center $\Gamma$(0,0,0) point. The arrows are guidelines for timescales in ps from which it can be inferred that the position of the peak matches well with MA reorientation timescale of $\sim$1.86 ps;   
 (b) MA rotational barrier in a cell without and with scissoring distortion. Inside, a 2$\times$2$\times$2 CsPbBr$_3$ supercell with one Cs replaced by MA is shown, where two situation $\Omega$$>$0 and $\Omega$$<$0 are indicated.}} 
 \label{scissor_freq}
\end{figure*}

The reorientation times of MA speed up along the series MAPbI$_3$ ($\sim$3.0 ps) to MAPbBr$_3$ ($\sim$1.5 ps) to MAPbI$_3$ ($\sim$1.2 ps)~\cite{selig2017organic}. It has been suggested that one of the reason could be a reduction in the energetic penalty for N-H$^{...}$X hydrogen bond elongation which brings down activation barrier due to decrease in unit-cell size~\cite{selig2017organic,gallop2018rotational}. It has been shown in MAPbI$_3$ by Carignano \textit{et al.} that barrier of approximately 13 meV/f.u. exists between $bd$ and $ed$ orientation which was calculated at room temperature by Boltzman inversion of MA orientation probability as a function of $\phi$ for 0.4 $\leq$ $|\cos\theta|$ $\leq$ 0.7~\cite{carignano2015thermal}. A similar barrier ($\sim$20 meV/f.u.) was found at 0 K by Motta \textit{et al.}~\cite{motta2015revealing} in MAPbI$_3$. Our calculated barrier for MAPbBr$_3$  is $\sim$16.9 meV/f.u  at 300 K and $\sim$24 meV/f.u. at 0 K indicating that barrier height remains similar from MAPbI$_3$ to MAPbBr$_3$ and, hence,  does not correlate well with the drastic change of their corresponding MA reorientation time. Thus we discard the aforementioned hypothesis. However, deviations from cubicity during the dynamics at 300 K can induce dynamic changes in barrier. Hence, the reorientation of MA would be influenced particularly by low-frequency deformation modes of inorganic-lattice. It is well known that most lattice vibrational frequencies increase along the series MAPbI$_3$ to MAPbBr$_3$ to MAPbCl$_3$~\cite{niemann2016halogen,leguy2016dynamic} which correlates well with faster MA reorientation along the series. In fact, in MAPbBr$_3$ we find that the scissoring distortions at 300 K (coupled to MA) has timescale (see Figure~\ref{scissor_freq}(a)) in resonance with calculated MA reorientation time of $\sim$1.86 ps. Assuming this type of lattice-MA coupling exists in other MAPbX$_3$, which should be verified, we can say that MA reorientation timescales are governed by lattice scissoring frequency.

\begin{table}[]
\begin{tabular}{|c|c|c|c|c|c|c|c|c|c|}
\hline
Temperature  & aa     & ab     & ac     & ba     & bb     & bc     & ca     & cb     & cc    \\
\hline
40 K  & 0.3529 & 0.3248  & 0.3048  & -0.0264 & 0.8333 & -0.0256 & 0.3507 & 0.3055 & 0.3689 \\
180 K & 0.3401 & 0.1939 & 0.1905  & 0.0263 & 0.4002 & 0.1696 & 0.1916 & 0.113 & 0.3349 \\
300 K & 0.0233 & 0.1314 & 0.051 & -0.0362  & 0.3248 & 0.0105 & -0.1678  & -0.0859 & 0.2124 \\
\hline
\end{tabular}
\caption{\footnotesize{C$^{kl}_T$ coupling between Pb-Br bond lengths and MA translation with temperature. In the table the k and l indices are mentioned.}}
  \label{octa_dis_T}
\end{table}


\begin{table}[]
\begin{tabular}{|c|c|c|c|c|c|c|c|c|c|}
\hline
Temperature  & aa     & ab     & ac     & ba     & bb     & bc     & ca     & cb     & cc    \\
\hline
40 K  & -0.025 & 0.0502 & -0.072 & -0.02  & 0.086  & -0.035 & -0.147 & 0.0482 & 0.081  \\
180 K & -0.008 & 0.1537 & -0.027 & -0.018 & 0.0869 & 0.1094 & 0.0098 & -0.008 & -0.017 \\
300 K & -0.099 & 0.1524 & 0.1469 & 0.0077 & -0.149 & -0.07  & 0.0648 & -0.042 & -0.09  \\
\hline
\end{tabular}
\caption{\footnotesize{C$^{kl}_{R}$ coupling between Pb-Br bond lengths and MA rotations with temperature. In the table the k and l indices are mentioned.}}
  \label{octa_dis_R4}
\end{table}

\begin{table}[]
\begin{tabular}{|c|c|c|c|}
\hline
Temperature	& ab & bc & ca \\
\hline
40 K & 0.342 & -0.270 & 0.924 \\
180	K &	0.403 &	0.089 &	0.685 \\
300	K &	0.621 &	0.727 &	0.548 \\
\hline
\end{tabular}
\caption{\footnotesize{F$^{kl}_T$ coupling between octahedral scissoring ($\Omega_{kl}$) and MA translation component parallel to the $k+l$ direction with temperature. In the table the k and l indices are mentioned.}}
  \label{scissor_T}
\end{table}

\begin{figure*}[tbh!]
 \centering
 \subfigure[]{\includegraphics[width=0.4\textwidth]{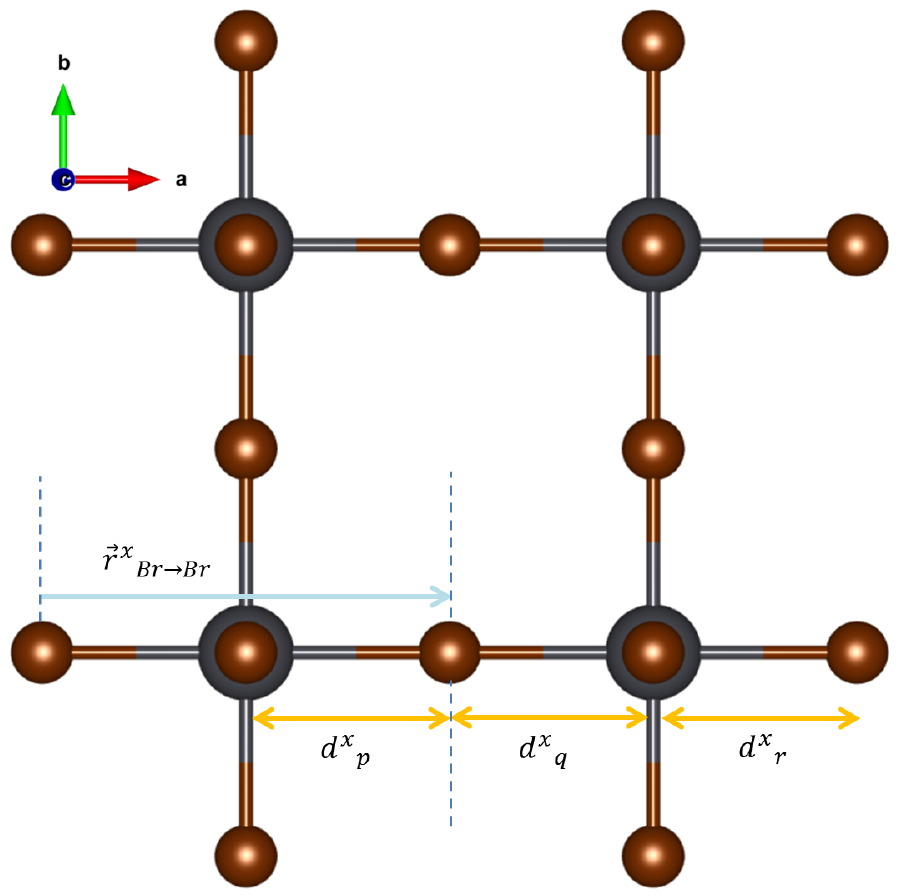}} 
 \subfigure[]{\includegraphics[width=0.56\textwidth]{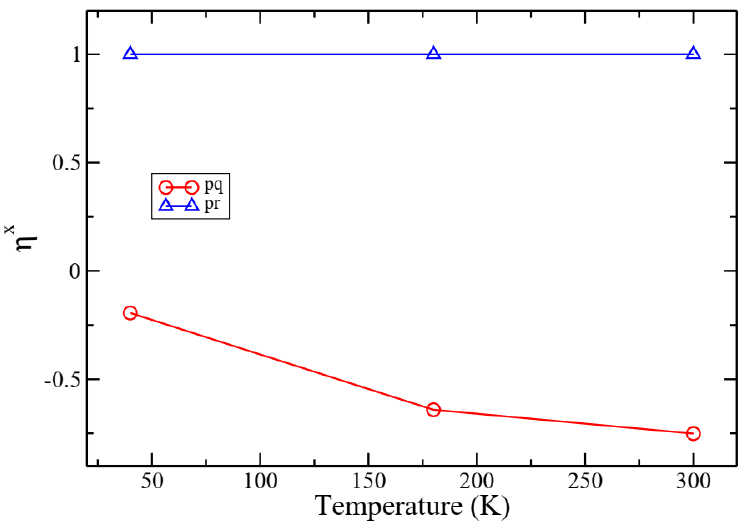}} 
 \caption{\footnotesize{ (a) Along $x$ direction Br-Br vector for characterizing octahedral rotation and different Pb-Br bond (p, q and r) lengths; (b) Correlation between p,q and q,r bond lengths along x direction with temperature.}} 
 \label{schematic_eta}
\end{figure*}

\begin{table}[hbt!]
\begin{tabular}{|c|c|c|c|}
\hline
Temperature  & x        & y        & z        \\
\hline
40 K  & -0.19366 & 0.392776 & -0.08023 \\
180 K & -0.64096 & -0.49703 & -0.61565 \\
300 K & -0.75108 & -0.75751 & -0.62608 \\
\hline
\end{tabular}
\caption{\footnotesize{$\eta^{n}_{pq}$ correlation strength with temperature. Here n indicates direction of Pb-Br bonds.}}
  \label{eta_pq}
\end{table}

\begin{table}[hbt!]
\begin{tabular}{|c|c|c|c|}
\hline
Temperature  & x        & y        & z        \\
\hline
40 K                     & 1.0 & 1.0 & 1.0 \\
180 K                    & 1.0 & 1.0 & 1.0 \\
300 K                    & 1.0 & 1.0 & 1.0 \\
\hline
\end{tabular}
\caption{\footnotesize{$\eta^{n}_{pr}$ correlation strength with temperature. Here n indicates direction of Pb-Br bonds.}}
  \label{eta_pr}
\end{table}

\subsection{Role and dynamics of H-bonds}

\begin{figure*}[tbh!]
 \centering
 \subfigure[]{\includegraphics[width=0.33\textwidth]{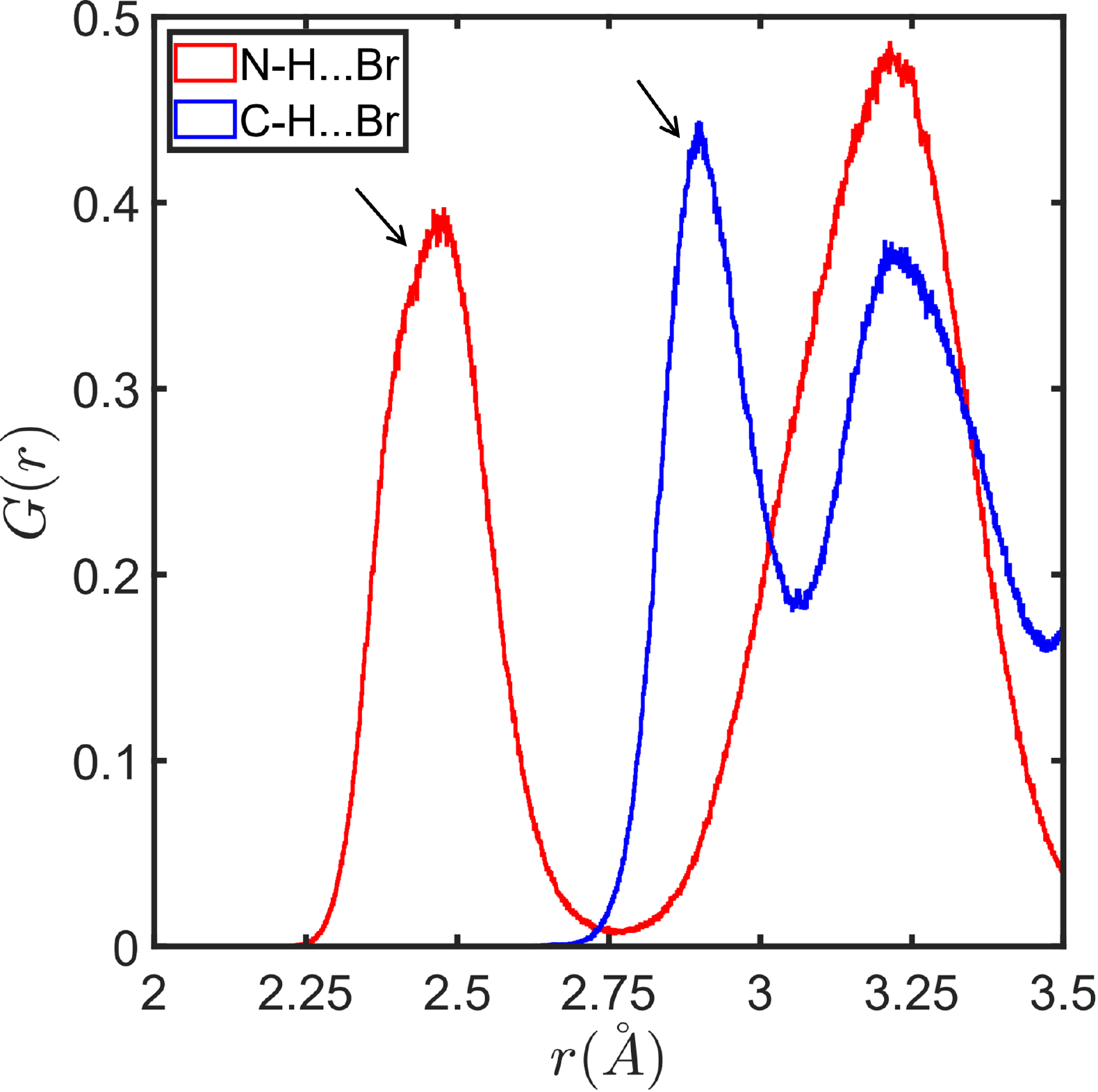}} 
 \subfigure[]{\includegraphics[width=0.33\textwidth]{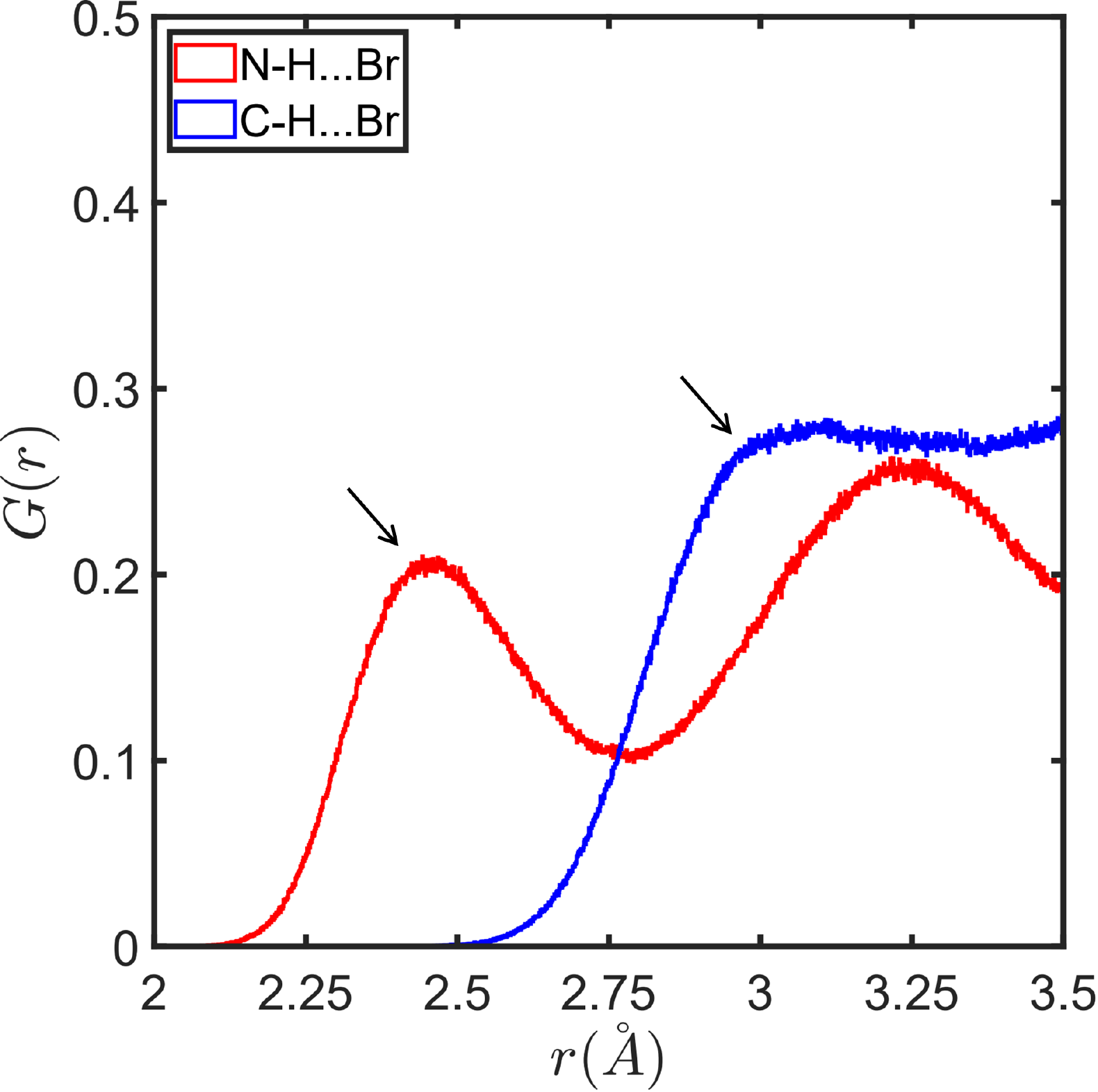}} 
 \subfigure[]{\includegraphics[width=0.33\textwidth]{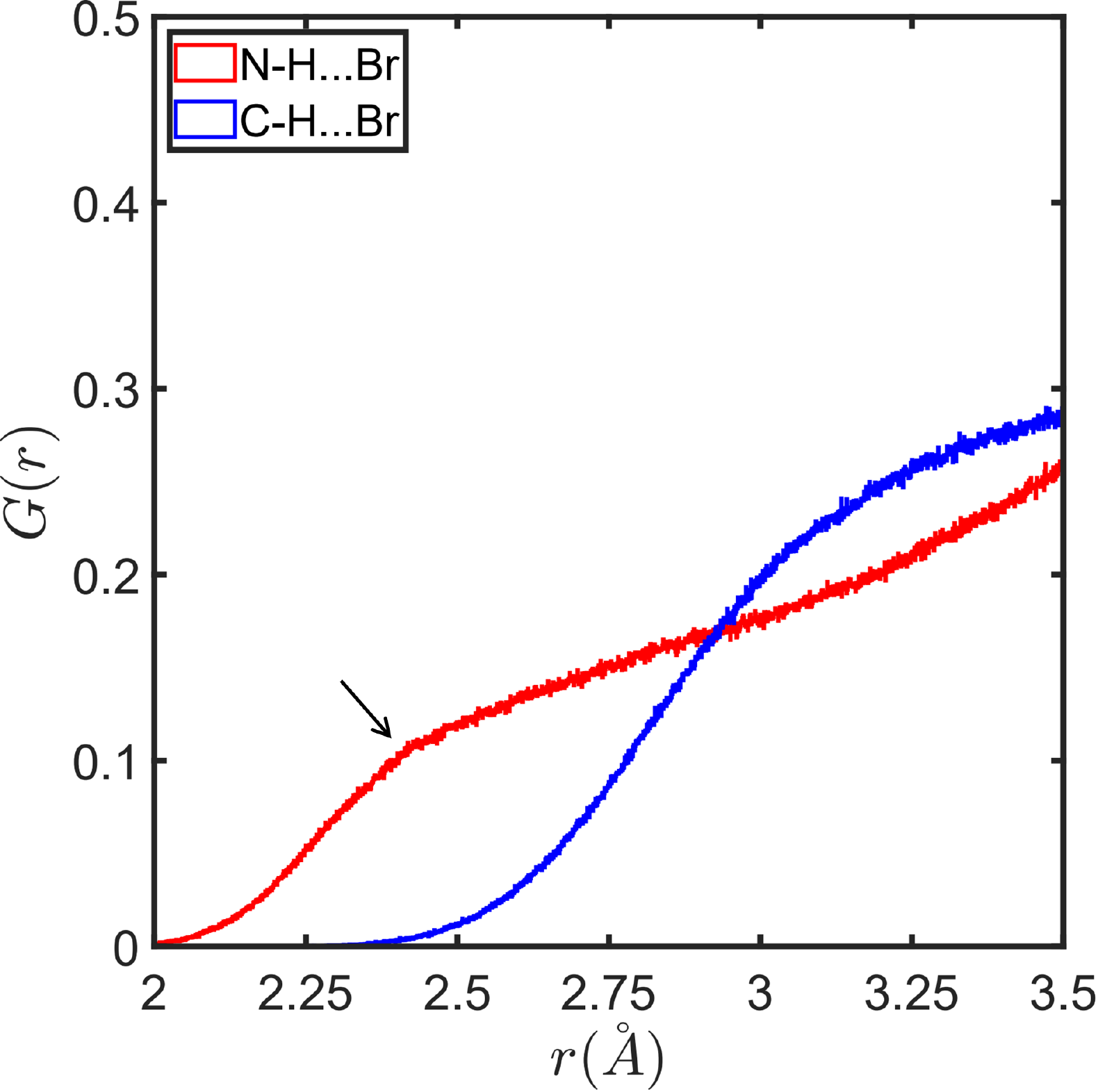}}
 \caption{\footnotesize{Radial distribution functions of N-H$^{...}$Br and C-H$^{...}$Br distances within a pseudo-cubic unit cell at (a) 40 K, (b) 180 K and, (c) 300 K. The arrows indicate the peaks where H-bonding interactions are maximum.}} 
 \label{rdf_H_Br}
\end{figure*}

Bernasconi \textit{et al.} suggested that distortion of octahedral in MAPbCl$_3$ is correlated to presence of strong H-bonds~\cite{bernasconi2018ubiquitous}. 
MD simulations revealed that H-bond lengths are temperature dependent in MAPbCl$_3$ (N-H$^{...}$Cl =2.2~\AA~at low temperature and 2.45~\AA~at high temperature) which is against the results of NPDF (neutron pair distribution function) experiments where the peak corresponding to the aforementioned equilibrium H-bond distance at 2.5~${\rm \AA}$ never shifts with temperature in both MAPbCl$_3$ and MAPbBr$_3$~\cite{bernasconi2018ubiquitous}.  In our simulations of MAPBBr$_3$ indeed we find peak-like feature at $\sim$2.46~\AA~in N-H$^{...}$Br PDF for all unit-cells irrespective of simulation temperatures (40, 180 and 300 K) (see Figure~\ref{rdf_H_Br}) agreeing with the experiment. We attribute the mismatch in the MD results of MAPbCl$_3$ to the usage of constant volume condition where cells were fixed which we allow to relax during our simulations. It is also interesting to note that same H-bond lengths are observed in different MAPbX$_3$ (X=I,Br,Cl) OIHPs.

\begin{figure*}[tbh!]
 \centering
 \subfigure[]{\includegraphics[width=0.37\textheight]{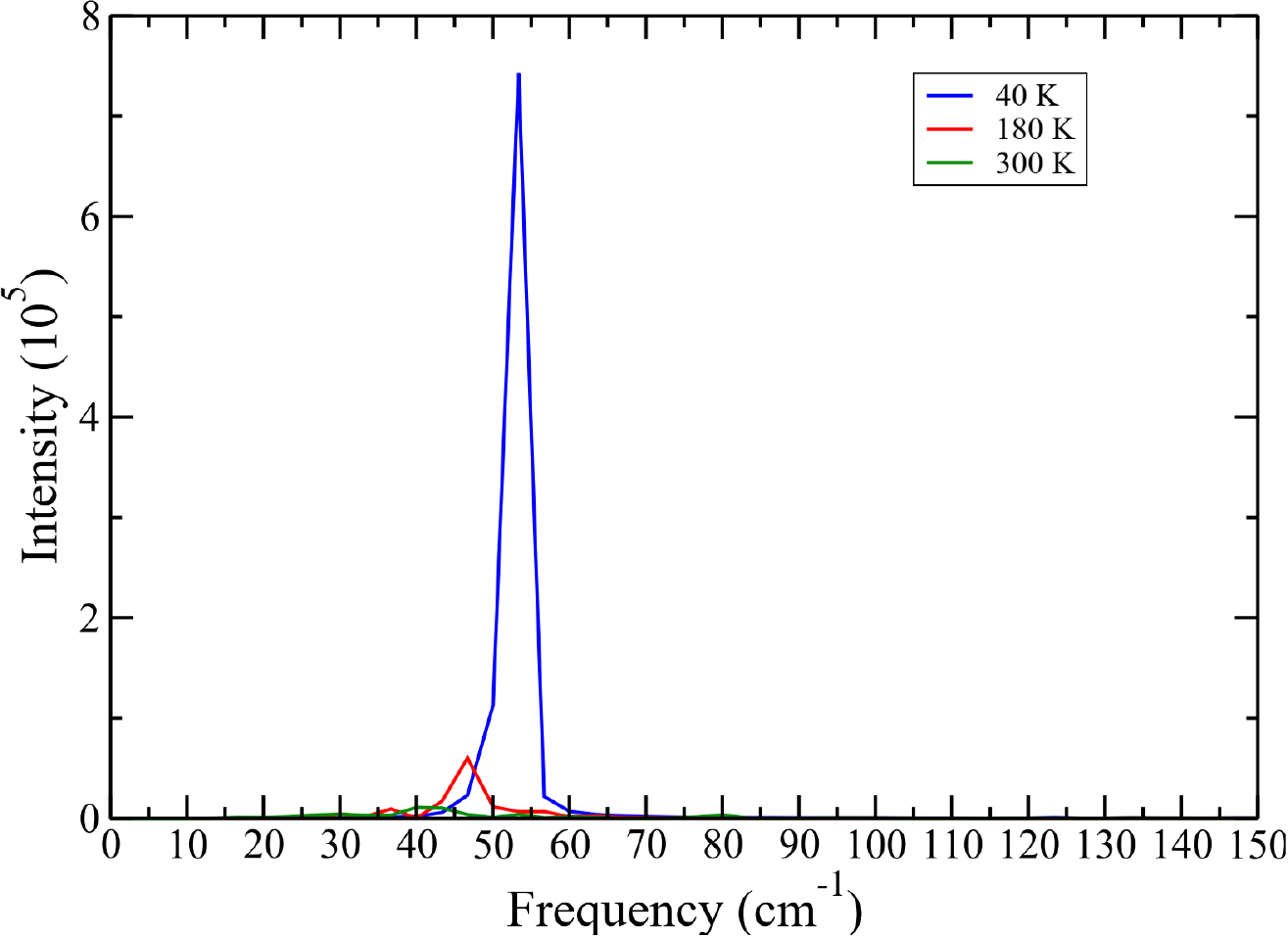}} 
 \subfigure[]{\includegraphics[width=0.38\textheight]{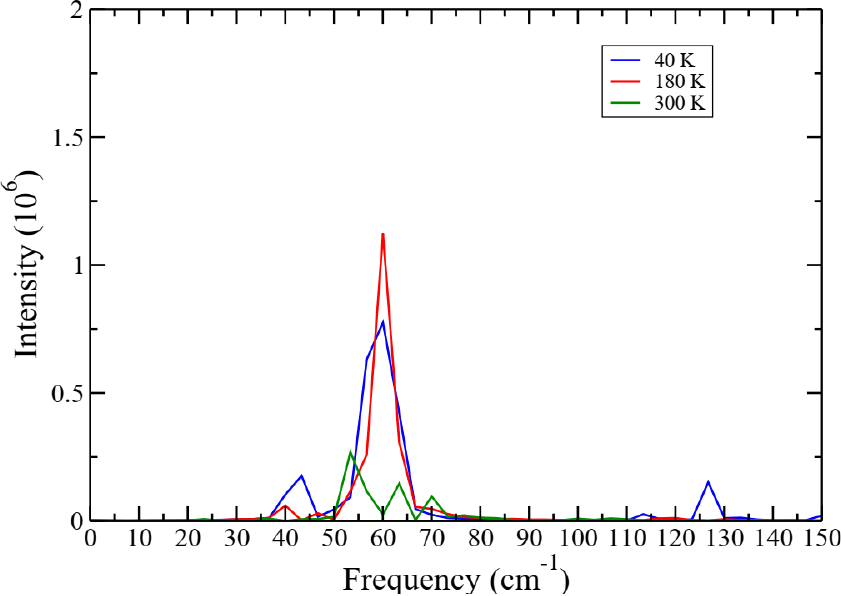}} 
 \subfigure[]{\includegraphics[width=0.37\textheight]{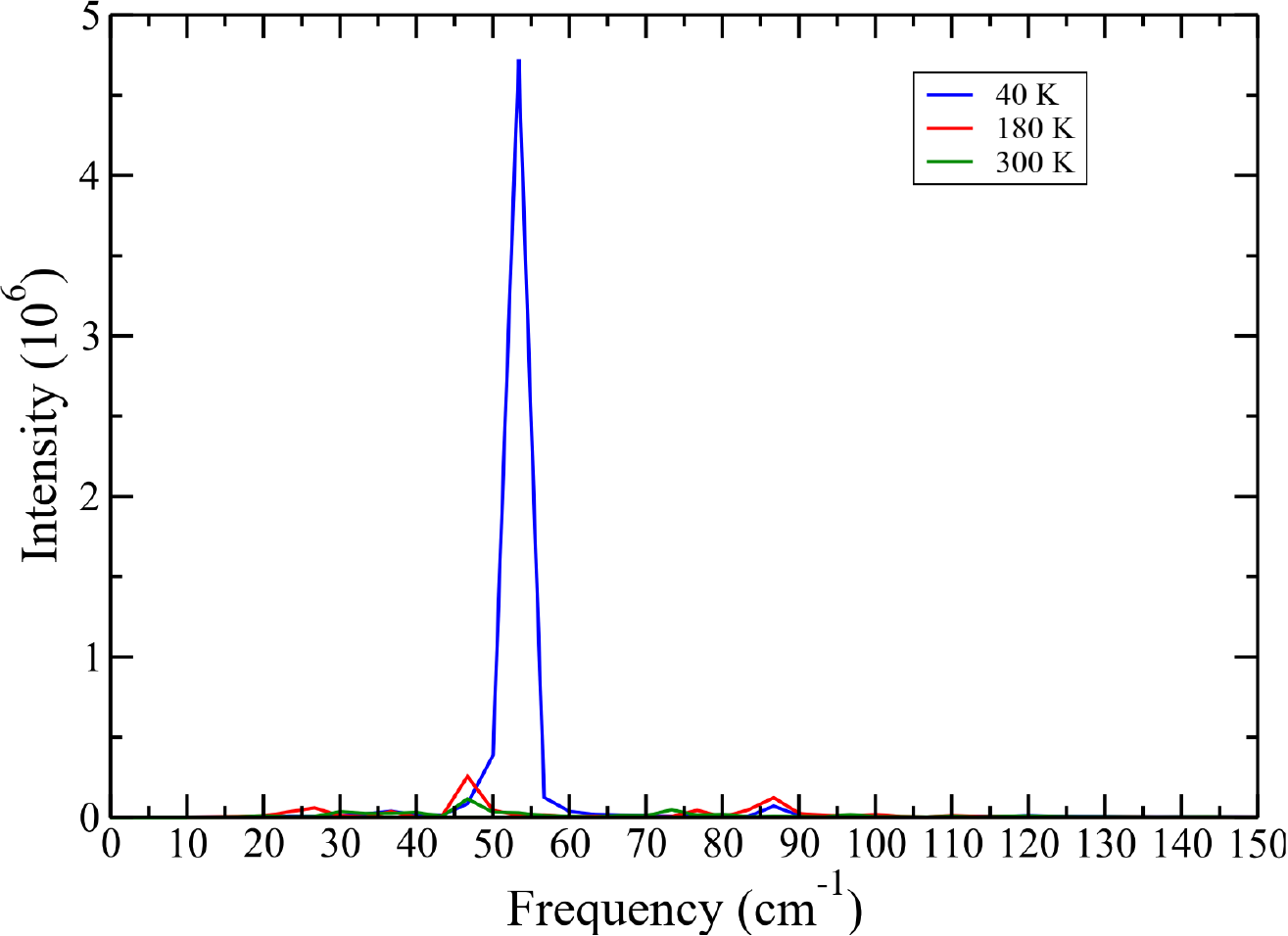}} 
 \subfigure[]{\includegraphics[width=0.38\textheight]{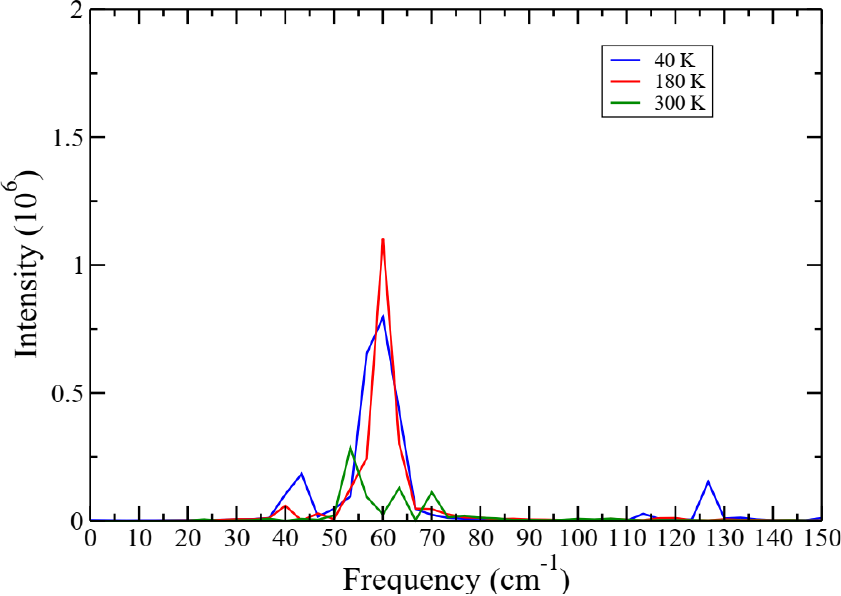}} 
 \caption{\footnotesize{The power spectra of scissoring mode in ca plane calculated using $\Omega_{ca}$ at zone-boundary (a) $M$(1/2,1/2,0) and (b) $R$ (1/2,1/2,1/2) points; The power spectra of scissoring mode in ca plane calculated using $\Omega^{\prime}_{ca}$ at zone-boundary (c) $M$(1/2,1/2,0) and (d) $R$ (1/2,1/2,1/2) points; It can be seen that using both method almost same spectra is obtained.}} 
 \label{scissor_method_comparison}
\end{figure*}

\begin{figure*}[tbh!]
 \centering
 \subfigure[]{\includegraphics[width=0.37\textheight]{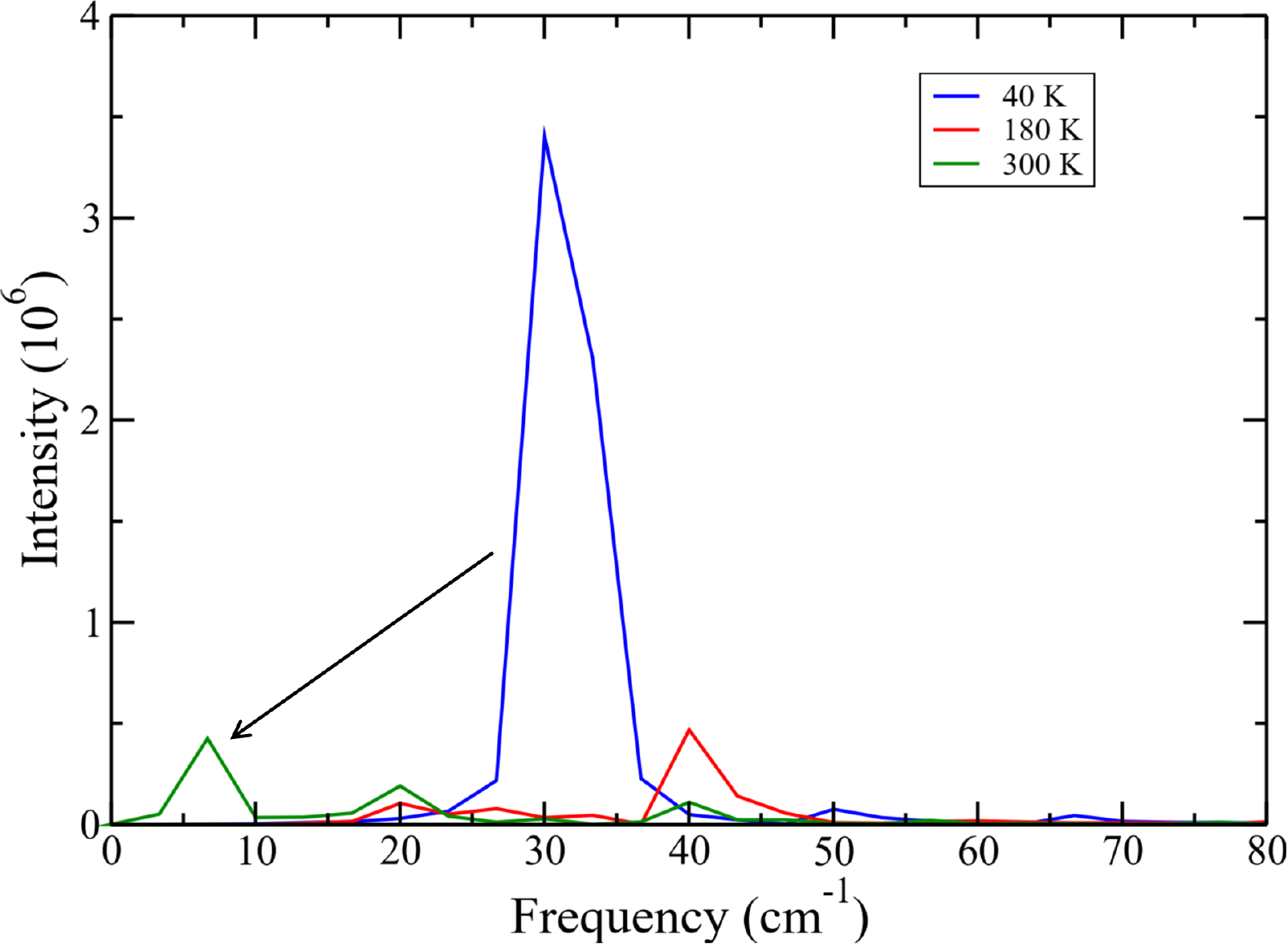}} 
 \subfigure[]{\includegraphics[width=0.37\textheight]{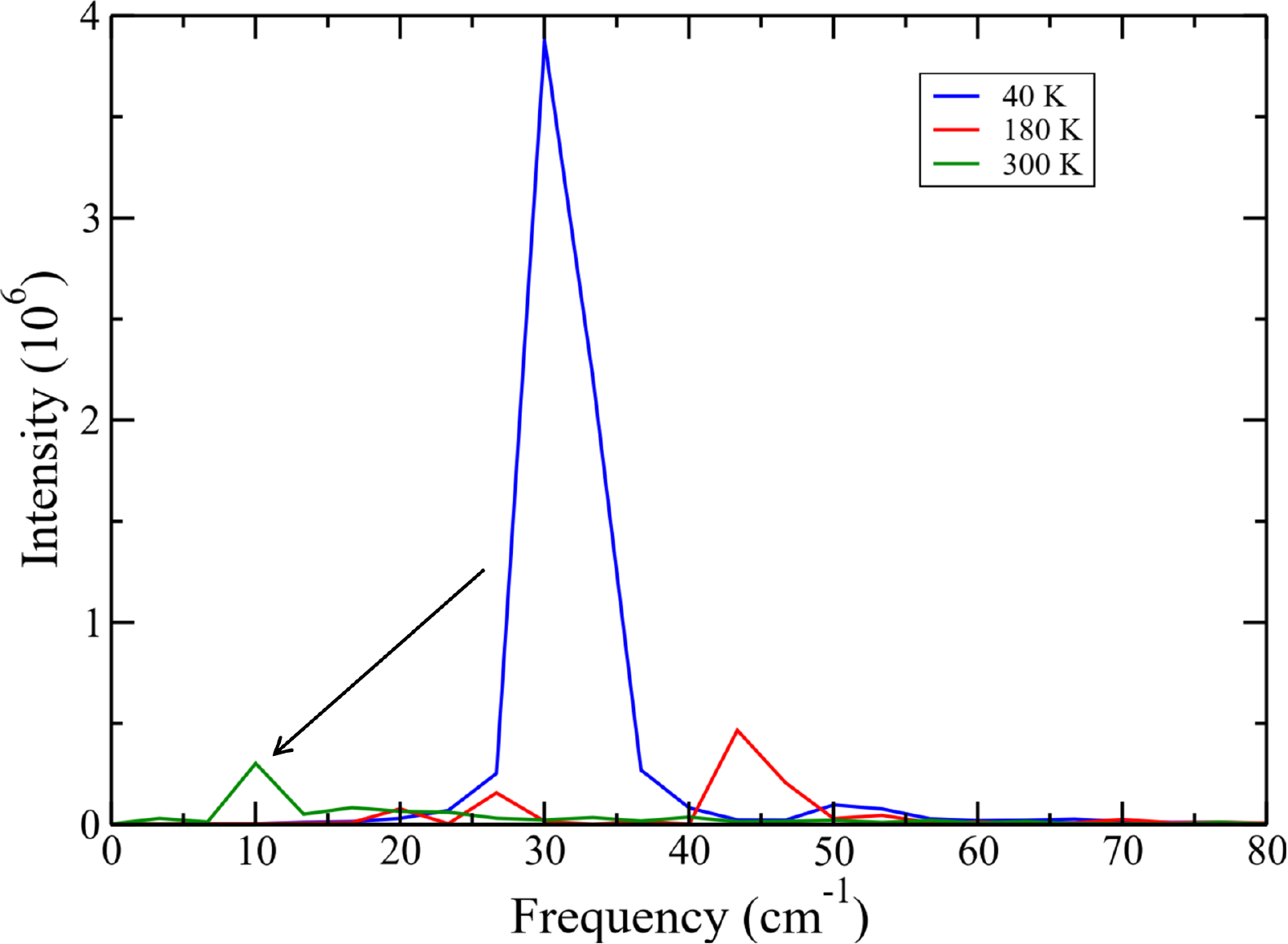}} 
 \subfigure[]{\includegraphics[width=0.37\textheight]{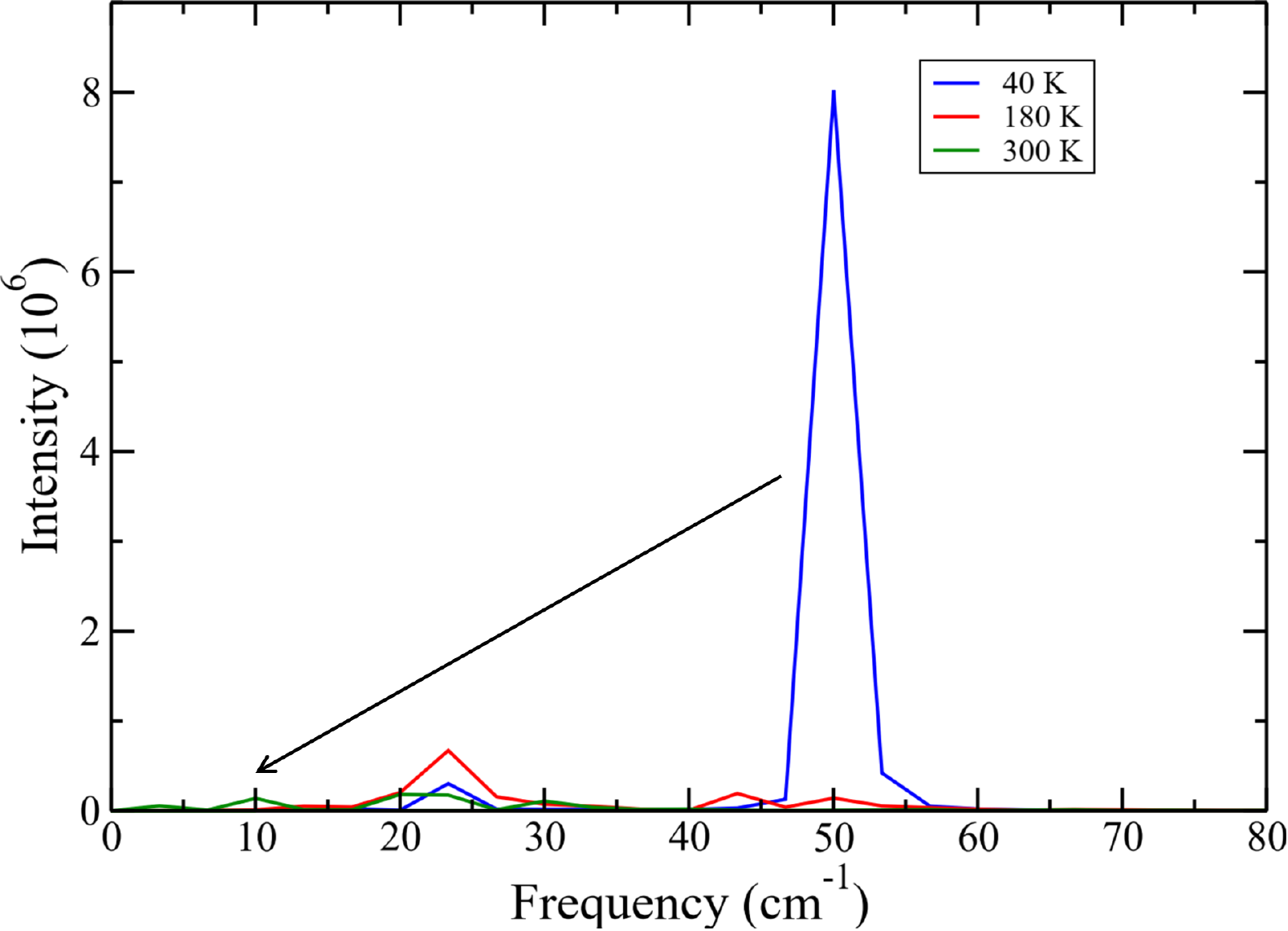}} 
 \caption{\footnotesize{The power spectra of rocking mode calculated at zone-boundary $R$ (1/2,1/2,1/2) point in (a) ab ($\xi_{ab}$); (b) bc ($\xi_{bc}$) and (c) ca ($\xi_{ca}$) plane.}} 
 \label{rocking_freq}
\end{figure*}

\section{Nature of phase transition?}
Apart from contribution of MA orientational degrees of freedom (order-disorder) in phase-transition mechanism we also investigate existence of displacive character. To this end the power spectra of scissoring ($\Omega_{\mu\nu}$) and rocking ($\xi_{\mu\nu}$) modes are calculated and shown in Figure~\ref{scissor_method_comparison} and~\ref{rocking_freq}. The power spectra of any such function is obtained by doing a time Fourier transform of velocity-autocorrelation at a $\vec{q}$-point ($\langle\dot{f}(\vec{q},t)^{*}\cdot\dot{f}(\vec{q},t+\tau)\rangle_{\omega}$) where -
\begin{equation}
\dot{f}(\vec{q},t)=\sum_{\vec{r}}\dot{f}(\vec{r},t)e^{-i\vec{q}\cdot\vec{r}}
\end{equation}
Here $\vec{r}$ refers to lattice points and the spatial Fourier transform is normalized by the total number of lattice points which is 4$\times$4$\times$4. From the softening of rocking modes at R point existence of displacive character is confirmed (Figure~\ref{rocking_freq}). 

\clearpage

\bibliography{reference}